\documentclass[aps,prx,reprint,superscriptaddress]{revtex4-1}

	\usepackage{amsmath}
	\usepackage{makeidx}
	\usepackage{amsfonts}
	\usepackage[ansinew]{inputenc}
	\usepackage[usenames,dvipsnames]{pstricks}
	\usepackage{subfigure}
	\usepackage{multirow}
	\usepackage{array}
	\usepackage{epsfig}
	\usepackage{tabularx}
	\usepackage{pst-grad}
	\usepackage{pst-plot}
	\usepackage[colorlinks,hyperindex]{hyperref}
	\hypersetup
	{
		colorlinks,%
		citecolor=black,%
		linkcolor=black,%
		urlcolor=black,%
	}

	\renewcommand{\vec}[1]{  \boldsymbol{\mathbf{#1}}  }
	\newcommand{\real}[1]{\mathrm{Re}\,#1}
	\newcommand{\imag}[1]{\mathrm{Im}\,#1}
	\newcommand{\diag}[1]{\mathrm{diag}\{#1\}}
	\newcommand{\Ams}{\mathrm{\AA}}

	\newcommand{\kk}{\mathbf{k}}
	\newcommand{\qq}{\mathbf{q}}
	\newcommand{\QQ}{\mathbf{Q}}
	\newcommand{\KK}{\mathbf{K}}
	
	\newcommand{\rr}{\mathbf{r}}
	\newcommand{\RR}{\mathbf{R}}

	\newcommand{\abs}[1]{\left| #1 \right|}

\makeindex

\begin{document}

\title{Hybrid $\kk\cdot\vec{p}$-tight binding model for subbands and infrared intersubband optics in few-layer films of transition metal dichalcogenides: MoS$_2$, MoSe$_2$, WS$_2$, and WSe$_2$}

\author{David A. Ruiz-Tijerina}
\affiliation{National Graphene Institute, University of Manchester, Booth St E, Manchester M13 9PL, UK}

\author{Mark Danovich}
\affiliation{National Graphene Institute, University of Manchester, Booth St E, Manchester M13 9PL, UK}

\author{Celal Yelgel}
\affiliation{National Graphene Institute, University of Manchester, Booth St E, Manchester M13 9PL, UK}

\author{Viktor Z\'olyomi}
\affiliation{National Graphene Institute, University of Manchester, Booth St E, Manchester M13 9PL, UK}

\author{Vladimir I.\ Fal'ko}
\affiliation{National Graphene Institute, University of Manchester, Booth St E, Manchester M13 9PL, UK}

\date{\today}

\begin{abstract}
We present a density functional theory parametrized hybrid k$\cdot$p tight binding model for electronic properties of atomically thin films of transition-metal dichalcogenides, 2H-$MX_2$ ($M$=Mo, W; $X$=S, Se).  We use this model to analyze intersubband transitions in $p$- and $n$-doped $2{\rm H}-MX_2$ films and predict the line shapes of the intersubband excitations, determined by the subband-dependent two-dimensional electron and hole masses, as well as excitation lifetimes due to emission and absorption of optical phonons. We find that the intersubband spectra of atomically thin films of the 2H-${MX_2}$ family with thicknesses of $N=2$ to $7$ layers densely cover the infrared spectral range of wavelengths between $2$ and $30\ {\rm \mu m}$. The detailed analysis presented in this paper shows that for thin $n$-doped films, the electronic dispersion and spin-valley degeneracy of the lowest-energy subbands oscillate between odd and even number of layers, which may also offer interesting opportunities for quantum Hall effect studies in these systems.
\end{abstract}

\maketitle

\section{Introduction}\label{sec:introduction}

The 2H-${MX_2}$ transition metal dichalcogenide compounds ($M$=Mo, W; $X$=S, Se)  are layered materials, where chalcogens and metal atoms form covalent bonds within two-dimensional (2D) layers with hexagonal lattice structure, and neighboring layers couple weakly through electrical quadrupole and van der Waals interactions.
This feature of chemical bonding makes atomically thin films of ${MX_2}$ sufficiently stable for extensive experimental studies aimed at
their implementation in various optoelectronic devices \cite{jariwala_applications, tmds_applications, hetro_tmds_bands}. In those recent studies, the closest attention has been paid to the inter-band optical properties of the monolayer transition-metal dichalcogenide (TMD) crystals \cite{wang_tmds_props,wangyao_review,mak_optoelectronics_tmds,tmds_light_matter}, due to their direct band 
gap \cite{mak_prl_2010}, valley-spin coupling \cite{pseudospin_control,breaking_valley_mose2,spinvalleystudy}, and long spin and valley memory of photo-excited carriers \cite{longlived}, spiced up by the Berry curvature effects for electrons and excitons in these two-dimensional semiconductors \cite{berryexcitons, pccp2016}. This is because in monolayer ${\rm MoS_2, MoSe_2, WS_2}$, and ${\rm WSe_2}$ the valence and conduction band edges both appear at the Brillouin zone (BZ) corners $K$ and $K'$, where the electronic Bloch states carry intrinsic angular momentum. 
\begin{figure}[!t]
	\begin{center}
		\includegraphics[width=1\columnwidth]{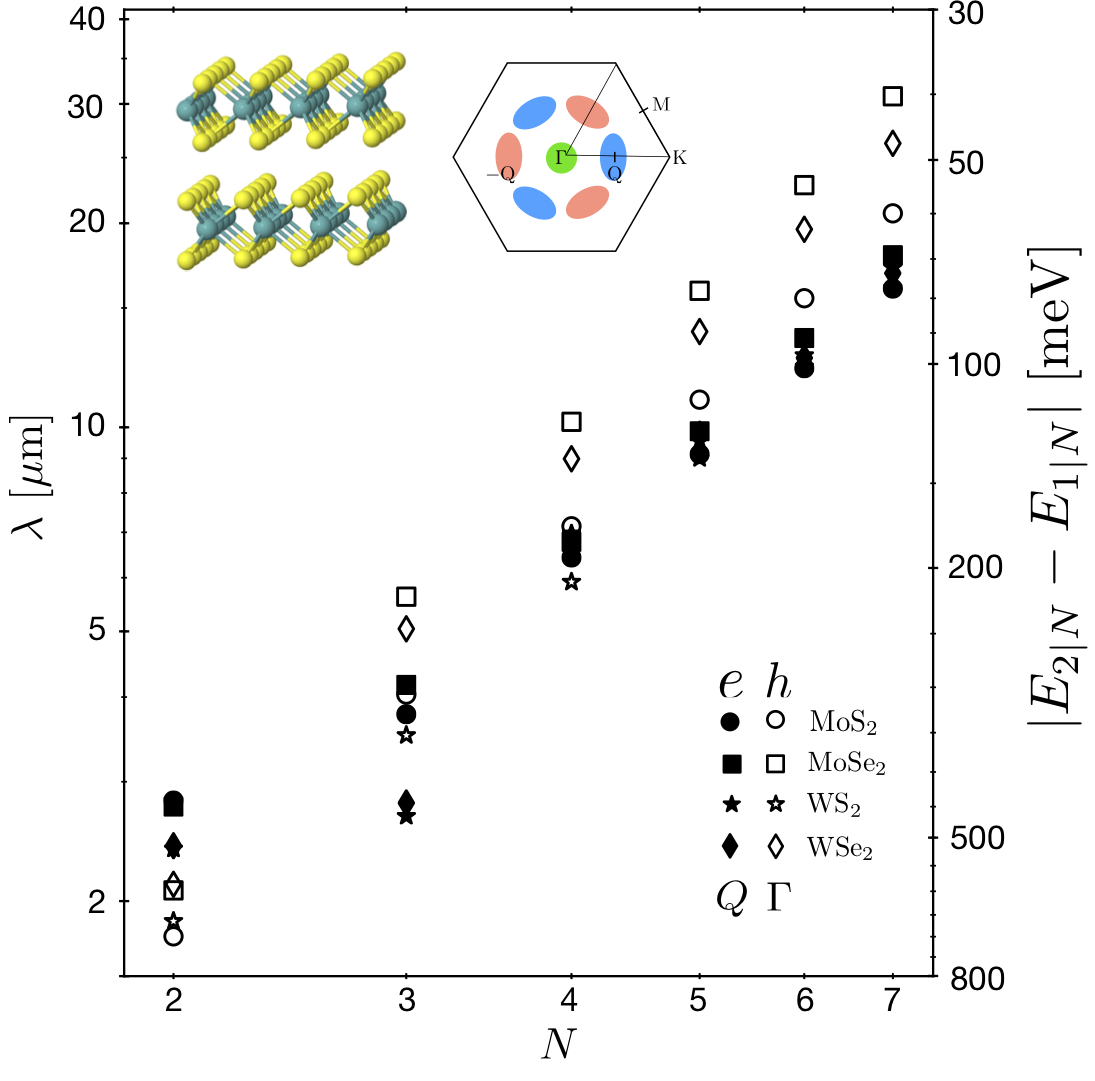}
		\caption{Energy spacings between the first two conduction (filled symbols, solid lines) and valence (empty symbols, dashed lines) subbands $1|N$ and $2|N$ as a function of number of layers $N$ of the four TMDs, corresponding to $n$ and $p$ doping, respectively, for $2\le N \le 7$. Both axes are in log scale, showing the approximate quadratic dependence of the spacings on the number of layers.
			The left vertical axis shows the wavelength $\lambda$ in ${\rm \mu m}$, corresponding to the energy spacings shown along the right vertical axis in ${\rm meV}$. The inset shows the building block of 2H-${MX_2}$ bulk crystals, composed of two monolayers with metal atoms in the middle and  chalcogens in the outer sublayers of each monolayer, and the Brillouin zone with the $\Gamma$ point and $Q$ valleys highlighted, corresponding to the conduction and valence band edges.}
		\label{fig:figure1_spacings}
	\end{center}
\end{figure}

Thicker crystals of 2H-${MX_2}$ quickly lose the direct band gap property upon increasing the film thickness to two or three layers \cite{lambrecht_prb_2012,cappelluti_prb_2013,debbichi_prb_2014,padilha_prb_2014,chang_scirep_2014,abinitioTB_prb_2015,bradley_nanolett_2015,sun_jchemphys_2016}. Density functional theory (DFT) of few-layer transition metal dichalcogenides predicts \cite{padilha_prb_2014,abinitioTB_prb_2015} that for holes the band edge relocates to the $\Gamma$ point, whereas for electrons it appears at six points situated somewhere near the $Q$ points at the middle of each $\overline{\Gamma K}$ segment (see inset in Fig.\ \ref{fig:figure1_spacings}). This has been demonstrated by studies of Shubnikov--de Haas oscillations in $n$-doped ${\rm MoS_2}$ \cite{shubnikov}. While the indirect character of few-layer TMD band structures suppresses their inter-band photo response, the multiplicity of subbands $n|N$ ($1\le n\le N$) at the conduction and valence band edges of the $N$-layer crystal, open a new avenue for optical studies of atomically thin TMD films.
 
 Here, we analyze theoretically intersubband transitions in few-layer ${\rm MoS_2, MoSe_2, WS_2}$ and ${\rm WSe_2}$, and show that the absorption/emission spectra of the primary transitions in $p$- and $n$-doped crystals (Fig.\ \ref{fig:figure1_spacings}) densely cover the infrared (IR)  spectrum down to the far-infrared range (FIR) of photon energies.  The analysis of subband properties of few-layer 2H-${MX_2}$ presented in this paper is based on the hybrid k$\cdot$p theory tight-binding model (HkpTB) approach, recently applied to the description of multilayer films of post-transitional-metal chalcogenides (such as ${\rm InSe}$ and  ${\rm GaSe}$) \cite{sam_inse, bandurin}. This approach consists of minimal $\kk\cdot {\bf p}$ theory Hamiltonians for 2H-${MX_2}$ monolayers \cite{kormanyos_prb_2013,kdotp}, supplemented by a $\kk \cdot {\bf p}$ expansion of the interlayer hopping near the relevant point (here, $\Gamma$ or $Q$) in the BZ, with all parameters fitted to DFT calculated few-layer dispersions, and $k_z$ dispersions in bulk crystals.

First, in Section \ref{sec:struct} we describe the lattice structure and discuss symmetries of few-layer 2D crystals of 2H-$MX_2$, especially the difference between films with odd and even numbers of layers and the corresponding degeneracies in their band structures. The DFT-parametrized HkpTB models for few-layer TMDs are formulated in Sections \ref{sec:Gmodel} and \ref{sec:Qmodel} for the valence band edge (holes) near the $\Gamma$ point and for conduction band (electrons) near the $Q$ points, respectively. 

In the case of $n$-doped films, our models predict a multi-valley subband structure with two valley triads connected by time-reversal symmetry, each consisting of three valleys related by $C_3$ rotations. We find that the lowest-energy subbands alternate between spin-split and spin-degenerate with number of layers $N$, consequence of the $\sigma_h$ mirror and full inversion symmetries of films with odd and even $N$, respectively. These band structure features open a variety of new possibilities for studies of quantum Hall physics in multi-layer TMD films.

We calculate subband energies, dispersions, and wave functions of electron/hole subbands in $N$-layer crystals of all four 2H-${MX_2}$ compounds, and optical oscillator strengths for radiative intersubband transitions. In particular, we take into account $n$- and $p$-type doping using a self-consistent analysis of charge and potential distributions across the film. Then, we analyze the inelastic broadening due to optical phonon emission, and the resulting spectral line shapes of IR/FIR absorption by $p$- and $n$-doped 2H-${MX_2}$ films. We find that the intersubband relaxation rates, determined by electron-phonon interactions, are much slower (one to two orders of magnitude) than the intra-subband relaxation in the same materials \cite{danovich_phonons,relax_exp}, and also these are an order of magnitude slower than intersubband relaxation of electrons and holes in III-V semiconductor quantum wells \cite{qw1}. Also, in Section \ref{sec:VBselectionrules} we show that the difference between  the two-dimensional masses $m_{1|N}$ and $m_{2|N}$ of electrons and holes in consecutive subbands leads to an additional temperature-dependent broadening of the intersubband transitions, $\sigma \sim \left|1-\frac{m_{1|N}}{m_{2|N}}\right|\max\{k_BT,\epsilon_F\}$, which appears to be the dominant intrinsic broadening effect for the IR/FIR absorption by 2H-${MX_2}$ films at room temperature.

\begin{figure}[!t]
	\begin{center}
\includegraphics[width=0.9\columnwidth]{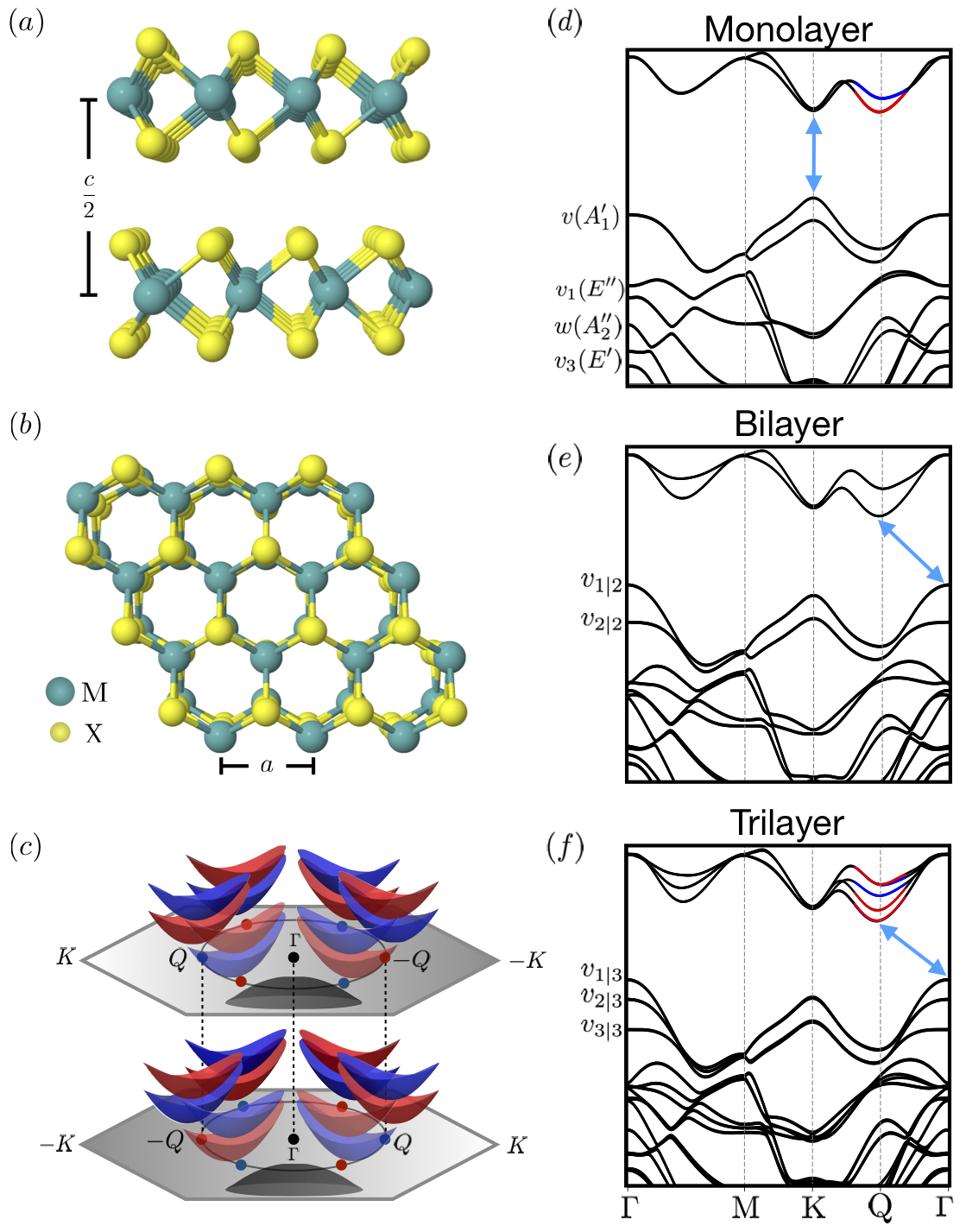}
		\caption{(a) Crystal structure of a bilayer 2H-stacked TMD ($MX_2$), the building block of multilayer TMDs, viewed from the side and (b) top. $M$ and $X$ represent the metal and chalcogen atoms, respectively. (c) Brillouin zones of the two monolayers in the 2H-stacked bilayer rotated by $180^{\circ}$ relative to each other, and schematics of the band dispersions at the valence band $\Gamma$ point and conduction band $Q$ point, including the six symmetry-related $Q$ valleys and spin-orbit splitting.
			(d)--(f) DFT band structures of monolayer, bilayer, and trilayer ${\rm WS_2}$, showing the transition from direct-gap ($K-K$)  monolayer, to indirect-gap ($\Gamma-Q$) multilayer semiconductor. At the $\Gamma$ point, we label the different valence bands in the monolayer and the irreducible representations of the $D_{3h}$ point group \cite{kormanyos_prb_2013}. The bilayer and trilayer valence subbands are further labeled according to the subscript notation $n|N$, where $n$ is the subband number and $N$ the number of layers.
			For odd number of layers ($N=1,3$), where spin-orbit splitting is present, the conduction subbands near the $Q$ point are colored according to the spin projection quantum number, with red (blue) corresponding to $s=\uparrow$ $(s=\downarrow)$, giving a total of $2N$ spin-polarized states. Kramer's doublets are given by $E_s(\kk)=E_{-s}(-\kk)$. For $N=2$ all bands are spin degenerate, resulting in $N$ doubly degenerate states, $E_s(\kk)=E_{-s}(\kk)$.
		}
		\label{fig:figure1}
	\end{center}
\end{figure}

\section{Multilayers of hexagonal transition metal dichalcogenides: overview}
\label{sec:struct}
The lattice structure of monolayer TMDs ${MX_2}\, {\rm (M=Mo, W; X=S, Se)}$ contains two hexagonal sublattices of metal and chalcogen atoms in its unit cell, as shown in Fig.\ \ref{fig:figure1}(b). The chalcogens form two sublayers, one above and one below the metal sublayer, forming a trigonal prismatic structure with the metal atom connected to three chalcogens above and below. The monolayer point-group symmetry is $D_{3h}$, consisting of $C_3$ rotations, $\sigma_v$ in-plane mirror reflections, and $\sigma_h$ out-of-plane mirror reflections. The most common bulk allotrope for these transition-metal dichalcogenides has 2H stacking \cite{mattheiss_tdms_1973}, built by adding subsequent layers rotated by $180^{\circ}$ with respect to the center of the hexagon, resulting in a structure where the chalcogen atoms from one layer are directly above or below metal atoms in the other layer [see Figs.\ \ref{fig:figure1}(a) and \ref{fig:figure1}(b)].
The interlayer distance $\tfrac{c}{2}$, with $c$ the out-of-plane lattice constant, is shown in Fig.\ \ref{fig:figure1}(a), and Fig.\ \ref{fig:figure1}(b) shows the in-plane lattice constant $a$. The resulting 3D layered crystal has a bipartite structure with two monolayers in the unit cell, belonging to the space group $P63/mmc$.

Multilayer TMDs with an even number of layers belong to the point group $D_{3d}$, which contains spatial inversion (${\bf r}\rightarrow -{\bf r})$ but lacks out-of-plane mirror symmetry. The combination of spatial inversion and time-reversal symmetry prescribed at zero magnetic field results in a constraint on the spin splitting of the electronic states for even number of layers, $E_{s}(\kk)=E_{-s}(\kk)$, where $s$ is the spin projection quantum number, such that all states throughout the BZ must be spin degenerate.  Similarly to the monolayer case, multilayer films with an odd number of layers belong to the point group $D_{3h}$, which contains the $z\rightarrow -z$ mirror symmetry $\sigma_h$ but lacks spatial inversion symmetry. Therefore, $s$ is a good quantum number for which spin degeneracy (present in films with even number of layers) can be lifted by spin-orbit (SO) coupling. 
While the SO splitting is absent for bands based on $p_{z}$ and $d_{z^2}$ orbitals  at the $\Gamma$ point, it is substantial near the $Q$ points, leading to the alternation of subband properties. That is, the subbands are spin degenerate for even numbers of layers, resulting in a six-fold degeneracy of dispersion along the $\overline{\Gamma K}$ line. For odd number of layers, subband spectra are three-fold degenerate, but with $E_{s}({\bf k})=E_{-s}(-{\bf k})$.

In Figs.\ \ref{fig:figure1}(d)--(f) we show how the DFT calculated band structure of ${\rm WS_2}$, representative of all four TMDs, evolves from monolayer to trilayer \cite{lambrecht_prb_2012,debbichi_prb_2014,padilha_prb_2014,chang_scirep_2014,abinitioTB_prb_2015,bradley_nanolett_2015,sun_jchemphys_2016} (DFT band structures of all four TMDs can be found in Ref.\ \cite{supplement}). The DFT calculations were performed using a plane-wave basis within the local
density approximation (LDA), with the \textsc{quantum espresso} \cite{quantum_espresso} plane-wave self-consistent field (PWSCF) \emph{ab initio} package. We considered the Perdew-Zunger exchange correlation
scheme \cite{pbe}, with fully-relativistic norm-conserving
pseudo-potentials, including non-collinear corrections. Pseudopotentials for Mo, W, S, and Se atoms were
generated using atomic code  ld1.x of the PWSCF package \cite{pseudo}. The cutoff energy in the plane-wave expansion
was set to $60\ {\rm Ry}$, and the BZ sampling of electronic states was
approximated using a Monkhorst-Pack uniform $k$ grid of
$24\times24\times1$ for all structures \cite{kspace}. We adopted a Methfessel-Paxton
smearing \cite{smearing} of $0.005\ {\rm Ry}$ and set the total energy convergence to less than $10^{-6}\ {\rm eV}$ in all calculations. Spin-orbit coupling was included in all electronic
band structure calculations. To eliminate spurious interactions between adjacent supercells, a $20-{\rm \AA}$ vacuum buffer space was inserted
in the out-of-plane direction. We used experimental values for the interlayer separations \cite{mos2_d,mose2_d,ws2_d,wse2_d}
and LDA optimized in-plane lattice constants for all four TMDs \cite{supplement}.

Using WS$_2$ as an example, Fig.\ \ref{fig:figure1} illustrates that a monolayer $MX_2$ has a direct band gap at the $K$ point of the BZ. The $z\rightarrow -z$ mirror symmetry and lack of inversion symmetry result in 
SO-split conduction and valence bands, classified by their spin projection quantum number [Fig.\ \ref{fig:figure1}(d)].  The large SO splitting at the valence band (VB) $K$ point and conduction band (CB) $Q$ point results from their metal $d_{xy}$ and $d_{x^2-y^2}$ orbital compositions. This is in contrast to the CB $K$ point, which is primarily made of metal $d_{z^2}$ orbitals, resulting in weaker SO splitting \cite{kdotp,wangyao_review}. In a 2H-${MX_2}$ bilayer, the combination of spatial inversion and time reversal symmetry forbids SO splitting, resulting in two spin-degenerate subbands (four bands in total) in the CB and VB, split by the interlayer coupling [Fig.\ \ref{fig:figure1}(e)]. Additionally, the interlayer coupling shifts the band edges 
to the $\Gamma$ point (VB) and in the vicinity of the $Q$ point (CB), making indirect gap semiconductors.

In the trilayer 2H-$MX_2$, the valence and conduction band edges remain at the $\Gamma$ and near $Q$ points.
As shown in Fig.\ \ref{fig:figure1}(f), the CB subbands are split by SO coupling at the $Q$ point due to the lack of spatial inversion symmetry in the case of odd numbers of layers. The resulting spectrum consists of two SO-split subbands in the middle, and two pairs of nearly spin-degenerate subbands above and below (see Appendix \ref{app:Nodd} for details). For the valence subbands, however, SO splitting is forbidden exactly at the $\Gamma$ point, due to it being its own time reversal counterpart, resulting in three nearly spin degenerate subbands [exact degeneracy for even $N$, and spin-splitting $E_{s}({\bf k})-E_{-s}({\bf k})\propto k^3$ for odd $N$]. This trend, which consists of the alternation of SO-split (for odd $N$) and spin-degenerate (for even $N$) subbands persists for TMD films with a larger number of layers, and all the same features are present in the spectra of all four 2H-$MX_2$ shown in Ref.\ \cite{supplement}.
Finally, we note that the in-plane (2D) carrier dispersions in different subbands $n|N$ (both on the VB and CB sides) are different, which affects the intersubband absorption line shapes, as we discuss in Secs.\ \ref{sec:Gmodel} and \ref{sec:Qmodel}.

\section{Hole subbands in $p$-doped few-layer TMDs}\label{sec:Gmodel}
Figure \ref{fig:figure1}(d) shows the monolayer valence bands relevant for the multilayer description, based on symmetry and energy considerations. The $v$ and $w$ valence bands are non-degenerate at the $\Gamma$ point, with the $v$ band composed of the metal $d_{z^2}$ orbital and chalcogen $p_z$ orbitals, whereas the $w$ band is composed of metal and chalcogen $p_z$ orbitals. Bands $v_1$ and $v_3$ belong to two-dimensional irreducible representations (Irreps), with the $v_1$ band composed of chalcogen $p_{x}, p_y$ and metal $d_{xz}, d_{yz}$ orbitals, and the $v_3$-band formed by chalcogen $p_x,\, p_y$ and metal $d_{xy},\, d_{x^2-y^2}$ orbitals \cite{kormanyos_prb_2013,kdotp,wangyao_review}. In the multilayer case, the $w$ and $v$ bands strongly repel as the $w$ band gets closer in energy to the $v$ band. The $v_1$ and $v_3$ bands, on the other hand, are weakly split with a narrow spread due to their orbital characters, and are pushed downwards relative to the $v$ band edge. The two-dimensional Irreps of $v_1$ and $v_3$ allow their coupling with the VB being only through SO interactions (see Appendix \ref{app:socG}). These features involving the symmetry, orbital composition, and proximity of the valence bands, supported by our numerical calculations, indicate that the VB is most strongly hybridized with the $w$ band, while the other valence bands $v_1$, $v_3$, provide corrections in second-order perturbation theory to the model parameters through the action of SO coupling (see Appendix \ref{app:socG}). Additionally, as pointed out in Sec.\ \ref{sec:struct}, in two-dimensional 2H-$MX_2$ crystals the CB and VB are almost spin degenerate at the $\Gamma$ point, despite the fact that atomic SO coupling in TMD compounds is strong. Therefore, to describe the valence subbands at the $\Gamma$ point, we construct a spinless two-band model including the $v$ and $w$ bands, fitting the band parameters and interlayer hopping terms to the DFT calculated band structure, where SO coupling is included implicitly. As indicated in Fig.\ \ref{fig:figure1}(d), these bands belong to the $A_1'$ and $A_2''$ Irreps of the $D_{3h}$ group of the $\Gamma$ point \cite{kormanyos_prb_2013,kdotp}, respectively. Therefore, bands $v$ and $w$
are, respectively, even and odd under $\sigma_h$ transformations, and do not mix in the monolayer case. However, in multilayers, band mixing across consecutive layers is allowed by symmetry. 

\begin{figure}[!t]
	\begin{center}
		\includegraphics[width=0.95\columnwidth]{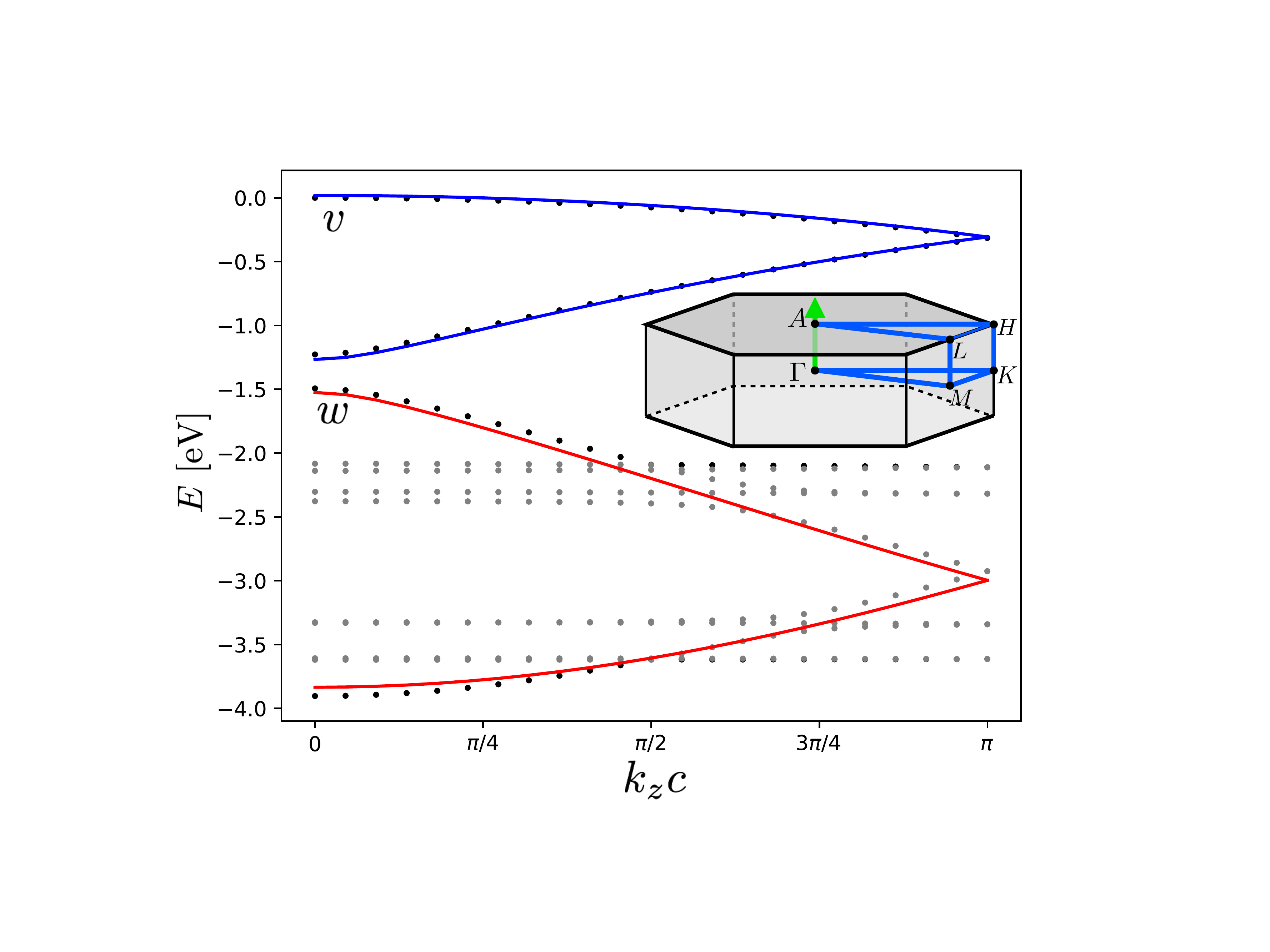}
		\caption{Bulk dispersion of 2H-stacked WS${}_2$ along the $\overline{\Gamma A}$ line. The DFT data (points) are well fitted by the two-band model Eq.\ (\ref{eq:VB_bulk}) (solid lines). The gray points correspond to the $v_1$ and $v_3$ valence bands. Inset: the first Brillouin zone of bulk TMDs.}
		\label{fig:VB_bulk}
	\end{center}
\end{figure}

\subsection{HkpTB for the $\Gamma$-point valence band edge}\label{sec:Gmodel2}
The monolayer dispersions of the valence bands $\sigma=v$ and $w$, are described by isotropic parabolic dispersions with band-dependent effective masses
\begin{equation}
E_\sigma(\kk)=E_\sigma^0 - \frac{\hbar^2 k^2}{2m_\sigma}.
\end{equation}
To construct the multilayer Hamiltonian, we include symmetry-constrained interlayer couplings, given to lowest orders in $\kk$ by
\begin{equation}
\label{eq:coupG}
t_\sigma(\kk)=t_\sigma^{(0)} + t_\sigma^{(2)}k^2; \quad
t_{vw}(\kk)=t_{vw}^{(0)} + t_{vw}^{(2)}k^2,
\end{equation}
where $t_v$ and $t_w$ are interlayer intra-band hopping terms, and $t_{vw}$ couples different bands in two consecutive layers. 

The multilayer Hamiltonian is given by
\begin{widetext}
\begin{equation}
{\small
\begin{split}
&\hat{H}_{N\Gamma}(\kk)=\sum_{s=\uparrow,\downarrow}\sum_{\sigma=v,w}\sum_{n=1}^{\lceil N/2 \rceil}
\Big[E_{\sigma}(\kk) + 2\delta_\sigma + 2\mu_\sigma(\kk) \Big]  \Big[a^{\dagger}_{ns\sigma}(\kk)a_{ns\sigma}(\kk)+\Theta(\tfrac{N}{2}-n)b^{\dagger}_{ns\sigma}(\kk)b_{ns\sigma}(\kk)\Big]\\
&-\sum_{s=\uparrow,\downarrow}\sum_{\sigma=v,w}\Big[\delta_\sigma + \mu_\sigma(\kk) \Big]\left[a^{\dagger}_{1s\sigma}(\kk)a_{1s\sigma}(\kk)+\left(1-\tfrac{\vartheta_N}{2}\right)b^{\dagger}_{N/2,s,\sigma}(\kk)b_{N/2,s,\sigma}(\kk)+\tfrac{\vartheta_N}{2}a^{\dagger}_{(N+1)/2,s,\sigma}(\kk)a_{(N+1)/2,s,\sigma}(\kk)\right]\\
&+\sum_{s=\uparrow,\downarrow}\sum_{\sigma=v,w}\Bigg(\sum_{n=1}^{\lceil N/2 \rceil}t_{\sigma}(\kk)\Theta(\tfrac{N}{2}-n)\Big[a^{\dagger}_{ns\sigma}(\kk)b_{ns\sigma}(\kk)+\text{H.c.}\Big]+\sum_{n=1}^{\lceil N/2\rceil-1} t_{\sigma}(\kk)\left[a^{\dagger}_{n+1,s,\sigma}(\kk) b_{ns\sigma}(\kk)+\text{H.c.}\right]\Bigg)\\
&+\sum_{s=\uparrow,\downarrow}\sum_{\sigma=v,w}\sum_{n=1}^{\lceil N/2\rceil} t_{vw}(\kk)\Theta(\tfrac{N}{2}-n)\Big[a^{\dagger}_{nsv}(\kk)b_{nsw}(\kk)-a^{\dagger}_{nsw}(\kk)b_{nv}(\kk)+\text{H.c.}\Big]\\
&+\sum_{s=\uparrow,\downarrow}\sum_{\sigma=v,w}\sum_{n=1}^{\lceil N/2 \rceil-1} t_{vw}(\kk)\left[a^{\dagger}_{n+1,s,w}(\kk)b_{nsv}(\kk)-a^{\dagger}_{n+1,s,v}(\kk)b_{nsw}(\kk)+\text{H.c.}\right]\\
&+\sum_{s=\uparrow,\downarrow}\sum_{\sigma=v,w}\sum_{n=1}^{\lceil N/2 \rceil}\Big[ U_{2n-1}a^{\dagger}_{ns\sigma}(\kk)a_{ns\sigma}(\kk)+U_{2n}\Theta(\tfrac{N}{2}-n)b^{\dagger}_{ns\sigma}(\kk)b_{ns\sigma}(\kk)  \Big],
\end{split}
}
\label{eq:VB_bilayer}
\end{equation}
\end{widetext}
where we have defined $\vartheta_N\equiv1-(-1)^N$. $a_{ns\sigma}^{(\dagger)}(\kk)$ and $b_{ns\sigma}^{(\dagger)}(\kk)$ annihilate (create) a band-$\sigma$ electron with spin projection $s$ and in-plane wave vector $\kk$ in the odd and even layers of the $n^{\rm th}$ unit cell, respectively. Additional model parameters include the on-site energy corrections $\delta_v$ and $\delta_w$, and $k$-dependent corrections of the form $\mu_{\sigma}(\kk)=\mu_\sigma k^2$, for the $v$ and $w$ bands, respectively, which take into account both the pseudo-interlayer potentials, as well as the spin-flip-induced interband-interlayer hopping (Appendix \ref{app:socG}). For odd $N$ the system has $\lfloor N/2 \rfloor$ complete unit cells and a truncated last unit cell $n = \lceil N/2 \rceil$, where $\lfloor A \rfloor$ and $\lceil A \rceil$ are the floor and ceiling functions, respectively.  This case is considered in Eq.\ (\ref{eq:VB_bilayer}) through the Heaviside function $\Theta(\tfrac{N}{2}-n)$, which removes the operators $b_{N/2,\sigma}^{(\dagger)}(\kk)$ when $N$ is odd. The minus sign in the last row of Eq.\ (\ref{eq:VB_bilayer}) for the interband interlayer hoppings ($t_{vw}$) is due to the opposite parity under $z\rightarrow -z$ of the $v$ and $w$ bands, described in Sec.\ \ref{sec:Gmodel}. Finally, we include on-site potential energy shifts $U_n$, for $1\le n \le N$, to take into account the effects of an electric field applied along the TMD film's $z$ axis. Under experimental conditions, such an electric field may originate from a negative back gate, which in addition dopes the system with a finite hole density $\rho_h$. This density is related to the potential profile $\{ U_n \}$ as \cite{sam_subbands_2018},
\begin{equation}\label{eq:SelfConsistentGamma}
\begin{split}
	&U_n = U_1 + e d\sum_{m=2}^N \mathcal{E}_{m-1,m};\quad n>1,\\
	&\mathcal{E}_{m-1,m} = \frac{2e}{\varepsilon_0}\sum_{l=m}^{N}\rho_{l},
\end{split}
\end{equation}
where $e$ is the (positive) fundamental charge, $d=c/2$ is the interlayer distance, and $\rho_l$ is the electrostatically induced density of holes in layer $l$, such that $\rho_h=\sum_{l=1}^N\rho_l$. The factor of 2 in the second equation comes from the spin degeneracy at the $\Gamma$ point. The potential profile and hole density must be determined self-consistently to satisfy Eq.\ \eqref{eq:SelfConsistentGamma}.

\begin{figure}[t!]
	\begin{center}
		\includegraphics[width=\columnwidth]{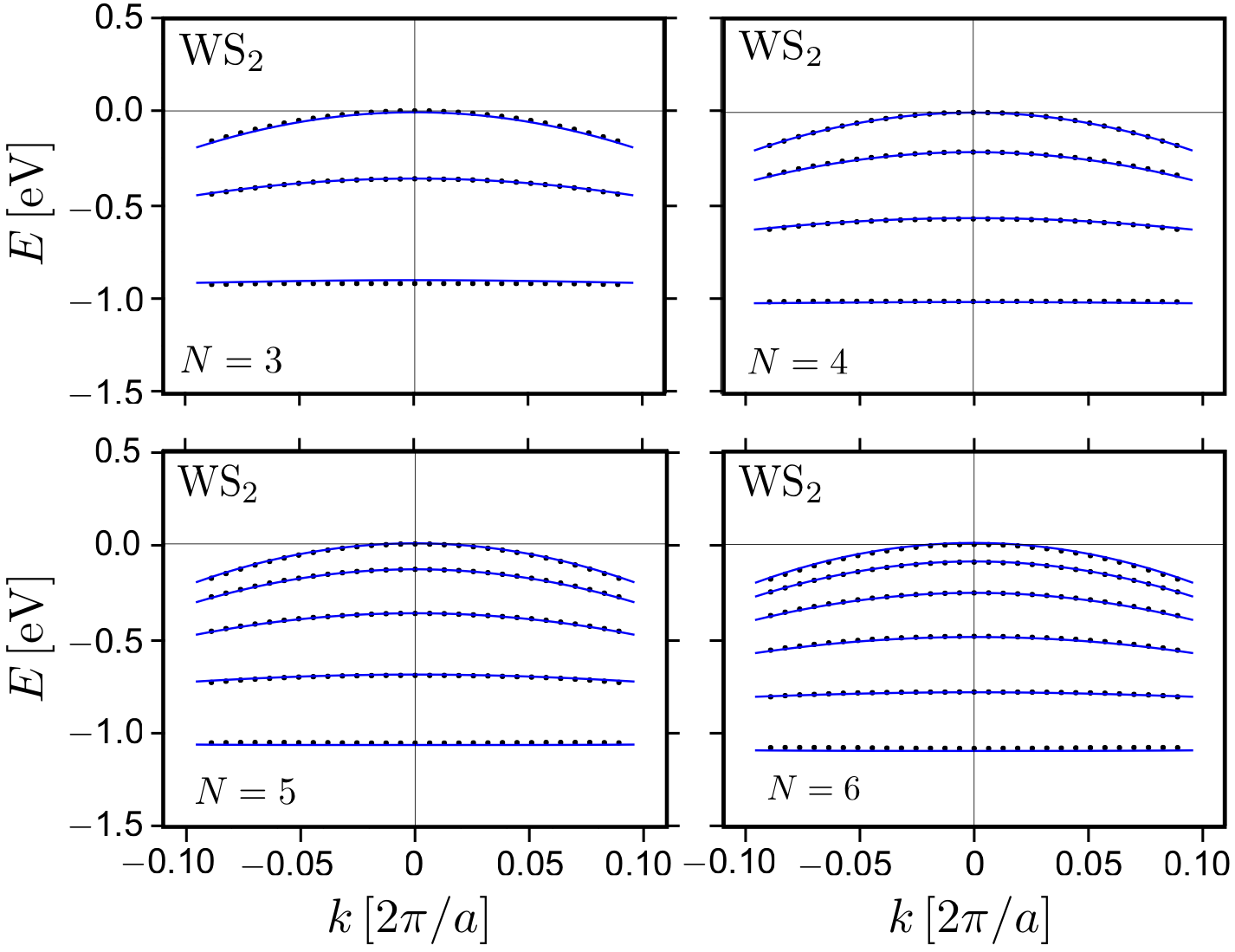}
		\caption{HkpTB model dispersions (solid lines) fitted to DFT results for the WS${}_2$ valence subbands near the $\Gamma$ point, representative of all four TMDs. Results are shown for number of layers $N=3$ to $6$. Fittings for all four TMDs can be found in Ref.\ \cite{supplement}.}
		\label{fig:Gfits}
	\end{center}
\end{figure}

Setting $U_n=0$, we obtain the model parameters in Eq.\ (\ref{eq:VB_bilayer}) for each 2H-$MX_2$ by fitting the results of numerical diagonalization of Eq.\ (\ref{eq:VB_bilayer}) to DFT calculations of bulk and few-layer dispersions. For example, the DFT bulk $k_z$ dispersion of ${\rm WS_2}$ is shown in Fig.\ \ref{fig:VB_bulk} for the $\overline{\Gamma A}$ cut through the 3D BZ. The solid lines in Fig.\ \ref{fig:VB_bulk} correspond to the bands of the bipartite Bloch Hamiltonian
\begin{widetext}
	\begin{equation}\label{eq:VB_bulk}
	H_{\Gamma}(\kk,k_z)=\begin{pmatrix}
	E_v(\kk)+2\delta_v+2\mu_v(\kk) & 0 & 2t_v(\kk)\cos{(\tfrac{k_z c}{2})} & 2it_{vw}(\kk)\sin{(\tfrac{k_z c}{2})}\\
	0 & E_w(\kk)+2\delta_w+2\mu_w(\kk) & -2it_{vw}(\kk)\sin{(\tfrac{k_z c}{2})} & 2t_w(\kk)\cos{(\tfrac{k_z c}{2})}\\
	2t_v(\kk)\cos{(\tfrac{k_z c}{2})} & 2it_{vw}(\kk)\sin{(\tfrac{k_z c}{2})} & E_v(\kk)+2\delta_v+2\mu_v(\kk) &0\\
	-2it_{vw}(\kk)\sin{(\tfrac{k_z c}{2})} & 2t_{w}(\kk)\cos{(\tfrac{k_z c}{2})} & 0 & E_w(\kk)+2\delta_w+2\mu_w(\kk)
	\end{pmatrix},
	\end{equation}
\end{widetext}
obtained from the model Eq.\ (\ref{eq:VB_bilayer}). Equation (\ref{eq:VB_bulk}) is written in the basis of the $v$ and $w$ bands of layers one and two of the bulk 2H crystal unit cell.
The fitted parameters for the four TMDs are given in Tables \ref{tab:VBFittings} and \ref{tab:VBHoppings}, and sample comparisons between the HkpTB model and DFT results for WS${}_2$, representative of all four TMDs, are shown in Fig.\ \ref{fig:Gfits}. Detailed comparisons for few-layer films of all four materials are available in Ref.\ \cite{supplement}.

Noting that the bulk VB edge is located at the $\Gamma$ point (Fig.\ \ref{fig:VB_bulk}), the dispersion near the band edge can be obtained from Eq.\ (\ref{eq:VB_bulk}) as
\begin{equation}
E_{\Gamma}(k_z,{\bf k})\approx-\frac{\hbar^2k_z^2}{2m_{v,z}}-\frac{\hbar^2k^2}{2m_{v,xy}}\left(1+\zeta k_z^2\right),
\label{eq:bulk_disp}
\end{equation}
where the bulk parameters are given in terms of the HkpTB model parameters,
\begin{subequations}
\begin{equation}
m^{-1}_{v,z}=\frac{\hbar^2}{2d^2}\Bigg( \frac{4{t^{(0)}_{vw}}^2}{\Delta E}+t_{v}^{(0)} \Bigg)
\end{equation}
is the out-of-plane bulk effective mass, with $d=c/2$ the interlayer distance and $\Delta E=E_v-E_w-2t^{(0)}_{v}+2t^{(0)}_{w}+2\delta_v-2\delta_w$ the bulk gap between the topmost $v$ and lowest $w$ bands at the $\Gamma$ point.
\begin{equation}
m^{-1}_{v,xy}=\left[ 1+\frac{4m_v}{\hbar^2}\left(t^{(2)}_{v}-\mu_v\right)\right]m_v^{-1},
\end{equation} 
is the in-plane bulk effective mass, and 
\begin{equation}
\begin{split}
&\zeta=
-\frac{\hbar^{-2}m_v d^2}{[1+\frac{4m_v}{\hbar^2}(t_v^{(2)}-\mu_v)]}\left\{2t_v^{(2)}
+\frac{4\hbar^2 t^2_{vw}}{\Delta E^2}\frac{m_w-m_v}{m_vm_w}
 \right.
\\
&\left.
+16t_{vw}\left(\frac{t_{vw}^{(2)}}{\Delta E}+\frac{t_{vw}}{\Delta E^2}(t_v^{(2)}-t_w^{(2)}+\mu_w-\mu_v)\right)\right\},
\end{split}
\end{equation}
\end{subequations}
is an anisotropic non-linearity factor. These parameter values, obtained by fitting to DFT calculations, can be found in Table \ref{tab:VBFittings}.

\begin{table}[t!]
	\centering
	\caption{Model parameters fitted to DFT data for the monolayer valence bands $E_{v}(\kk)$ and $E_{w}(\kk)$, and bulk valence band dispersion for the four TMDs. The monolayer parameters include the band edges energy difference  $E_v^0-E_w^0$, and the effective masses $m_v, m_w$ given in terms of the free electron mass $m_0$. The 3D bulk parameters include the out-of-plane and in-plane effective masses $m_{v,z}, m_{v,xy}$, respectively,  and the in-plane dispersion non-linearity parameter $\zeta$.}
	\label{tab:VBFittings}
	\begin{tabular*}{\columnwidth}{@{\extracolsep{\stretch{1}}}*{7}{lccc}@{}}
		\hline\hline
		& $E_v^0 - E_w^0\,[{\rm eV}]$ & $m_v\,[m_0]$ & $m_w\,[m_0]$ \\
		& $m_{v,z}\ [m_0]$ & $m_{v,xy}\ [m_0]$ & $\zeta\ [{\rm \AA^2}]$ \\
		\hline
		${\rm MoS_2}$ &1.75			& 3.726	& 0.304 \\
		& 1.04 & 0.693 & -5.24 \\
		${\rm MoSe_2}$& 1.56		& 5.575	& 0.505 \\
		& 1.42 & 0.786 & -5.99 \\
		${\rm WS_2}$	& 2.08		& 2.885	& 0.353 \\
		& 0.840 & 0.615 & -5.86 \\
		${\rm WSe_2}$&1.81			& 3.420	& 0.760 \\
		& 1.08 & 0.700 & -5.45 \\
		\hline\hline
	\end{tabular*}
\end{table}

\begin{table}[t!]
	\centering
	\caption{Model parameters fitted to DFT data for the valence band interlayer hopping terms $t_v(\kk)$, $t_w(\kk)$ and $t_{vw}(\kk)$. $\delta_v$, $\delta_w$, $\mu_v$, and $\mu_w$ are the on-site energy offsets due to the pseudo-interlayer potential and spin-flip coupling terms.}
	\label{tab:VBHoppings}
	\begin{tabular*}{\columnwidth}{@{\extracolsep{\stretch{1}}}*{7}{lcccc}@{}}
		\hline\hline
		& $t_v^{(0)}\,[{\rm eV}]$ & $t_w^{(0)}\,[{\rm eV}]$ & $t_v^{(2)}\,[{\rm eV\Ams^2}]$ & $t_w^{(2)}\,[{\rm eV\Ams^2}]$ \\
		& $t_{vw}^{(0)}\ [{\rm eV}]$& $t^{(2)}_{vw}\,[{\rm eV\AA^2}]$ & $\delta_v\,[{\rm meV}]$ & $\delta_w\,[{\rm meV}]$ \\
		& & &  $\mu_v\ [{\rm eV\AA^2}]$ & $\mu_w\ [{\rm eV\AA^2}]$\\
		\hline
		${\rm MoS_2}$	& -0.333	& 0.592	& 1.744	& 2.684 \\
		&   0.432     & -1.206    & -62.18 &  -41.43 \\
		& & & -0.351 & 6.770 \\
		${\rm MoSe_2}$	& -0.307	&0.657	& 1.830	& 2.626 \\
		&    0.453   & -1.140     & -29.13    & -10.85 \\
		& & & -0.261 & 2.736 \\
		${\rm WS_2}$	&-0.322	& 0.574& 1.718	& 3.205 \\
		&  0.404          & -1.226     & -36.98    & -48.89 \\
		& & & -0.614 & 5.834 \\
		${\rm WSe_2}$	& -0.291	& 0.649	& 1.814	& -1.382 \\
		&    0.4309       &  -0.049   &-25.27   &-69.93 \\
		& & & -0.519 & 0.192 \\
		
		\hline\hline
	\end{tabular*}
\end{table}

Then, the subband energies and dispersions in TMD films with $N\gg1$ can be analyzed by quantizing hole states with dispersions described by Eq.\ (\ref{eq:bulk_disp}) in a slab of thickness $L=Nd$. When doing so, one has to complement Eq.\ (\ref{eq:bulk_disp}) with the general Dirichlet-Neumann boundary condition for the standing waves of holes at both film surfaces
\begin{equation}
\left[\pm\nu d\partial_z \psi(z)+\psi(z)\right]_{z=\pm \frac{L}{2}}=0,
\label{eq:boundary}
\end{equation}
where the $\pm$ correspond to the top and bottom layers, respectively, and $\nu$ is a dimensionless parameter.
Assuming a solution of the form $\psi(z) = u e^{ik_z z}+ve^{-ik_z z}$, one finds from Eq.\ (\ref{eq:boundary}) that $k_z$ in Eq.\ (\ref{eq:bulk_disp}) obeys
\begin{equation}\label{eq:quant_condition}
L k_z +2\arctan (\nu k_z d)=\pi n,
\end{equation}
where the integer $n$ is the subband index.
For large number of layers and for subbands near the band edge, $k_z\sim \tfrac{1}{L}\ll \tfrac{1}{d}$, and  $\arctan(\nu k_z d)\approx \nu k_z d$, so that we can approximate
\begin{equation}
k_z\approx\frac{\pi n }{d(N+2\nu)},
\label{eq:kz_quan}
\end{equation}
leading to the subband energies and dispersions
\begin{equation}
\label{eq:sub_disp}
\begin{split}
E_{n\ll N|N}({\bf k})&=-\frac{\hbar^2}{2m_{v,z}}\frac{\pi^2n^2}{d^2(N+2\nu)^2}
\\
&-\frac{\hbar^2k^2}{2m_{v,xy}}\left[1+ \frac{\zeta\pi^2n^2}{d^2(N+2\nu)^2}\right].
\end{split}
\end{equation}

\begin{figure}[!t]
	\begin{center}
				\includegraphics[width=0.98\columnwidth]{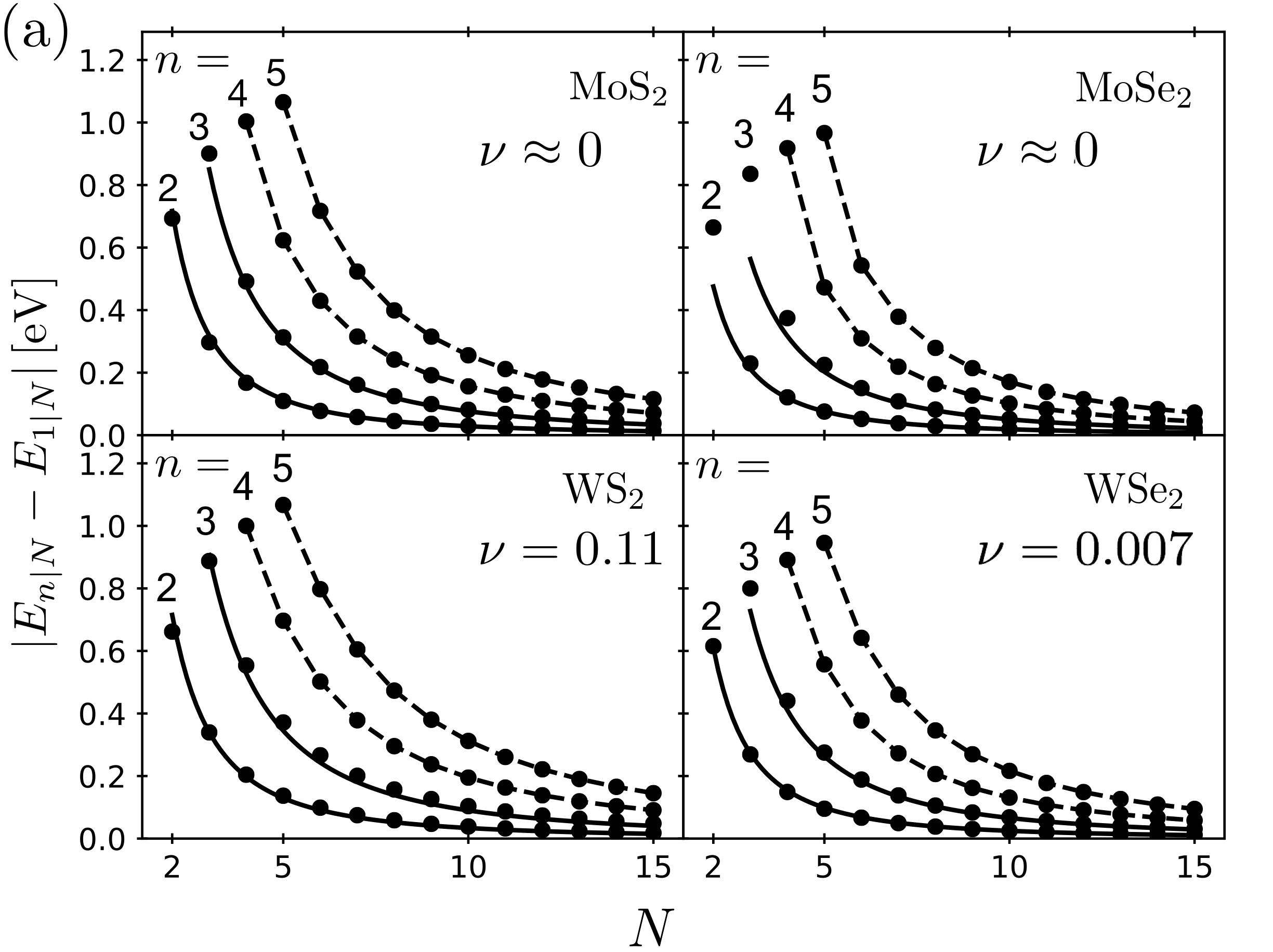}
				\includegraphics[width=\columnwidth]{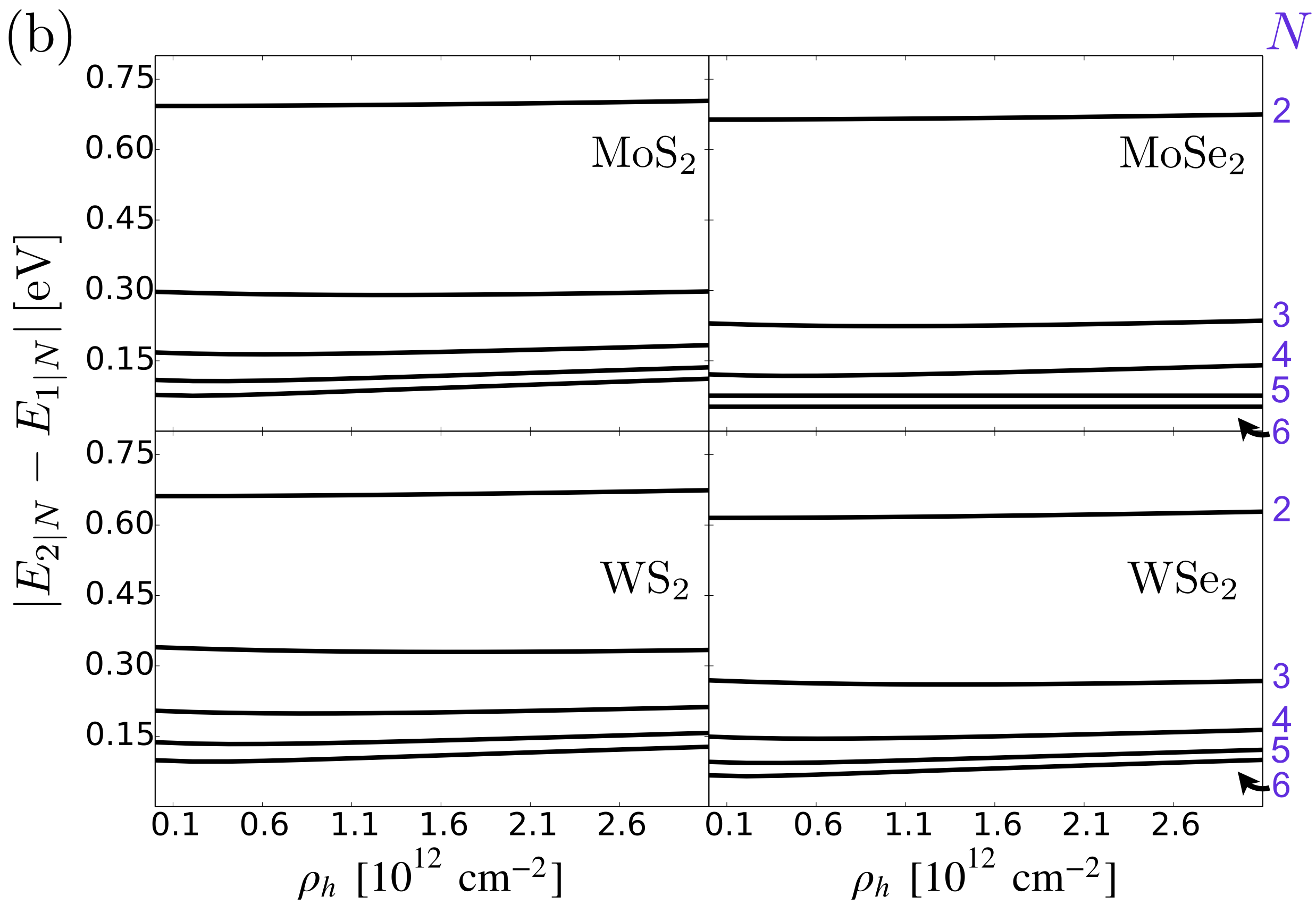}
		\caption{(a) Energy spacings between the first and $n$th valence subband ($n=2$ to $5$) for the four TMDs, as a function of the number of layers $N$. The $\nu$ parameter corresponding to each TMD is given in each panel.
			The solid lines for the first two transitions are obtained using Eq.\ (\ref{eq:sub_disp}), showing a good fit between the model and DFT. (b) $1|N \rightarrow 2|N$ transition energy as a function of the hole density $\rho_h$. Each curve corresponds to the value of $N$ indicated on the right.}
		\label{fig:spacing_vs_N_G}
	\end{center}
\end{figure}

The large-$N$ asymptotics of the separation between the lowest two subbands, $|E_{1|N}-E_{2|N}|$, was used to determine the value of the boundary parameter $\nu$ for holes in each TMD, resulting in $\nu\approx 0$ for ${\rm MoS_2}$ and ${\rm MoSe_2}$, $\nu=0.11$ for ${\rm WS_2}$, and $\nu=0.007$ for ${\rm WSe_2}$, using the dispersions and lowest intersubband splittings shown in Figs.\ \ref{fig:Gfits} and \ref{fig:spacing_vs_N_G}(a). The good agreement between the full HkpTB model and the asymptotic analysis shown in Fig.\ \ref{fig:spacing_vs_N_G}(a) enables us to describe the main intersubband transition $1|N\rightarrow 2|N$ in $p$-doped $N$-layer 2H-$MX_2$ as
\begin{equation}
\label{eq:first_trans}
|E_{1|N}-E_{2|N}|=\frac{3\pi^2\hbar^2}{2m_{v,z}d^2(N+2\nu)^2}.
\end{equation}
Furthermore, the hole subband effective masses
\begin{equation}
m_{n|N}^{-1}=m_{v,xy}^{-1}\left[1+\frac{\zeta\pi^2n^2}{d^2(N+2\nu)^2}\right],
\label{eq:subm}
\end{equation}
obtained from Eq.\ (\ref{eq:sub_disp}), describe well the subband dependence of the in-plane masses, as seen in Fig.\ \ref{fig:spacing_vs_N_G}(a).

Fig.\ \ref{fig:spacing_vs_N_G}(b) shows the effects of a doping-induced potential profile in the film on the first subband transition energy $|E_{2|N}-E_{1|N}|$, as a function of the total gate-induced hole density $\rho_h$. Results are shown for all four TMDs with film thicknesses of $N=2$ to 6 layers, up to moderate doping levels. Following a slight decrease for weak doping, the transition energies grow monotonically, but with only a weak blue shift.

\subsection{Selection rules for intersubband transitions, and dispersion-induced line broadening}\label{sec:VBselectionrules}

Next, we use the model developed above for the description of hole subbands to study intersubband optical transitions, electron-phonon relaxation, and absorption line shapes of IR/FIR light.

\begin{figure}[t!]
	\begin{center}
		\includegraphics[width=1\columnwidth]{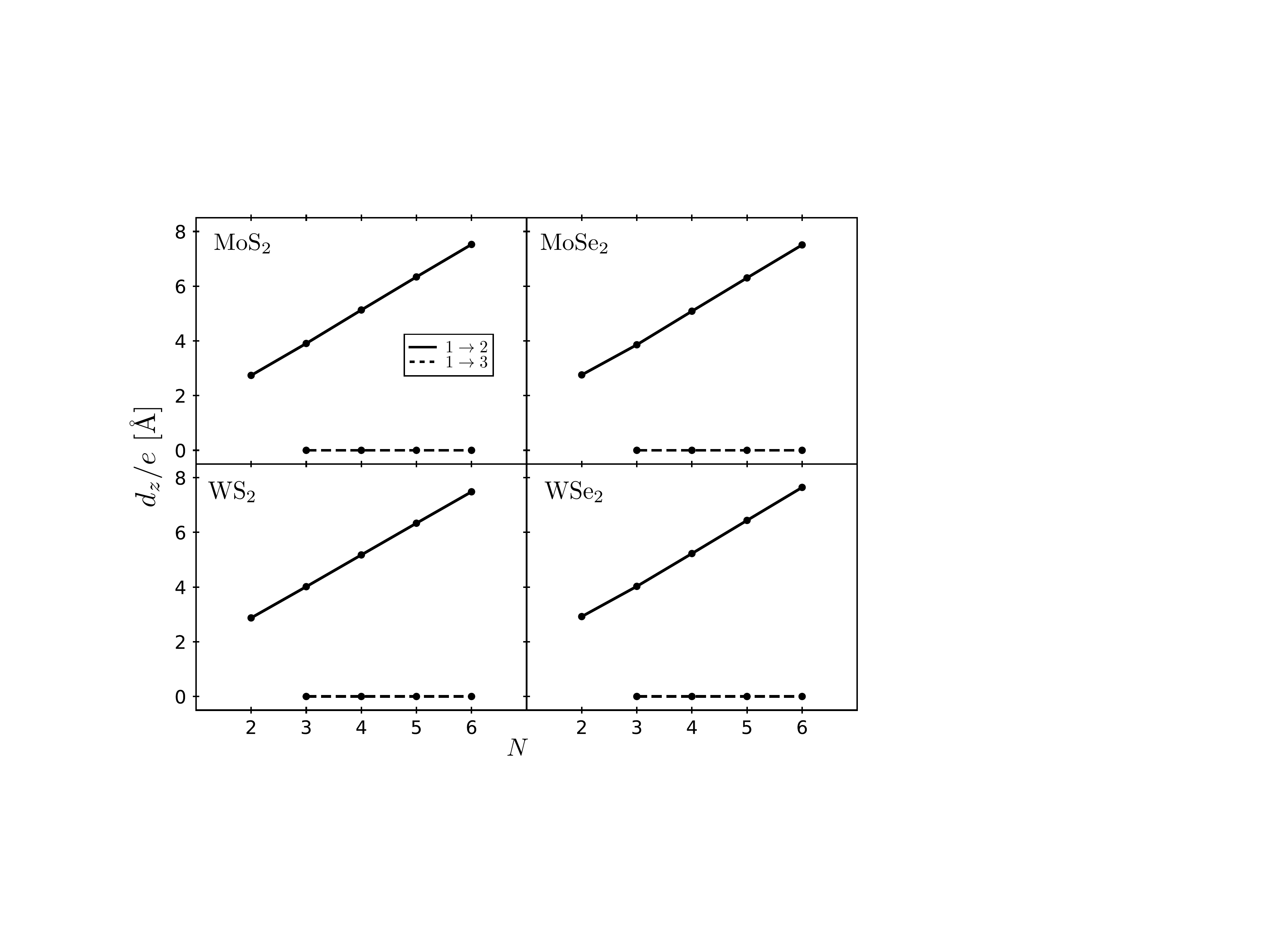}
		\caption{Out-of-plane dipole moment matrix elements for the VB subbands, for the first two intersubband transitions $1\rightarrow 2$ (solid) and $1\rightarrow3$ (dashed).}
		\label{fig:dz_plots_G}
	\end{center}
\end{figure}

 The optical transition amplitude between two given subbands $n$ and $n'$  is determined by the out-of-plane dipole moment
\begin{equation}\label{eq:dipole}
\begin{split}
d^{n,n'}_z({\bf k}) &= e \langle n,\kk | z | n',\kk \rangle  
\\
&= e\sum_{j=1}^N\sum_{\sigma=v,w} z_j C^*_{n,j,\sigma}({\bf k})C_{n',j,\sigma}({\bf k}),
\end{split}
\end{equation}
where $N$ is the total number of layers, $z_j$ denotes the $z$ coordinate of layer $j$, and $C_{n,j,\sigma}(\kk)$ are the components of the $n^{\rm th}$ subband eigenstate. 
The calculated dipole moment matrix element for the first two intersubband transitions is plotted in Fig.\ \ref{fig:dz_plots_G} as a function of the number of layers.
The selection rules for intersubband transitions driven by out-of-plane polarized light are determined by the odd parity of $z$ under both spatial inversion and mirror reflection ($\sigma_h$).
The subband states for even and odd number of layers also have a definite parity under spatial inversion and mirror reflection, respectively, due to the crystal's symmetry. Therefore, intersubband transitions between same parity subbands are forbidden, as shown in Fig.\ \ref{fig:dz_plots_G} for the first two intersubband transitions. All this makes the $1|N\rightarrow 2|N$ transition the dominant feature in the IR/FIR absorption by thin TMD films.

\begin{figure}[t!]
	\begin{center}
		\includegraphics[width=0.98\columnwidth]{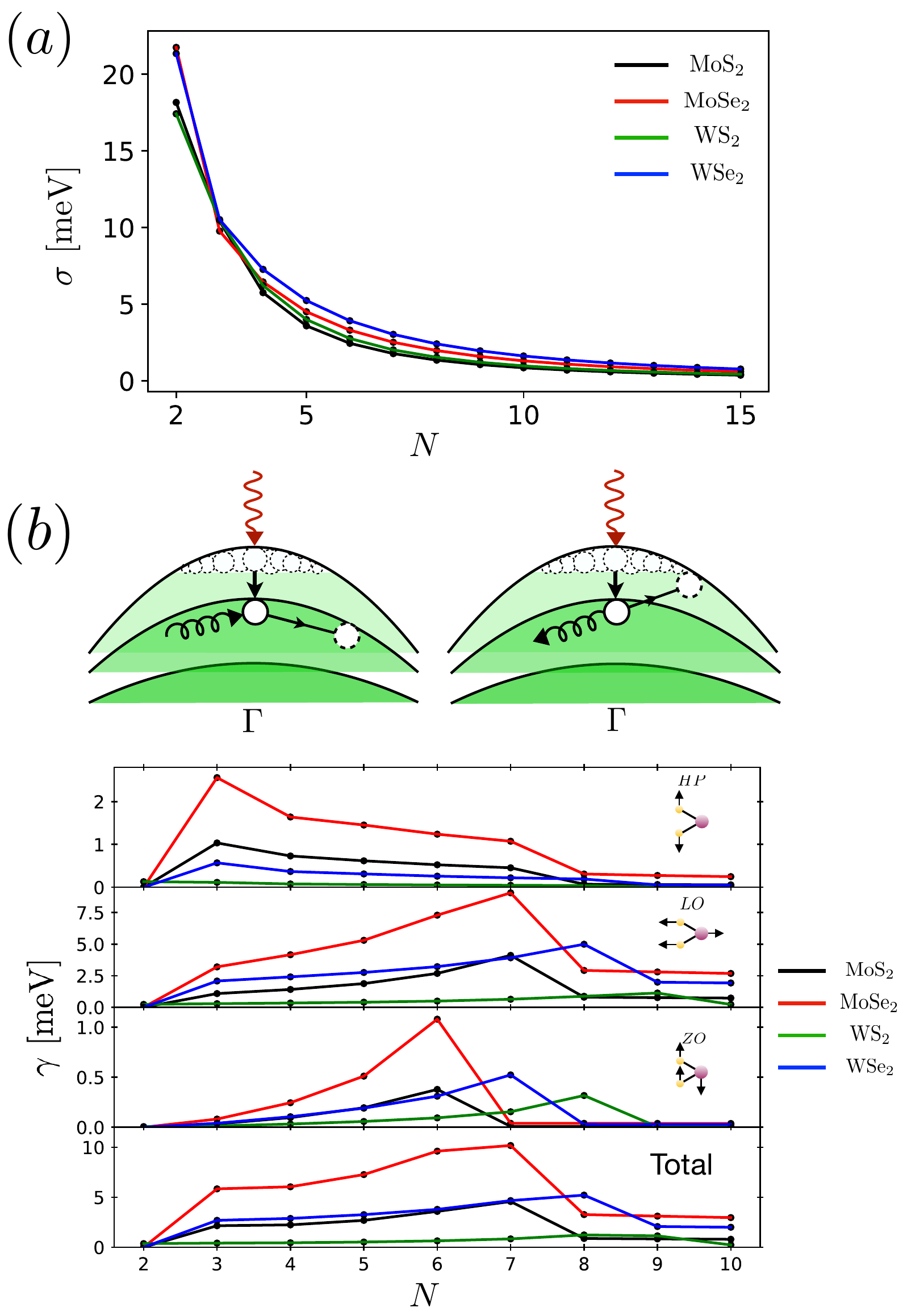}
		\caption{(a) Absorption line widths for VB subbands at room temperature ($300\ {\rm K}$) as a function of number of layers for the four TMDs, considering only DOS broadening. 
		(b) Phonon-induced broadening at room temperature ($T=300$ K) due to intersubband emission and intrasubband absorption of optical phonons modes (top to bottom) HP, LO and ZO, for the four TMDs as a function of number of layers $N$, with the combined broadening shown in the bottom panel.}
		\label{fig:linewidths_phonons_G}
	\end{center}
\end{figure}

The intersubband absorption line shape is affected by the difference between the effective masses of subbands $1|N$ and $2|N$. The lighter in-plane hole mass in the initial state ($1|N$ subband) as compared to the final state ($2|N$ subband) spreads the absorption spectrum toward lower energies. Heavy $p$-doping of the TMD film or Boltzmann distribution of the holes in the case of light $p$-doping, sets the lower limit for the line width of the $1|N\rightarrow 2|N$ absorption line, which we call the density of states (DOS) broadening
\begin{equation}
\sigma = \left(1-\frac{m_{1|N}}{m_{2|N}}\right)\max\{\epsilon_F,k_{\rm B}T\}\log 2.
\end{equation}
Here, $k_{\rm B}$ is the Boltzmann constant, $T$ is the temperature, and $m_1,\, m_2$ are the effective masses of the first and second subbands.  This limit for the line width is illustrated in Fig.\ \ref{fig:linewidths_phonons_G}(a) for $N$-layer 2H-$MX_2$ at room temperature. We note here that DOS broadening is similar to the inhomogeneous broadening, in the sense that it can be overcome by placing the TMD film inside an optical resonator that would select intersubband modes with particular values of in-plane momentum ${\bf k}$.

\subsection{Broadening due to electron-phonon intra- and intersubband relaxation}
\label{sec:e-ph-G}

In contrast to the elastic DOS broadening, phonon-induced intra- and intersubband relaxation broaden the absorption line in a way that cannot be avoided by a clever choice of the electromagnetic environment. Below we consider emission and absorption of homopolar (HP), longitudinal (LO) and out-of-plane (ZO) optical phonons, which we assume to be dispersionless.
This choice is motivated by the fact that these are the strongest coupled modes in TMDs, as established by earlier studies \cite{danovich_phonons,calandra_lo}. Also, we take phonon modes of few-layer films as independent and degenerate. This approximation is justified by the fact that splittings due to hybridization between layers are much smaller than the monolayer phonon frequency \cite{multilayer_phonons}.

The hole-phonon couplings for a phonon in mode $\mu={\rm HP,\, LO}$, or ZO in layer $j$, interacting with a hole in layer $i$, are given by (see Appendices \ref{sec:lo_coup} and \ref{sec:zo_coup})
\begin{subequations}
	\label{eq:geph}
	\begin{equation}
 g^{j,i}_{\mathrm{HP}}(\qq)=\delta_{ij}\sqrt{\frac{\hbar}{2\rho\omega_{\mathrm{HP}}}}D_v,
	\end{equation}
	\smallskip
	\begin{equation}
\begin{split}
&g^{j,i}_{\mathrm{LO}}(\qq)=\sqrt{\frac{\hbar}{2\rho \frac{M_r}{M}\omega_{\rm LO}}}\frac{2\pi i e^2 Z (-1)^j}{A(1+r_*q)}
 e^{-qd|i-j|},
\end{split}
	\end{equation}
\smallskip	
	\begin{equation}
\begin{split}
&g^{j,i}_{{\rm ZO}}(\qq)=\sqrt{\frac{\hbar}{2\rho \frac{M_r}{M}\omega_{\text{ZO}}}}\frac{2\pi e^2 Z_z}{A}
e^{-q d|i-j|}\frac{i-j}{|i-j|},
\end{split}
	\end{equation}	
\end{subequations}
where $\omega_{\mu}$ denotes the corresponding phonon frequency; $\rho$ is the mass density of the material; $D_v$ is the deformation potential in the valence band; $A$ is the unit cell area; $M$ and $M_r$ are the total unit-cell mass and reduced mass of the metal and two chalcogens, respectively; $Z$ and $Z_z$ are the in-plane and out-of-plane Born effective charges, respectively; and $r_*$ is the screening length in the material. The various parameters taken from Refs.\ \onlinecite{danovich_phonons,calandra_lo,chinese_phonons,chinese_phonons2} are given in Table\ \ref{tab:phonon_params}.

\begin{table*}
	\caption{Electron-phonon coupling parameters for LO, HP, and ZO
		phonon modes.  $\omega_{\rm HP}$,  $\omega_{\rm LO}$, and $\omega_{\rm ZO}$ are the HP, LO, and ZO mode energies; $\rho$ is the mass density; $D_{v},\, D_{c}$ are the valence and conduction deformation potentials; $Z,\, Z_z$ are the in-plane and out-of-plane Born effective charges; $r_*$ is the screening length; $M_r/M$ is the ratio of the reduced mass of the metal and chalcogens to the total unit-cell mass; and $A$ is the unit-cell area.
	}
	\begin{tabular}{lccccccccccc}
		\hline \hline
		& $\hbar\omega_{\text{HP}}$ [meV] & $\hbar\omega_{\text{LO}}$ [meV] &  $\hbar\omega_{\rm ZO}$ [meV] & $\rho$ [g/cm$^2$] & $D_v$ [eV/{\AA}] & $D_c$ [eV/{\AA}]
		& $Z$ & $Z_z$ &  $r_*\ [{\rm \AA}]$    & $M_{\rm r}/M$ & $A\ [{\rm \AA^2}]$  \\
		\hline
		MoS${}_2$           & 51       &    49         & 59   & $3.1\times 10^{-7}$ & 3.5   & 7.1 & 1.08  & 0.1 & 41 & 0.24 & 8.65  \\
		MoSe${}_2$          & 30 & 37  & 44   & $4.5\times 10^{-7}$ &  3.8 & 7.8   &  1.8   &   0.15      & 52&0.249 &   9.37\\ 
		WS${}_2$          & 52 & 44  &  55  & $4.8\times 10^{-7}$   &  1.5  & 3.4 & 0.47   &    0.07      & 38 & 0.29 & 8.65  \\ 
		WSe${}_2$          & 31 & 31  & 39     & $6.1\times 10^{-7}$ &  2.2 & 2.7  &    1.08       & 0.12 & 45 & 0.25 & 9.37  \\ \hline		
		 \hline
	\end{tabular}
	\label{tab:phonon_params}
\end{table*}
The phonon-induced broadening is determined by the lifetime of the hole in the excited subband state, which includes contributions from intersubband relaxation due to emission (low and high temperature) and intrasubband absorption (high temperature) [Fig.\ \ref{fig:linewidths_phonons_G}(b)]. We note that intrasubband emission contributions are thermally activated, since they require carriers to be thermally excited to energies higher than the corresponding phonon energy. The typical energy of thermally distributed carriers at room temperature is $\frac{1}{2}k_BT\sim 13$ meV, whereas the phonon energies are of order $30$--$50$ meV (Table\ \ref{tab:phonon_params}), making this process irrelevant. Similarly, the process involving intersubband absorption from the second subband to the third is suppressed by the larger intersubband spacings, as compared to the phonon energies and the first intersubband spacings for $N\lesssim 10$, and therefore will not be considered.

The phonon-induced broadening is accounted for by 
\begin{equation}\label{eq:phononrate}
\begin{split}
\gamma =& 2\pi \sum_{\mu,\qq,j}
\left|\sum_{i}\sum_{\sigma=v,w}g^{j,i}_{\mu}(\qq)C^*_{n,i,\sigma}(\qq)C_{m,i,\sigma}(0)
\right|^2
\\
&\times \left\{
[1+n_T(\hbar\omega_{\mu})] \delta\left[E_{m}(0)-E_n(\qq)-\hbar\omega_{\mu}\right]
\right.
\\
&\quad\left.
+
\delta_{nm}n_T(\hbar\omega_{\mu})\delta\left[E_{m}(0)-E_{m}(\qq)+\hbar\omega_\mu\right]
\right\},
\end{split}
\end{equation}
where the sums are over the phonon modes $\mu={\rm HP,\,LO,\,ZO}$, the phonon wave vector $\qq$, and the layer number $1\le j\le N$. $C_{n,i,\sigma}$ are the components of the $n^{\rm th}$ subband eigenstate on layer $i$ in band $\sigma$, and $n_T(\hbar\omega_{\mu})$ is the Bose-Einstein distribution for a phonon in mode $\mu$ at temperature $T$.
The first term in curly brackets describes intersubband phonon emission, whereas the second term describes intrasubband phonon absorption.

The resulting phonon-induced broadenings at room temperature are shown in Fig.\ \ref{fig:linewidths_phonons_G}(b). The main contribution comes from intersubband relaxation, with  intrasubband absorption suppressed by the phonon occupation number. The intersubband LO phonon contribution dominates the broadening due to the strong coupling attributed to the large in-plane Born effective charge, and the long-range nature of the coupling. The reduced broadening for $N=2$ is due to the large intersubband spacing, which suppresses intersubband relaxation, and the fact that the second subband is almost flat, which suppresses intrasubband absorption.
 The peaks in the broadenings for certain numbers of layers correspond to near resonances between the phonon energies and the intersubband spacings. Phonon broadening is seen to be most detrimental for ${\rm MoSe_2}$ in particular, and in general for all TMDs with seven or eight layers. Beyond this number of layers, the phonon energies become larger than the intersubband spacings, thus preventing intersubband relaxation, however, intrasubband absorption is still present and dominates for $N>7$. Finally, we note that the broadening values are found to be smaller than those observed in III-V quantum wells  \cite{qw1}, implying a weaker detrimental effect on the absorption/emission line shape in these materials.

\subsection{Room-temperature absorption spectrum in $p$-doped TMD films}

\begin{figure}[t!]
	\begin{center}
		\includegraphics[width=1\columnwidth]{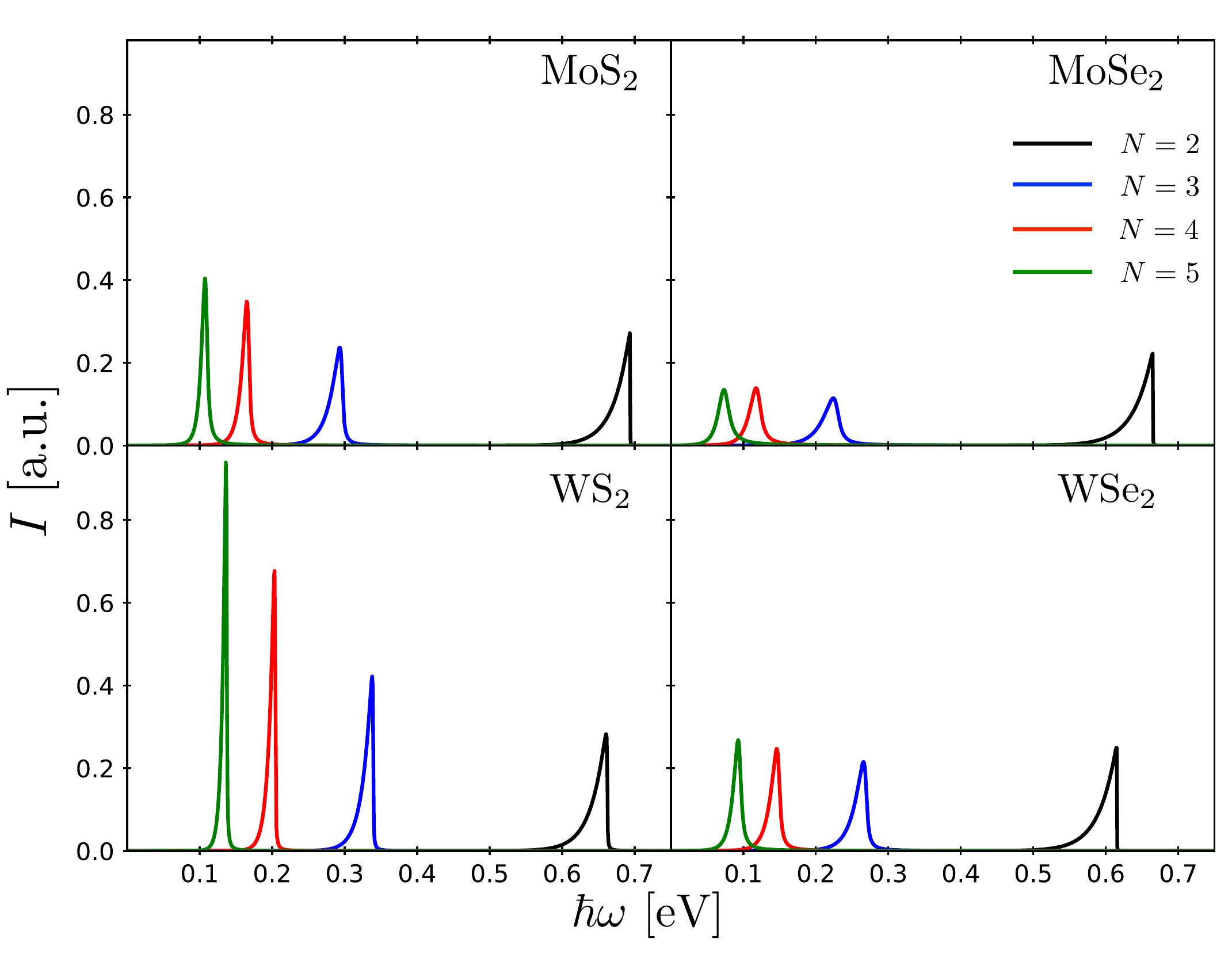}		
		\caption{Optical absorption lines for $N=2$ to $5$ layers of lightly $p$-doped ${\rm MoS_2,\, MoSe_2,\, WS_2}$, and ${\rm WSe_2}$ at room temperature ($T=300\ {\rm K}$).
		}
		\label{fig:figure2G}
	\end{center}
\end{figure} 

The cumulative effect of inelastic (e-ph) and elastic (DOS) broadening of the intersubband $1|N\rightarrow 2|N$ absorption spectra of lightly $p$-doped TMD films is described by
\begin{equation}\label{eq:lineshapeG}
\begin{split}
I(\hbar\omega)&=\frac{4\pi}{\hbar}\abs{E_z(\hbar\omega)}^2\sum_{\bf k}\abs{d_z^{1,2}(\kk)}^2 f_T({\bf k})
\\
&\qquad \times \frac{\gamma/\pi}{[E_{1}({\bf k})-E_{2}({\bf k})-\hbar\omega]^2+\gamma^2},
\end{split}
\end{equation}
where $f_T({\bf k})$ is the Fermi function for hole occupation in the lowest subband corresponding to hole density $n_h$, and temperature $T$ (we assume that all higher-energy hole subbands are empty). 
The resulting absorption spectra at room temperature for the four TMDs with different number of layers are shown in Fig.\ \ref{fig:figure2G}. 
The spectra show the combination of DOS broadening, which produces a tail towards lower photon energies, with the phonon-induced broadening, most relevant for $N>2$, which gives a small tail towards higher energies, making the lines more symmetric and reducing their amplitudes. The smaller phonon couplings in ${\rm WS_2}$ result in tall, narrow, and asymmetric line shapes, with intensity increasing with the number of layers, reflecting the growing dipole matrix element. This is in contrast to ${\rm MoSe_2}$, where the larger phonon-induced broadening results in smaller and more symmetric peaks for $N>2$.

\section{Electron subbands in $n$-doped few-layer TMDs}\label{sec:Qmodel}
\subsection{HkpTB for the conduction band near the $Q$ point}\label{sec:QmodelA}

\begin{table*}[t!]
	\centering
	\caption{Monolayer and bulk conduction band parameters fitted to DFT calculations of  the four TMDs. The effective masses are given in terms of the free electron mass $m_0$.  The band-edge energy $E_{0}$ is given relative to the valence band edge at the $\Gamma$ point, and $2\Delta_0$ is the spin-orbit splitting at the $Q$ point. The monolayer parameters include the effective masses in the $x$ and $y$ directions for the spin split bands, and the band minima offsets $q_{\downarrow}$ and $q_{\uparrow}$.
	The conduction band bulk dispersion parameters include the in-plane effective masses $m_{c,x}, m_{c,y}$, and out-of-plane mass $m_{c,z}$; band minima offsets $\kappa_0$ and $\beta$, and in-plane dispersion non-linearity parameters $\zeta_x$ and $\zeta_y$.}
	\label{tab:CBFits}
	\begin{tabular}{lcccccccc}
		\hline\hline
		& $m_{x,\uparrow}\,[m_0]$ & $m_{y,\uparrow}\,[m_0]$ & $q_{\uparrow}\,[10^{-3}\Ams^{-1}]$ & $m_{x,\downarrow}\,[m_0]$ & $m_{y,\downarrow}\,[m_0]$ & $q_{\downarrow}\,[10^{-3}\Ams^{-1}]$ & $E_0\,[\mathrm{eV}]$ & $2\Delta_0\ [{\rm meV}]$	\\
& $m_{c,z}\ [m_0]$ & $m_{c,x}\ [m_0]$ & $m_{c,y}\ [m_0]$ & $\zeta_x\ [{\rm \AA^2}]$ & $\zeta_y\ [\rm \AA^2]$& $\kappa_0\ [{\rm \AA^{-1}}]$ & $\beta\ [10^{-4}\,{\rm \AA}]$ & \\		
		\hline
		${\rm MoS_2}$		& 0.595	& 1.035	& 20.49	& 
		 0.666	& 1.105	& 7.16 & 1.994 & 67.0  \\
		& 0.525 	& 0.550 	& 0.735	 & -3.90 & -7.94 &0.0456 & -1.3 &\\
		${\rm MoSe_2}$	& 0.583	& 1.060	& 54.21	& 
		 0.518	& 1.106	& 26.93	& 1.891 & 21.0 \\ 
		& 0.500 	& 0.510	& 0.760	& -4.65 &  -4.26 &0.0663 & 0.32 & \\
		${\rm WS_2}$		& 0.529	& 0.722	& 13.74	& 
		 0.763	& 0.892	& -20.65	& 2.059 & 254 \\
		& 0.510  	& 0.528	& 0.596	 &-4.30&  -4.12&   0.0344 &  0.53 & \\
		${\rm WSe_2}$	& 0.468	& 0.753	& 49.63	& 
		 0.676	& 0.908	& 1.88	& 1.94 & 214 \\
		& 0.466 	& 0.479	& 0.608	& -4.19&  -5.80&  0.0599 & -0.9  &\\
		\hline\hline
	\end{tabular}
\end{table*}

The conduction band edges in monolayer MoS${}_2$, MoSe${}_2$, WS${}_2$, and WSe${}_2$ are located at the $K$ points, but accompanied by local dispersion minima that appear near the six inequivalent points $\tau\vec{Q}$, $\tau C_3\vec{Q}$ and $\tau C_3^2\vec{Q}$, where $\tau = \pm 1$, and $\vec{Q}=\tfrac{2\pi}{3a}\hat{x}$ is the midpoint between $\Gamma$ and $\KK$ ($a$ is the lattice constant). For a given value of $\tau$, there are three valleys connected by $C_3$ rotations about the BZ center [Fig.\ \ref{fig:figure1}(c)], such that we need only describe the dispersion near the two points $\tau \vec{Q}$, which are related by time reversal.

\begin{figure}[t!]
	\begin{center}
		\includegraphics[width=0.92\columnwidth]{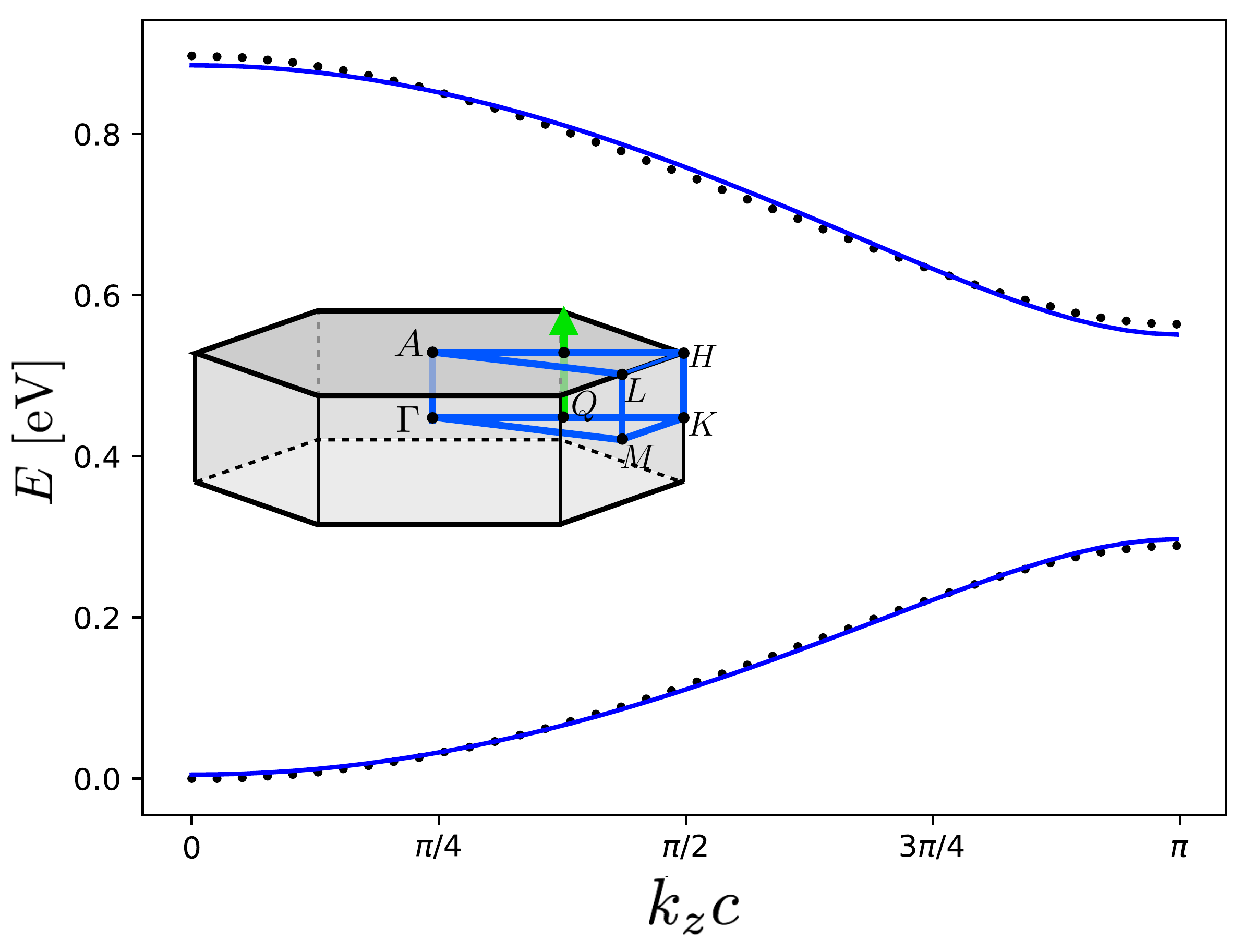}
		\caption{Bulk dispersion of 2H-stacked WS${}_2$ along the Brillouin zone path $\overline{Q A}$, defined by $k_x=k_y=0$ and $k_z \in [0;\,\pi/c]$, where $k_x$ and $k_y$ are measured relative to the $Q$ point, as shown in the inset. DFT data (points) are well fitted by the model Eq.\ (\ref{eq:QH_bulk}) (solid line). Inset: the first Brillouin zone of bulk TMDs.}
		\label{fig:CBbulk_main}
	\end{center}
\end{figure}

For spin projection $s$, the monolayer dispersion near the $\tau \QQ$ valley is given by \cite{kdotp}
\begin{equation}\label{eq:monolayer_conduction}
	E_{s}^{\tau}(\kk)=\frac{\hbar^2(k_x-q_{s}^\tau)^2}{2m_{x,s}^\tau} + \frac{\hbar^2 k_y^2}{2m_{y,s}^\tau} + E_{0}+\tau s \Delta_0,
\end{equation}
where $m_{x,s}^\tau$ and $m_{y,s}^\tau$ are effective masses; $E_0$ is a constant energy shift; $2\Delta_0$ is the spin-orbit splitting between the spin-up and down band edges; and $q_{s}^\tau$ is the band-edge momentum relative to the valley along the $\hat{x}$ axis. From time-reversal symmetry we obtain the dispersion for the opposite valley as $E_{s}^{\tau}(\kk)=E_{-s}^{-\tau}(-\kk)$, which requires ($\alpha = x,\,y$) $m_{\alpha,s}^{\tau}=m_{\alpha,-s}^{-\tau}$ and $q_s^{\tau}=-q_{-s}^{-\tau}$.

As described in Sec.\ \ref{sec:struct}, the 2H-stacked bilayer consists of subsequent layers rotated by $180^{\circ}$ with respect to each other. In reciprocal space, this means that a conduction-band state of spin projection $s$ and momentum $\tau\QQ+\kk$ of the first layer will hybridize with its in-plane inversion partner of spin $s$ and momentum $-\tau\QQ-\kk$ in the second one  [Fig.\ \ref{fig:figure1}(c)]. The multilayer Hamiltonian for the conduction subbands about $\tau\QQ$ is given by
\begin{widetext}
\begin{equation}\label{eq:Qhamil}
\begin{split}
H_{NQ}^{\tau}({\bf k}) &= \sum_{n=1}^{\lceil N/2 \rceil}\sum_{s=\uparrow,\downarrow}
\left[E_s^{\tau}({\bf k}) + \left(\delta_{n,1} + \delta_{n,\lceil N/2\rceil} \right)\delta E \right]
\left[a^{\dagger}_{n\tau s}({\bf k})a_{n\tau s}({\bf k})+\Theta(\tfrac{N}{2}-n)b^{\dagger}_{n,-\tau,-s}(-{\bf k})b_{n,-\tau,-s}(-{\bf k})\right]
\\
&+\sum_{n=1}^{\lceil N/2 \rceil}\sum_{s=\uparrow,\downarrow}t_{\tau}({\bf k})\Theta(\tfrac{N}{2}-n)\left[b^{\dagger}_{n,-\tau,s}(-{\bf k})a_{n\tau s}({\bf k})+\text{H.c.}\right]+\sum_{n=1}^{\lceil N/2 \rceil-1}\sum_{s=\uparrow,\downarrow}t_{\tau}^*({\bf k})\left[b^{\dagger}_{n,-\tau,s}(-{\bf k})a_{n+1,\tau, s}({\bf k})+\text{H.c.}\right]\\
&+\sum_{n=1}^{\lceil N/2 \rceil-1}\sum_{s=\uparrow,\downarrow}t'[a^{\dagger}_{n+1,\tau,s}({\bf k})a_{n\tau s}({\bf k})+b^{\dagger}_{n+1,-\tau,s}(-{\bf k})b_{n,-\tau,s}(-{\bf k})]\\
& + \sum_{n=1}^{\lceil N/2 \rceil}\sum_{s=\uparrow,\downarrow}\left[U_{2n-1}\,a_{n\tau s}^\dagger(\kk) a_{n\tau s}(\kk) + U_{2n}\Theta(\tfrac{N}{2}-n)b_{n,-\tau,-s}^\dagger(\kk)b_{n,-\tau,-s}(\kk) \right],
\end{split}
\end{equation}
\end{widetext}
where $a^{(\dagger)}_{n,\tau,s}({\bf k})$ and $b^{(\dagger)}_{n,\tau,s}({\bf k})$ annihilate (create) electrons of spin projection $s$, in-plane wave vector $\kk$ and valley quantum number $\tau$, on the odd and even layers of the $n^{\rm th}$ bulk unit cell.
The alternation of spin indices and hopping terms are a result of 2H stacking. The model is parameterized by the terms in $t_\tau(\kk)$ given in Eq.\ (\ref{eq:hopping}), the interlayer pseudo-potential $\delta E$, implemented as an on-site energy shift at the boundary layers, and the next-nearest-neighbor hopping $t'$ included to improve the fitting to DFT bands. The interlayer hopping has the form (see Appendix \ref{app:TR})
\begin{equation}\label{eq:hopping}
	t_\tau(\kk) = t_{0} +\tau t_{1}k_x + iu_{1} k_y + t_{2}k_x^2 + u_{2} k_y^2,
\end{equation}
up to second order in the in-plane crystal momentum. Given the lack of $\sigma_h$ symmetry for even $N$, the spin projection $s$ is, strictly speaking, not a good quantum number, and spin mixing is allowed. This is discussed in Appendix \ref{app:TR}. However, using an expansion about the $Q$ point in our DFT results shows that spin mixing is much weaker \cite{polini_prb_2014} than $t_{\tau}(\kk)$, and can be neglected. We also found $u_1$ to be several orders of magnitude smaller than $t_1$; as a result, we consider $t_\tau(\kk)$ to be real.

Finally, in the last term of Eq.\ \eqref{eq:Qhamil} we take into account electrostatic doping effects through the layer-dependent potential energy $U_n$ ($1\le n \le N$)  \cite{sam_subbands_2018}:
\begin{equation}\label{eq:SelfConsistent}
\begin{split}
	&U_n = U_1 + e d\sum_{m=2}^N \mathcal{E}_{m-1,m};\quad n>1,\\
	&\mathcal{E}_{m-1,m} = \frac{3e}{\varepsilon_0}\sum_{l=m}^{N}\sum_{\tau=\pm1}\sum_{s=\uparrow,\downarrow}\rho_{l}^{s\tau}.
\end{split}
\end{equation}
Here, $\rho_{l}^{s\tau}$ is the electron (number) density induced in layer $l$ belonging to the spin-$s$ subbands in valley $\tau$. The factor of 3 in the second equation comes from the valley degeneracy, which is preserved even in the presence of an out-of-plane electric field.

\begin{figure}[!t]
	\begin{center}
		\includegraphics[width=\columnwidth]{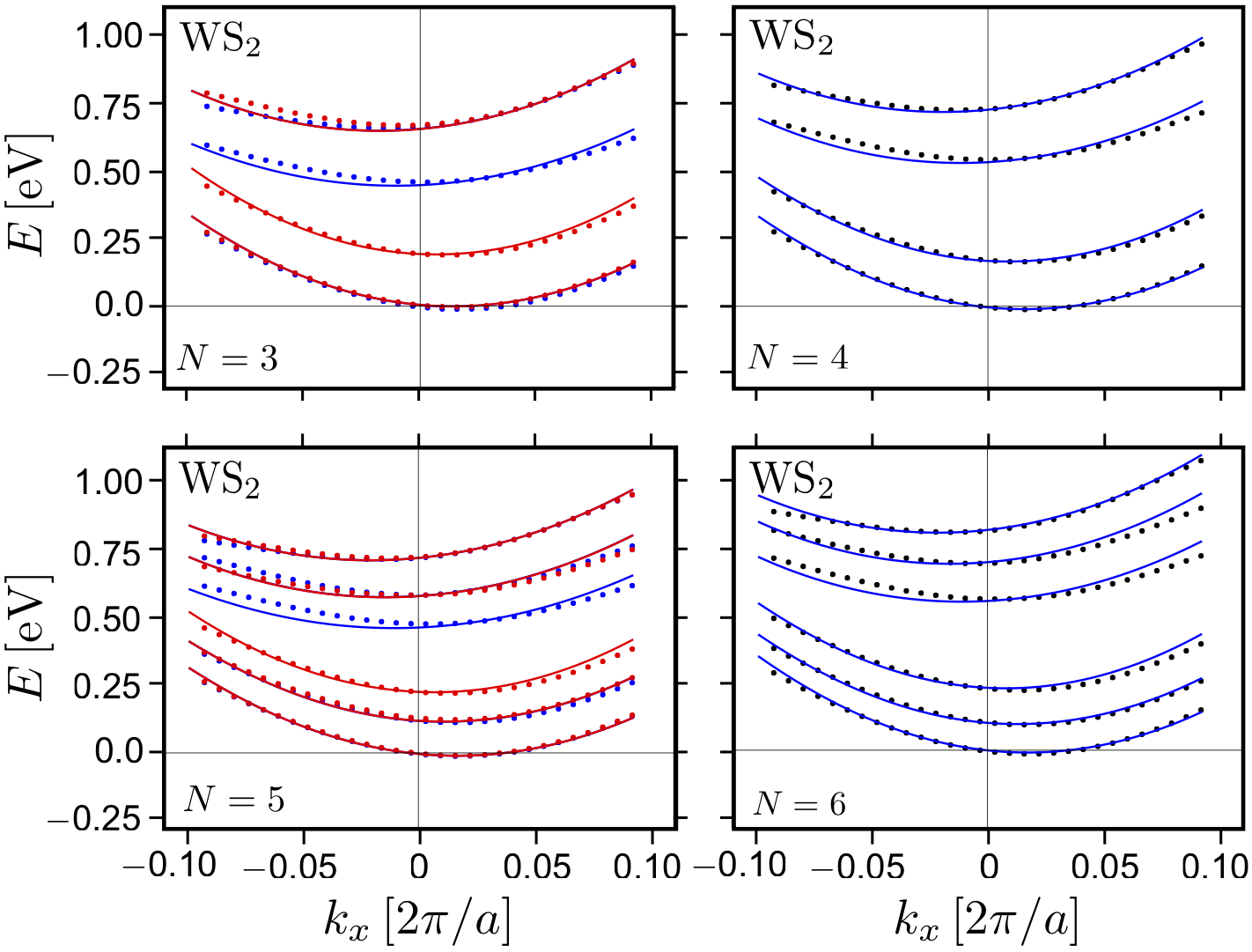}
		\caption{HkpTB model dispersions (solid lines) fitted to DFT results for the WS${}_2$ conduction subbands (points) near the $Q$ point ($k_x=0$), representative of all four TMDs. Results are shown for number of layers $N=3$ to $6$. For odd $N$, line and point colors indicate subbands with different spin projections, with blue (red) corresponding to spin down (spin up). For even $N$, spin-up and -down subbands are degenerate, and shown with black dots.  Fittings for all four TMDs can be found in Ref.\ \cite{supplement}.}
		\label{fig:Qfits}
	\end{center}
\end{figure}

\begin{table}[t!]
	\centering
	\caption{Model parameters fitted to DFT data for the conduction band interlayer hopping terms. $\delta E$ is an energy offset for the first and last layers of the structure that accounts for surface effects.}
	\label{tab:CBHoppingFits}
	\begin{tabular*}{\columnwidth}{@{\extracolsep{\stretch{1}}}*{7}{lccc}@{}}
		\hline\hline
		& $t_0\,[\mathrm{eV}]$ & $t_1\,[\mathrm{eV}\Ams]$ & $t_2\,[\mathrm{eV}\Ams^{2}]$ \\
		& $t'\ [{\rm meV}]$ & $u_2\,[\mathrm{eV}\Ams^{2}]$ & $\delta E\,[\mathrm{meV}]$ \\
		\hline
		${\rm MoS_2}$		& 0.203	&  0.213	& 0.0419\\
		&12.7	& -0.662 	& 8.90\\
		${\rm MoSe_2}$	& 0.215	& 0.180	& -0.145\\
		& 20.5	& -0.447	& -4.29 \\
		${\rm WS_2}$		& 0.210	& 0.233	& -0.123\\
		& 5.24	& -0.864	& -3.95\\
		${\rm WSe_2}$	& 0.211	& 0.209	& 0.231\\
		& 9.54	& -0.797	& 4.21\\
		
		\hline\hline
	\end{tabular*}
\end{table}
In the bulk limit, and in the absence of external electric fields ($U_n=0$), we have the bipartite Hamiltonian
\begin{equation}\label{eq:QH_bulk}
\begin{split}
	H_{Q}^{\tau}(\kk,k_z)=&\varepsilon_0(\kk,k_z)s_0\pi_0+ \tau\Delta(\kk)s_3\pi_3\\
	&+2t_\tau(\kk)\cos{\left(\frac{k_z c}{2} \right)}s_0\pi_1,\\
	\varepsilon_{0}(\kk,k_z) =& \frac{E_\uparrow^+(\kk)+E_\downarrow^+(\kk)}{2} + 2t'\cos{(k_z c)},\\
	\Delta(\kk) =& \frac{E_\uparrow^+(\kk)-E_\downarrow^+(\kk)}{2},
\end{split}
\end{equation}
where $\Delta(\kk)$ is the $\kk$-dependent monolayer spin-orbit splitting for wave vector ${\bf k}$ measured relative to the $Q$ point; $s_i$ and $\pi_i$ ($i=0$ to $3$) are Pauli matrices acting on the spin and layer degrees of freedom, respectively, and $s_0$ and $\pi_0$ are the identity in their corresponding subspaces. The model parameters for the four TMDs were fitted to the DFT-calculated monolayer and 3D bulk dispersions, and are presented in Tables\ \ref{tab:CBFits} and \ref{tab:CBHoppingFits}. A sample bulk fitting is shown in Fig.\ \ref{fig:CBbulk_main} for ${\rm WS_2}$, along the path defined by $k_x=k_y=0$ ($Q$ point) and $k_z\in [0,\pi/c]$, with the solid line corresponding to the model Eq.\ (\ref{eq:QH_bulk}). A sample comparison between our HkpTB model to DFT results for WS${}_2$ few-layer structures, representative of all four TMDs, is shown in Fig.\ \ref{fig:Qfits}. Detailed comparisons for few-layer films of all four materials are available in Ref.\ \cite{supplement}.

As discussed in Sec.\ \ref{sec:struct}, the global symmetry alternation between $\sigma_h$ for odd $N$, and spatial inversion symmetry for even $N$, results in the striking qualitative differences between the cases with even and odd number of layers in Fig.\ \ref{fig:Qfits}. 
The two-fold spin degeneracy observed for even $N$ is a consequence of spatial inversion and time reversal symmetry, resulting in $E^{\tau}_s(\kk)=E^{\tau}_{-s}(\kk)$. By contrast, $\sigma_h$ mirror symmetry for $N$ odd makes $s$ a good quantum number, while the lack of inversion symmetry allows for spin-orbit splitting. Notice also that the two middle spin-split subbands remain fixed for all odd values of $N$, while the rest of the bands are nearly spin-degenerate. As discussed in Appendix \ref{app:Nodd}, these features can be traced back to the SO splitting in the monolayer case, and the particular form of Hamiltonians $H_{NQ}^{\tau}(\kk)$ for odd $N$.

Expanding the lowest eigenvalue of Eq.\ (\ref{eq:QH_bulk}) for valley $\tau$ about $k_z=0$, corresponding to the bulk conduction band edge (Fig.\ \ref{fig:CBbulk_main}), the dispersion can be written as
\begin{equation}\label{eq:HQNn}
\begin{split}
	E_{Q}^\tau(\kk,k_z) \approx& \frac{\hbar^2}{2m_{c,x}}(k_x-\tau[\kappa_{0}-\beta k_z^2])^2\left(1+\zeta_x k_z^2\right)\\
	&+\frac{\hbar^2k_y^2}{2m_{c,y}}\left(1+\zeta_y k_z^2 \right) + \frac{\hbar^2k_z^2}{2m_{c,z}} + E_{Q}^{0},
\end{split}
\end{equation}
where $m_{c,z}$ is the out-of-plane bulk effective mass; $m_{c,x},\, m_{c,y}$ are the in-plane effective masses in the $x$ and $y$ directions, respectively; $\zeta_x, \,\zeta_y$ are anisotropic non-linearity factors; $\kappa_0$ and $\beta$ account for the band minimum offset from the $Q$ point along the $\hat{x}$ direction; and $E_Q^0$ is a constant energy shift. These constants are related to our HkpTB model parameters through the expressions [see Eq.\ (\ref{eq:monolayer_conduction})]
\begin{subequations}
\begin{equation}\label{eq:mcx}
\begin{split}
	m_{c,x}=& \frac{2m_{x,\uparrow}^\tau m_{x,\downarrow}^\tau}{m_{x,\uparrow}^\tau + m_{x,\downarrow}^\tau},
\end{split}
\end{equation}
\begin{equation}\label{eq:mcy}
\begin{split}
	m_{c,y}=& \frac{2m_{y,\uparrow}^\tau m_{y,\downarrow}^\tau}{m_{y,\uparrow}^\tau + m_{y,\downarrow}^\tau},
\end{split}
\end{equation}
\begin{equation}\label{eq:mcz}
\begin{split}
	m_{c,z}=& \frac{\hbar^2}{8d^2}\left[\frac{t_0^2}{4\Delta_0}\left(1 - \frac{t_1^2}{t_1^2 + 2t_0t_1} \right) - t' \right.\\
	& \,+ \left. \frac{t_1^2+2t_0t_2}{4\Delta_0} \left(\kappa_0+\frac{ t_0t_1}{t_1^2+2t_0t_2}\right)^2\right]^{-1},
\end{split}
\end{equation}
\begin{equation}\label{eq:kappa0}
\begin{split}
	\kappa_{0}=&\frac{m_{x,\uparrow}^+ q_{\downarrow}^+ + m_{x,\downarrow}^+ q_{\uparrow}^+}{m_{x,\uparrow}^+ + m_{x,\downarrow}^+},
\end{split}
\end{equation}
\begin{equation}\label{eq:kappa2}
	\beta=\frac{2 m_{c,x} d^2}{\hbar^2}\frac{t_1^2+2t_0t_2}{\Delta_0}\left(\kappa_0 + \frac{t_0t_1}{t_1^2+2t_0t_2} \right),
\end{equation}
\begin{equation}\label{eq:zetacx}
	\zeta_x = \frac{2 m_{c,x} d^2}{\hbar^2}\frac{t_1^2+2t_0t_2}{\Delta_0},
\end{equation}
\begin{equation}\label{eq:zetacy}
	\zeta_y = \frac{4m_{c,y} d^2}{\hbar^2}\frac{t_0u_2}{\Delta_0},
\end{equation}
\begin{equation}\label{eq:EQ0}
\begin{split}
	E_Q^{0} = E_0 - 2\Delta_0
	 +\frac{\hbar^2}{4}\frac{(q_{\uparrow}^\tau+q_{\downarrow}^\tau)^2}{m_{x,\uparrow}^\tau+m_{x,\downarrow}^\tau} + 2t'.
\end{split}
\end{equation}
\end{subequations}

Similarly to the subbands on the valence-band side (Sec.\ \ref{sec:Gmodel}), the conduction subbands in TMD films with $N\gg 1$ can be analyzed by quantizing the electron states in a slab of finite thickness $L=Nd$, with dispersions described by Eq.\ (\ref{eq:HQNn}). However, note that the coefficients of Eqs.\ (\ref{eq:mcx})--(\ref{eq:EQ0}) are independent of spin projection and valley, and thus not representative of the odd $N$ case. This is a consequence of the explicit inversion symmetry of the bulk model (\ref{eq:QH_bulk}). Nonetheless, the SO splitting resulting from the lack of inversion symmetry and the presence of $\sigma_h$ symmetry in a system with odd $N$, can be introduced through the TMD quantum well boundary conditions.

The unit cell for 2H crystals contains two layers, which below we label  A and B [see Eq.\ (\ref{eq:Qhamil})]. For odd $N$, inversion symmetry is broken in opposite ways for the two layers in the unit cell, given that, as discussed in Sec.\ \ref{sec:introduction}, they are rotated by $180^\circ$ with respect to each other. This results in different boundary conditions for electrons at a given termination of the TMD film, depending on whether the final layer is of type A or B. This generalizes the boundary conditions used for the valence band at the $\Gamma$ point [Eq.  (\ref{eq:boundary})] to
\begin{subequations}
\begin{equation}
\left[\pm\left(\nu_0 + s\tau\nu_1 \right)d\partial_z\psi_s^\tau(z) + \psi_s^\tau(z) \right]_{z=\pm\tfrac{L}{2}}=0,
\end{equation}
for the boundary at $z=\pm L/2$ when the film terminates on an A layer, and 
\begin{equation}
\left[\pm\left(\nu_0 - s\tau\nu_1 \right)d\partial_z\psi_s^\tau(z) + \psi_s^\tau(z) \right]_{z=\pm\tfrac{L}{2}}=0,
\end{equation}
\end{subequations}
when the final layer at position $z=\pm L/2$ is of type B. Here, $\nu_0\,,\nu_1\ll N$ are dimensionless parameters. This results in spin- and valley-dependent quantization conditions
\begin{equation}
	k^{s,\tau}_{z,n|N} \approx \frac{\pi n}{d[N+2\nu_0+s\tau\nu_1 \vartheta_N]},
\end{equation}
where $\vartheta_N\equiv1-(-1)^N$ gives $0$ for even $N$ and $2$ for odd $N$.
Overall, the low-energy spectrum of a thin film has the form
\begin{equation}\label{eq:QlargeNE}
\begin{split}
&E^{s,\tau}_{n\ll N|N}({\bf k})=\frac{\hbar^2}{2m_{c,z}}\frac{\pi^2 n^2}{d^{2}[N+2\nu_{0}+s\tau\nu_1\vartheta_N]^2}
\\
&+ \frac{\hbar^2}{2m^{s,\tau}_{c,x; n|N}}(k_x-\kappa_{n|N}^{s,\tau})^2+\frac{\hbar^2k_y^2}{2m^{s,\tau}_{c,y; n|N}}+E^0_Q,
\end{split}
\end{equation}
where the subband in-plane effective masses in the $\alpha=x, y$ directions are
\begin{subequations}
	\begin{equation}\label{eq:CBmasses_N}
\left[m^{s,\tau}_{\alpha,n|N} \right]^{-1} \approx m_{c,\alpha}^{-1}\left[1 + \frac{\zeta_\alpha \pi^2n^2}{d^2[N+2\nu_0+s\tau\nu_1\vartheta_N]^2} \right],
	\end{equation}
and the momentum offset from the $Q$ point is given by
\begin{equation}\label{eq:subbandmomenta}
\kappa^{s,\tau}_{n|N} \approx \tau\kappa_0 + \frac{\tau n^2\pi^2\,\beta }{d^2\left[N +2\nu_0+s\tau\nu_1\vartheta_N \right]^2}.
\end{equation}
\end{subequations}

As in the monolayer case, the low-energy subband dispersions described by Eq.\ (\ref{eq:QlargeNE}) near the six valleys at BZ points $\tau\QQ$, $\tau C_3\QQ$, and $\tau C_3^2 \QQ$, can be divided into two triads  related by time-reversal symmetry, with quantum numbers $\tau = \pm 1$. The three valleys for a given $\tau$ are connected by $C_3$ rotations, as sketched in the inset of Fig.\ \ref{fig:figure1}. As a consequence, for odd number of layers, where inversion symmetry is broken and SO splitting is parameterized by $\nu_1$, the spin and valley degrees of freedom of the bottom subband are locked, and the low-energy states have valley degeneracy of $g_{\rm odd}=6$. Conversely, for even number of layers the bottom subbands are spin degenerate, giving a total degeneracy of $g_{\rm even}=12$. These large subband degeneracies and multi-valley structures, together with the anisotropic dispersions found within each valley, may have important implications for the transport and quantum Hall properties of $n$-doped multilayer TMDs \cite{evenodd}.

\begin{figure}[t!]
\begin{center}
\includegraphics[width=0.88\columnwidth]{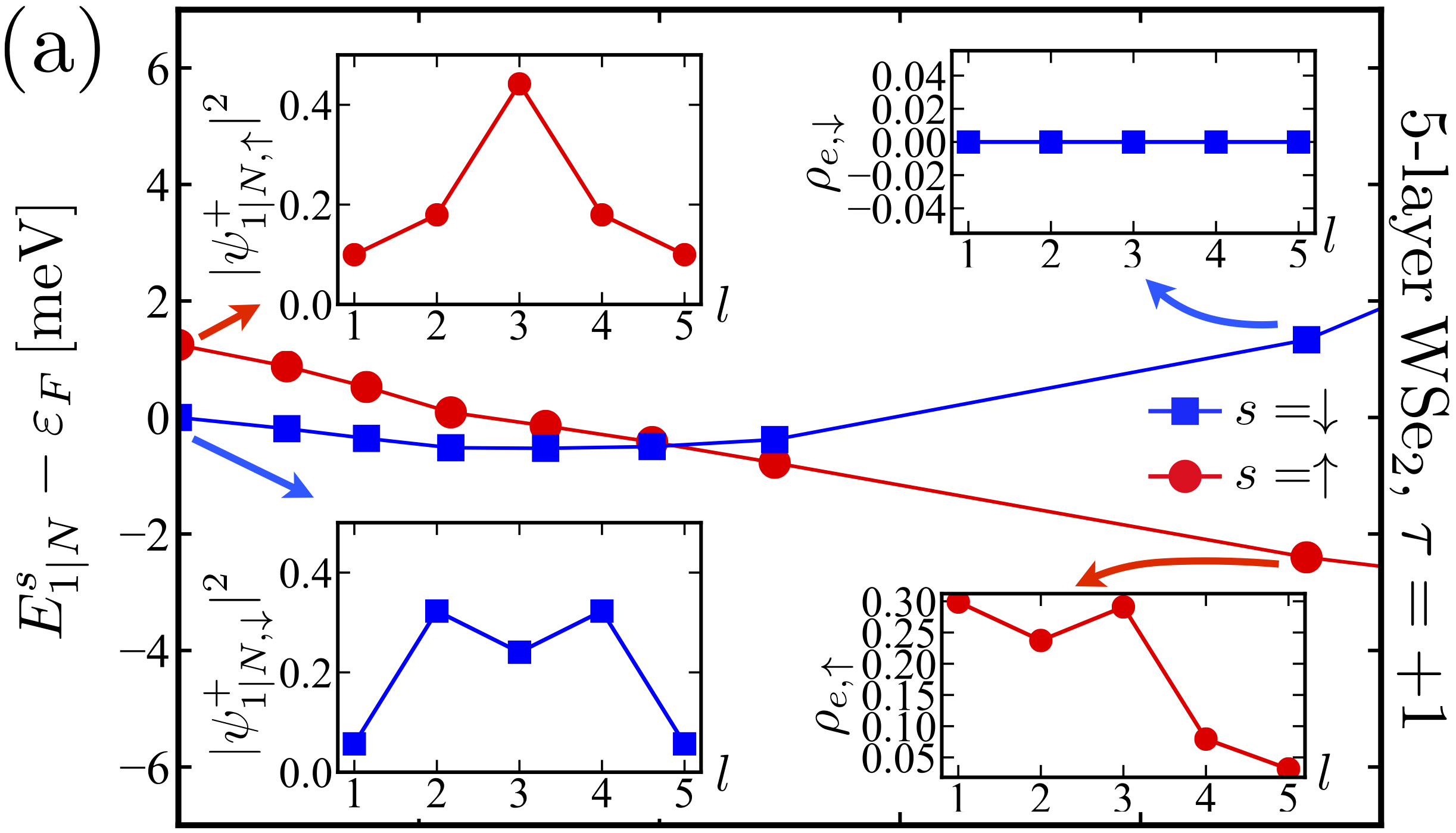}
\includegraphics[width=0.873\columnwidth]{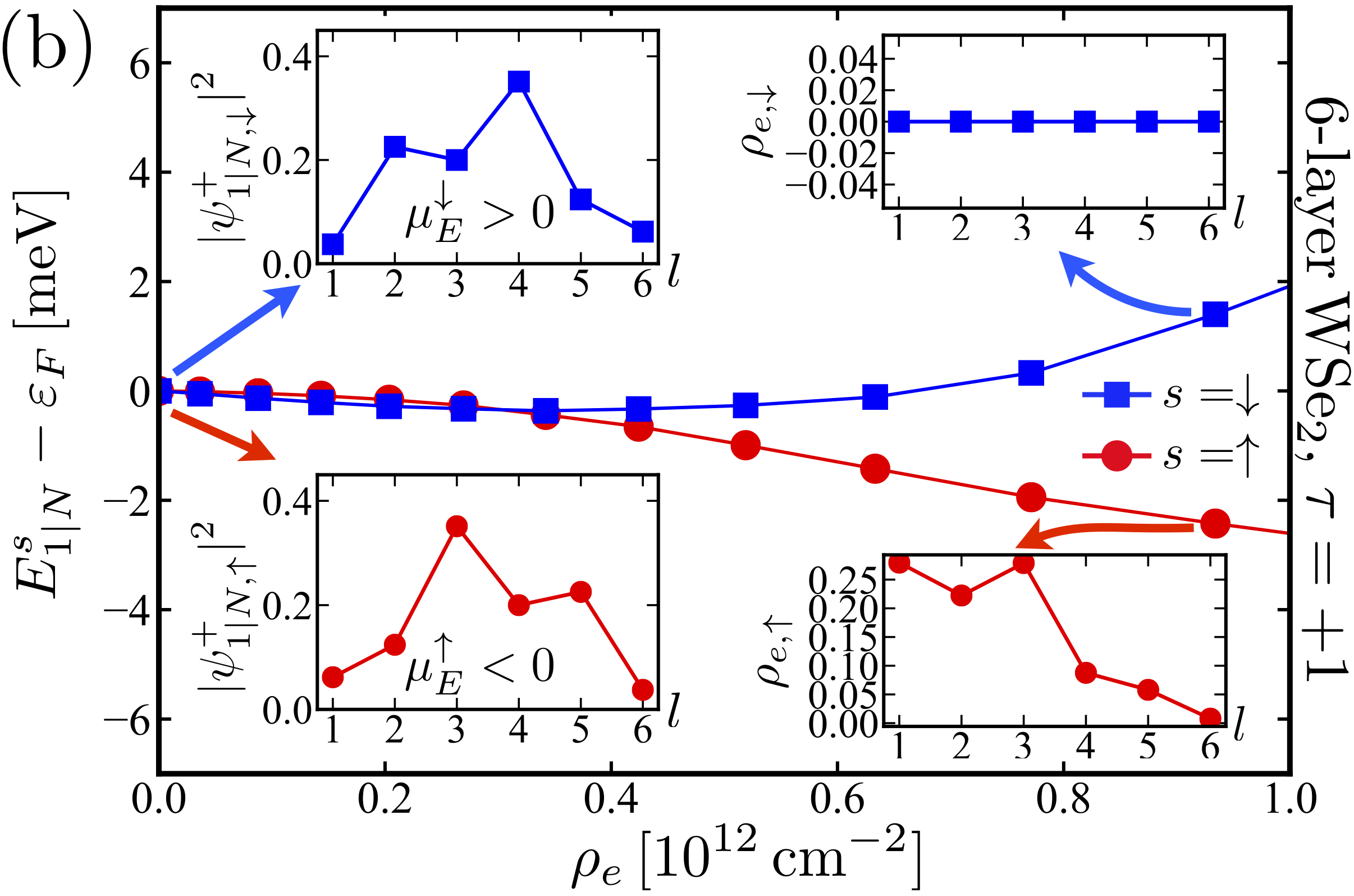}
\includegraphics[width=0.88\columnwidth]{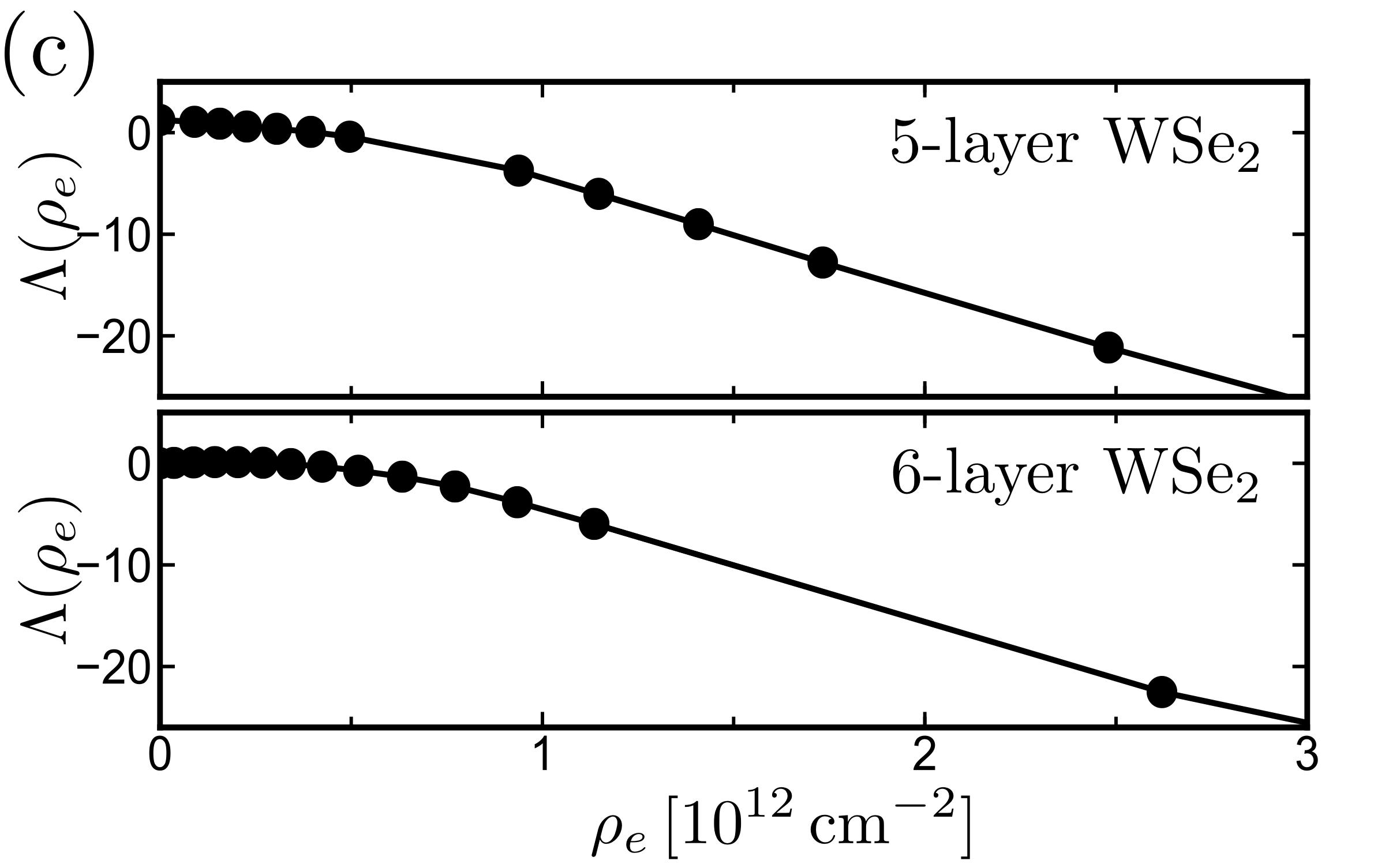}
\includegraphics[width=0.88\columnwidth]{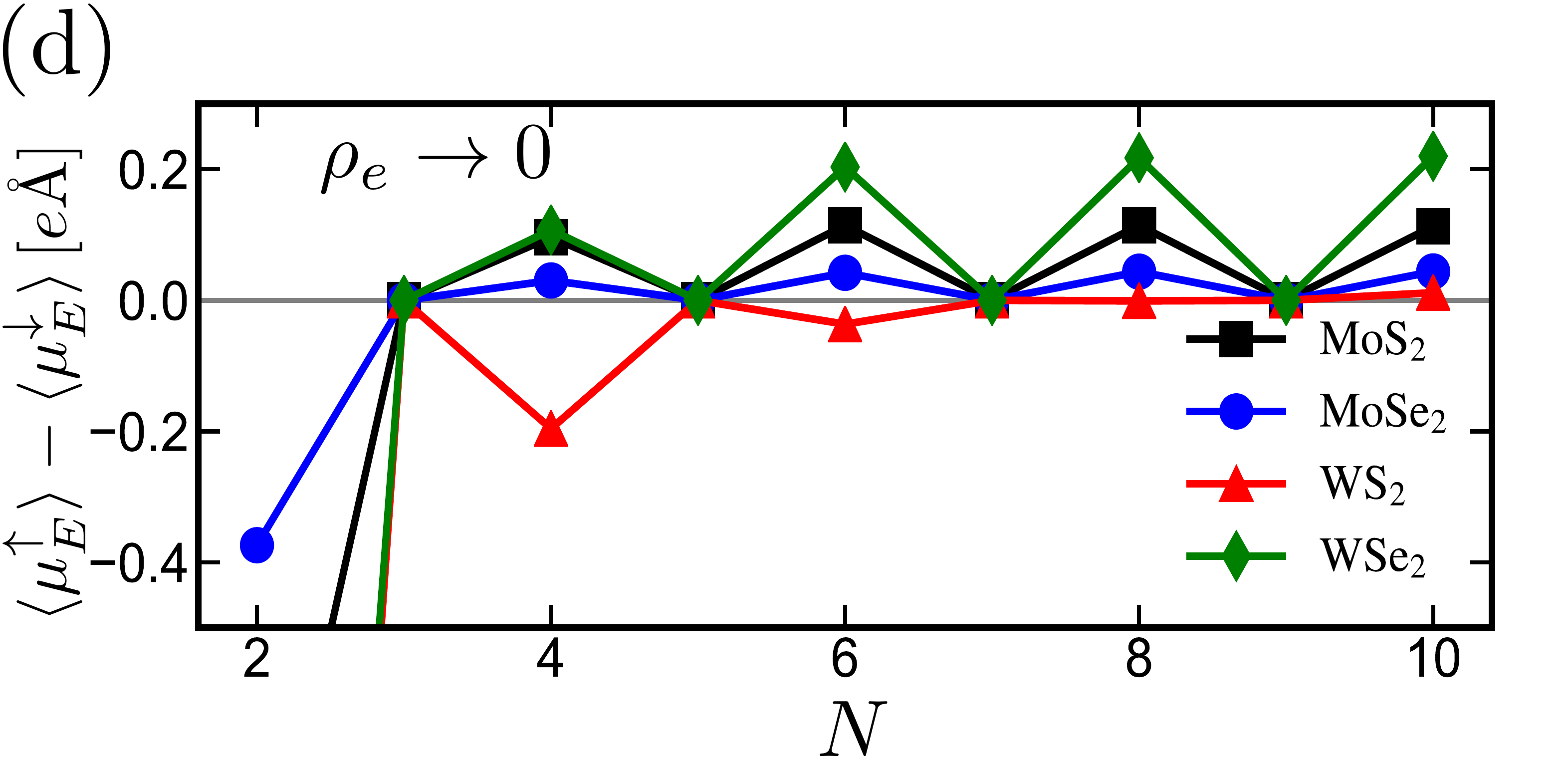}
\caption{Spin-up and -down subband edge energies $E_{1|N}^{s}$ at valley $\tau=+$, measured with respect to the Fermi level, for (a) five-layer and (b) six-layer WSe${}_2$, as a function of electron density $\rho_e$. The potential profiles across the TMD film were determined self-consistently for a temperature of $T=10\,{\rm K}$ for a single (positive) back gate. In both panels, left and right insets show the spin-up and -down charge distributions for $\rho_e\rightarrow 0$ and for finite $\rho_e$, respectively. (c) Corresponding gate-doping-induced spin splittings at valley $\tau=+$. (d) Imbalance between the $\hat{z}$-axis electric dipole moments of the spin-up and -down subband edges for $\rho_e\rightarrow 0$.}
\label{fig:5and6splittings}
\end{center}
\end{figure}

Both inversion and $\sigma_h$ symmetry are broken when an electric field is applied along the $\hat{z}$ axis of the film, leading to a modulation of the spin splittings in the case of odd $N$, and lifting of the spin degeneracy for even $N$. As a first approximation, this effect can be introduced into the lowest subband dispersion by substituting
\begin{equation}\label{eq:Lambda}
	E_{Q}^0\rightarrow E_{Q}^0(\rho_e)+s\tau \Lambda(\rho_e)
\end{equation}
in  Eq.\ \eqref{eq:QlargeNE} for $n=1$, where $\rho_e=3\sum_{s,\tau,n}\rho_n^{s\tau}$ is the total electron density induced by the field. These spin-dependent shifts of the band edges,  abbreviated $E_{1|N}^s\equiv E_{1|N}^{s+}(\kappa_{1|N}^{s+},0)$, parametrize the symmetry breaking by the potential profile. As an example, representative of all four TMDs, Figure \ref{fig:5and6splittings} shows this splitting for the lowest conduction subband of five- and six-layer WSe${}_2$, in the case where the field is induced by a single positive back gate near layer one. The corresponding induced electron densities and potential profiles were determined self-consistently to satisfy Eq.\ \eqref{eq:SelfConsistent}, and chosen inside a range that is easily accessible to experiments.

In the case of even $N$ we can interpret the spin-splitting strength in the weak doping regime in terms of the spatial charge distributions of the lowest-energy spin-up and -down subband states. The left insets in Fig.\ \ref{fig:5and6splittings}(b) show the spread of these two states along the film's $\hat{z}$ axis in the zero-doping limit ($\rho_e\rightarrow 0$). The two distributions are asymmetric with respect to the middle of the film, and related to each other by a $z \rightarrow -z$ mirror operation, resulting in opposite electric dipole moments $\langle \mu_E^\uparrow \rangle =- \langle\mu_E^\downarrow \rangle$ about the middle plane of the multilayer structure. The splitting in the zero-doping limit can be approximated as $\Lambda(\rho_e\rightarrow 0) \approx -(\langle\mu_E^\uparrow\rangle - \langle\mu_E^\downarrow\rangle)E_z>0$, where $E_z>0$ is the electric field in the $\hat{z}$ direction. This is shown in Fig.\ \ref{fig:5and6splittings}(b) for electron densities between 0 and $0.2\times 10^{12}\,{\rm cm}^{-2}$. This linear splitting contribution is weak: even at small doping, it is overcome by non-linear screening effects, which produce a large negative splitting $\Lambda<0$ [Fig.\ \ref{fig:5and6splittings}(c)] and quickly deplete the spin-down subband, as shown in the right insets of Fig.\ \ref{fig:5and6splittings}(b). For odd $N$, where in the zero-doping limit the states of both spin band edges are symmetric about the middle layer of the film [see insets of Fig.\ \ref{fig:5and6splittings}(a)], spin splitting is determined entirely by non-linear effects. This analysis for the $\rho_e \rightarrow 0$ limit can be generalized to all even and odd values of $N$ for all four TMDs, as shown in Fig.\ \ref{fig:5and6splittings}(d), where we show that, in all four cases, the spin-resolved electric dipole moments alternate between finite and zero for even and odd $N$, respectively. Also, note that according to Eq.\ \eqref{eq:Lambda}, the sign of the spin splitting is inverted for opposite valleys.

\subsection{Intersubband transitions and dispersion-induced line broadening in $n$-doped $N$-layer TMDs}\label{sec:Qdispbroad}

\begin{figure}[!t]
	\begin{center}
		\includegraphics[width=1\columnwidth]{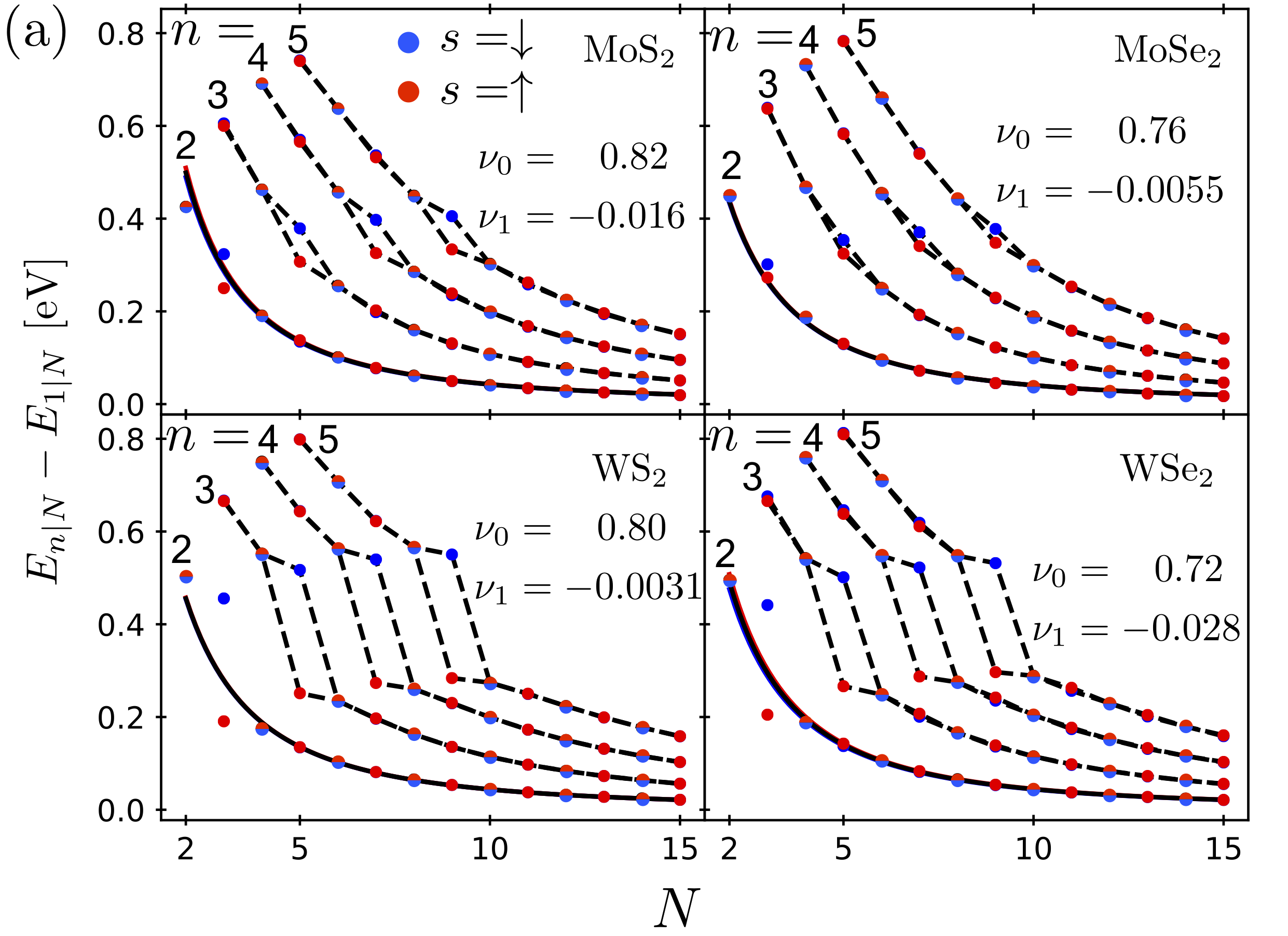}
		\includegraphics[width=0.98\columnwidth]{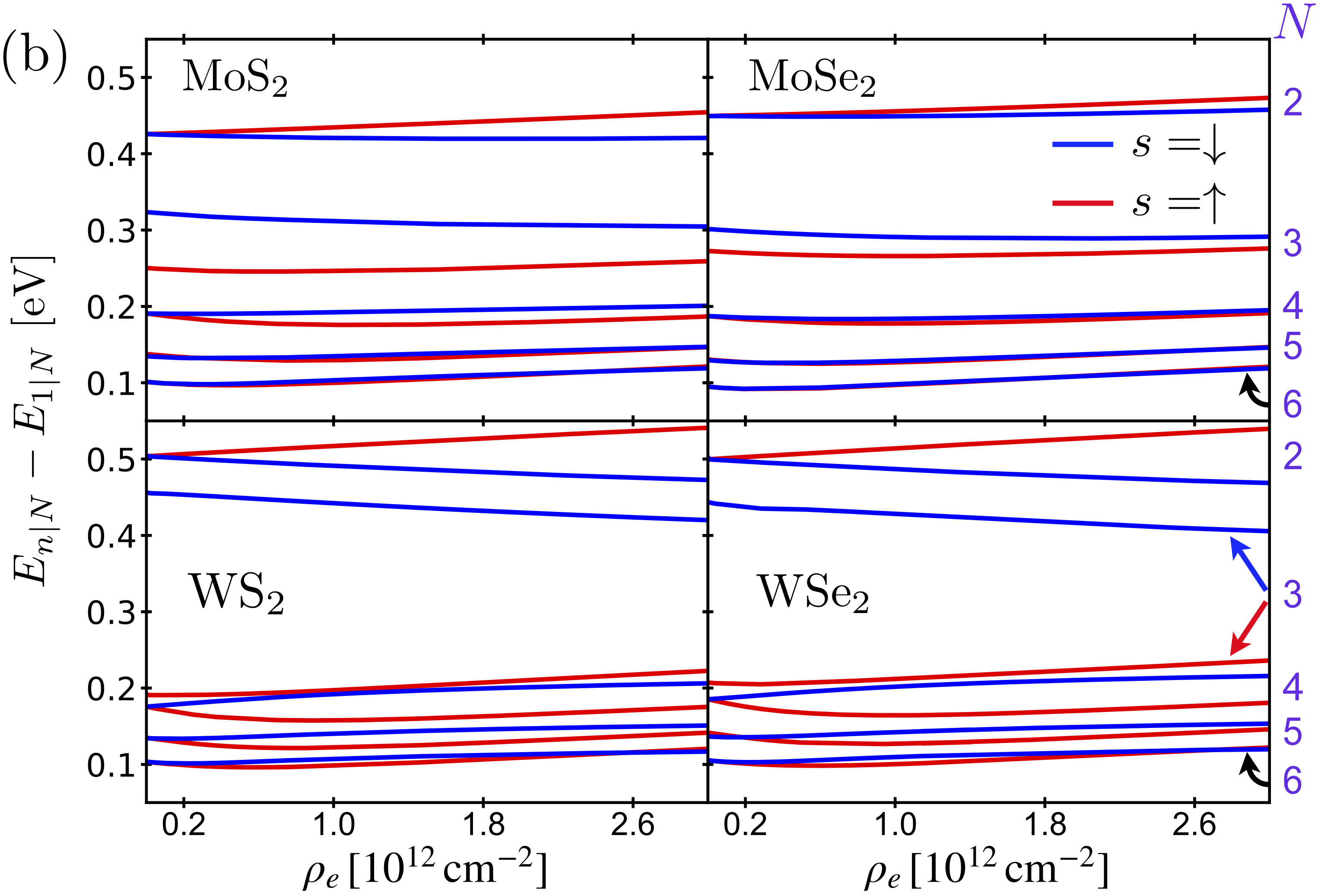}				
		\caption{(a) Energy spacings between the first and $n$th conduction subbands ($n=2$ to 5) for the four TMDs, as functions of the number of layers $N$. The solid lines in each panel corresponds to Eq.\ (\ref{eq:intersub_spacing_Q}) for the main transition between the first and second subbands using the DFT bulk parameters, showing good agreement between the HkpTB model and DFT. Parameters $\nu_0$ and $\nu_1$, fitted for $N\ge 4$, are given for the four TMDs in their corresponding panels. 
			Blue (red) points and solid lines in each panel represent spin-down (-up) polarized subband spacings and fittings.
			Black points and solid line correspond to subband spacings for even $N$ layers, where subbands are spin degenerate. (b) Lowest spin-up and -down subband transition as a function of total electron doping, for film thicknesses $N=2$ to 6 layers. The number of layers $N$ corresponding to each pair of curves is indicated on the right. For two layers, the doping level shifts the spin-up (spin-down) transition energy upward (downward) in energy, following a  linear trend within for moderate doping levels. For $N>2$ non-linear effects begin to appear already for weak doping, and the splitting of the spin-up and -down transitions is inverted with respect to $N=2$.}
		\label{fig:spacing_vs_N_Q}
	\end{center}
\end{figure}

Numerically diagonalizing the HkpTB Hamiltonian in Eq.\ (\ref{eq:Qhamil}) with the parameters of Tables \ref{tab:CBFits} and \ref{tab:CBHoppingFits}, we obtain the energy spacings between the first and next few subbands of TMD films shown in Fig.\ \ref{fig:spacing_vs_N_Q}.
Using Eq.\ (\ref{eq:QlargeNE}), we estimate the separation between the lowest two subbands of a given spin projection $s$ as
\begin{equation}\label{eq:intersub_spacing_Q}
\begin{split}
E^{s,\tau}_{2|N} - E^{s,\tau}_{1|N} &\approx \frac{15\pi^4\hbar^2\beta^2}{2m_{c,x}d^4(N+2\nu_0)^4}+\frac{3\pi^2\hbar^2}{2m_{c,z}d^2(N+2\nu_0)^2}\\
&\qquad-\vartheta_N s\tau\nu_1\frac{6\pi^2\hbar^2}{2m_{c,z}d^{2}(N+2\nu_0)^3}.
\end{split}
\end{equation}
Similarly, we estimate the splitting between the lowest subbands of opposite spin as
\begin{equation}\label{eq:Q11}
E^{-s,\tau}_{1|N} - E^{s,\tau}_{1|N} \approx s\tau\nu_1 \vartheta_N\frac{2\pi^2\hbar^2}{2m_{c,z}d^{2}(N+2\nu_0)^3}.
\end{equation}
We used Eqs.\ (\ref{eq:intersub_spacing_Q}) and (\ref{eq:Q11}) to determine the boundary parameters $\nu_0$ and $\nu_1$ for each of the considered TMDs (MoS${}_2$: $\nu_0=0.82, \nu_1=-0.016$, MoSe${}_2$: $\nu_0=0.76, \nu_1=-0.0055$, WS${}_2$: $\nu_0=0.80, \nu_1=-0.0031$, WSe${}_2$: $\nu_0=0.72, \nu_1=-0.028$). The results are shown with solid lines in Fig.\ \ref{fig:spacing_vs_N_Q}. In the absence of an electric field, the energy splitting between the lowest two spin polarized transitions is of order few ${\rm meV}$ for the four TMDs. However, this splitting is enhanced by electron doping, as shown in Fig.\ \ref{fig:spacing_vs_N_Q}(b) for two- to six-layer films of all four TMDs. For $N=2$, the spin-up (-down) transition energy grows (decreases) linearly with the doping level for small electron densities. The opposite trend is found for $N=3$ to $6$, where the spin-up subband transition appears at lower energy, and non-linear effects begin to appear already for low electron doping. We conclude that the electron doping level can be used as an additional tunable parameter to modify the energies and degeneracies of the lowest optical transitions in 2H-TMDs.

\begin{figure}[t!]
	\begin{center}
		\includegraphics[width=0.98\columnwidth]{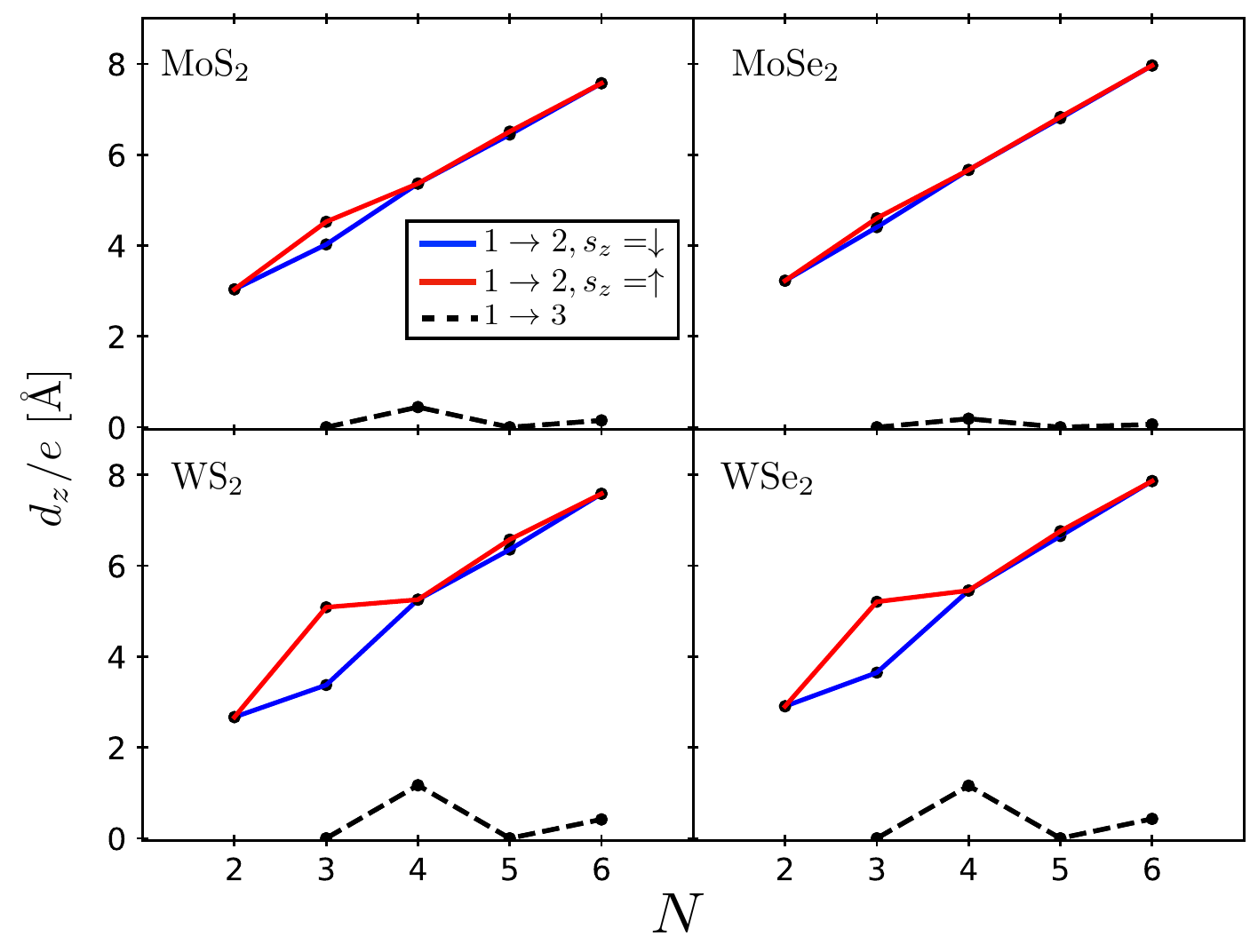}
		\caption{Out-of-plane dipole moment matrix elements for the first two conduction intersubband transitions, $1\rightarrow 2$ (solid line) and $1\rightarrow3$ (dashed line). Transitions between spin-down (-up) subbands are shown in blue (red).}
		\label{fig:dz_plots_Q}
	\end{center}
\end{figure}

Next, we use the model developed above for electron subbands to study intersubband optical transitions, intersubband electron-phonon relaxation, and the intersubband absorption line shapes for IR/FIR light. As discussed in Sec.\ \ref{sec:VBselectionrules}, the optical transition amplitude between two given subbands $n,\, n'$  is determined by the out-of-plane dipole moment
\begin{equation}\label{eq:dipole}
\begin{split}
d_{\tau,s;z}^{n,n'}({\bf k}) =& e \langle n,s;\tau,\kk | z | n',s;\tau,\kk \rangle\\
 & = e\sum_{j=1}^N z_j C^{\tau,s *}_{n,j}({\bf k})C_{n',j}^{\tau,s}({\bf k}),
\end{split}
\end{equation}
where $N$ is the total number of layers, $z_j$ denotes the $z$ coordinate of layer $j$, and $C_{n,j}^{\tau,s}(\kk)$ are the components of the $n^{\rm th}$ subband eigenstate of spin projection $s$ and valley quantum number $\tau$. The calculated dipole moment matrix element as a function of number of layers for the first two intersubband transitions is plotted in Fig.\ \ref{fig:dz_plots_Q}.

Similarly to the valence subbands case, optical transitions in films with odd number of layers $N$ are allowed only between states with opposite-parity subband indices, corresponding to opposite parity under $\sigma_h$ transformation.
The spin-orbit splitting present for odd $N$ results in a spin selection rule, allowing transitions only between subbands with the same out-of-plane spin projection $s$.
For even $N$, where $\sigma_h$ symmetry is absent, transitions between subbands with same-parity indices are allowed. This is in contrast to the VB at the $\Gamma$ point, and is a consequence of the multiple-valley structure of the CB, which makes it possible to form degenerate even and odd (under inversion) combinations of states, giving a finite dipole moment, as shown in  Fig.\ \ref{fig:dz_plots_Q} for the first two intersubband transitions, considering both spin-down and -up polarized subbands. This makes $1|N\rightarrow2|N$ transition the dominant feature in the IR/FIR absorption by thin $n$-doped TMD films.

Similarly to the holes in $p$-doped TMDs, the line shape of the electron intersubband absorption in $n$-doped films is also affected by the difference between the effective masses of subbands $1|N$ and $2|N$. However, in contrast to the case of holes, for electrons the line shapes depend also on the relative in-plane wave vectors of the conduction subband minima, as well as the anisotropic subband dispersions. The resulting broadening for $N$-layer 2H-$MX_2$ films at room temperature, obtained numerically from the calculated line shapes, is shown in Fig.\ \ref{fig:linewidth_phonons_Q}(a). Our calculations show that the aforementioned DOS broadening factors result in a typically larger broadening, which spreads the absorption spectrum towards both lower and higher energies from the main transition.

\subsection{Electron-phonon relaxation and room-temperature absorption spectra in $n$-doped TMD films}\label{sec:Qrelax}

\begin{figure}[t!]
	\begin{center}				
		\includegraphics[width=1\columnwidth]{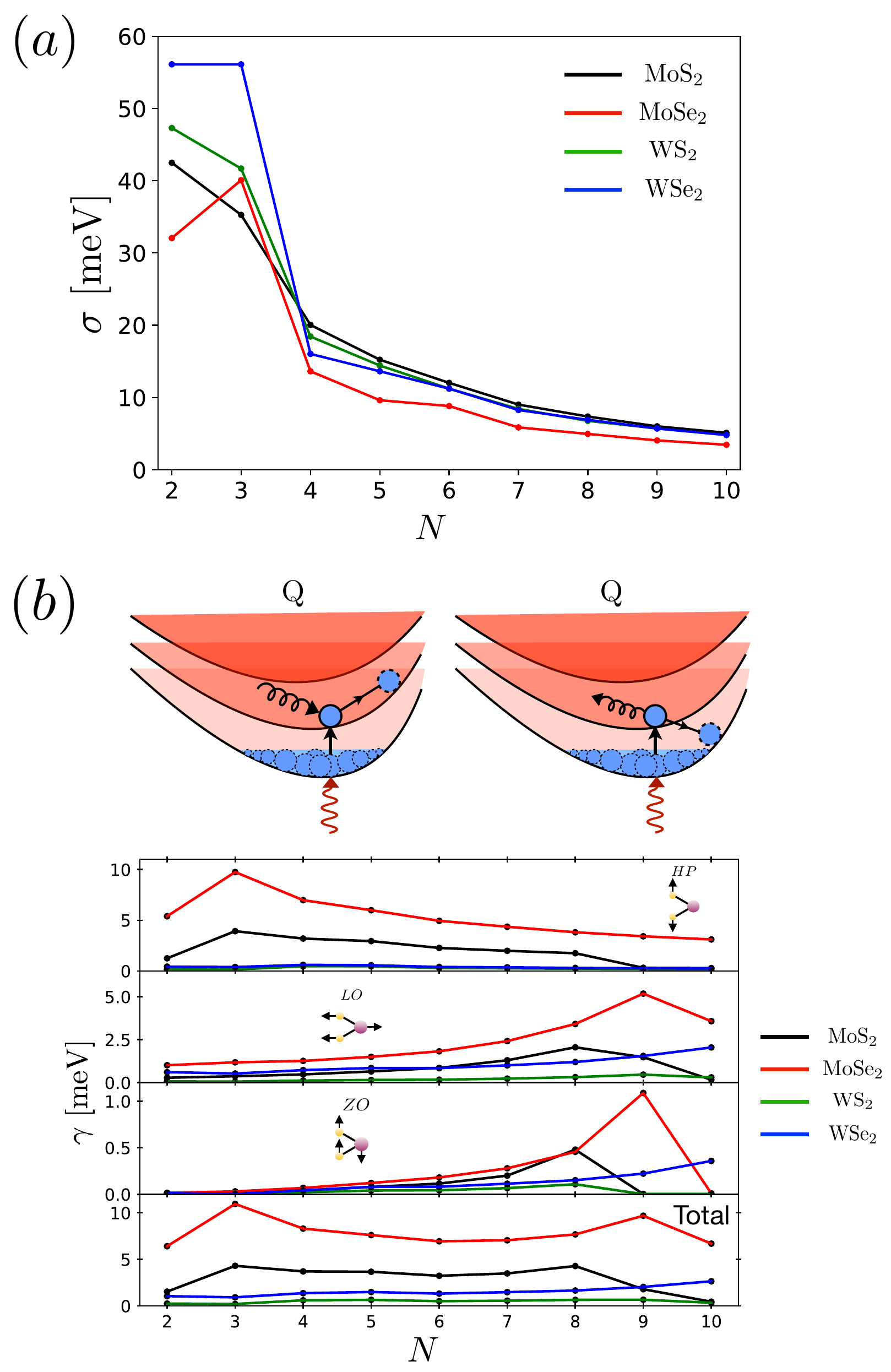}				
		\caption{
			(a) Absorption line widths for CB subbands at room temperature ($T=300\ {\rm K}$) as a function of number of layers for the four TMDs, considering only DOS broadening.
			(b) Phonon-induced broadening at room temperature ($T=300\ {\rm K}$) due to intersubband emission and intrasubband absorption of optical phonons in modes (top to bottom) HP, LO and ZO. Results are shown for the four TMDs as a function of number of layers $N$, with the total broadening shown in the bottom panel.}
		\label{fig:linewidth_phonons_Q}
	\end{center}
\end{figure}

The phonon-induced broadening for conduction subbands $m$ and $n$, generated by intrasubband and intersubband relaxation, is accounted for by
\begin{equation}\label{eq:phononrateQ}
\begin{split}
&\gamma_{n,m}^{\tau,s} = 2\pi \sum_{\mu,\qq,j}
\left|\sum_{i}g^{j,i}_{\mu}(\qq)C_{n,i}^{\tau,s*}(\kappa_{m}^{\tau,s}\hat{x}+\qq)C_{m,i}^{\tau,s}(\kappa_{m}^{\tau,s}\hat{x})
\right|^2
\\
&\times \left\{
[1+n_T(\hbar\omega_{\mu})] \delta\left[E_{m}(\kappa_{m}^{\tau,s}\hat{x})-E_n(\kappa_{m}^{\tau,s}\hat{x}+\qq)-\hbar\omega_{\mu}\right]
\right.
\\
&\left.
+
\delta_{nm}n_T(\hbar\omega_{\mu})\delta\left[E_{m}(\kappa_{m}^{\tau,s}\hat{x})-E_{m}(\kappa_{m}^{\tau,s}\hat{x}+\qq)+\hbar\omega_\mu\right]
\right\},
\end{split}
\end{equation}
where $\kappa_m^{\tau,s}$ is the subband edge offset from the $Q$ point of subband $m$ with spin projection $s$, and $g_{\mu}^{j,i}$ are the electron-phonon couplings for the three phonon modes $\mu$= HP, LO, and ZO given in Eq.\ (\ref{eq:geph}), with $D_v$ replaced by $D_c$ for the HP phonon. The first term in Eq.\ (\ref{eq:phononrateQ}) describes intersubband relaxation due to phonon emission, whereas the second describes intrasubband phonon absorption in the excited subband. The phonon induced broadening for the four TMDs obtained using the electron-phonon coupling parameters in Table\ \ref{tab:phonon_params} are shown in Fig.\ \ref{fig:linewidth_phonons_Q}(b). The dominant contribution comes from intersubband relaxation due to HP and LO phonon modes, with a smaller contribution from the thermally suppressed intrasubband absorption. The large HP phonon deformation potential at the $Q$ point, as compared to its value at the $\Gamma$ point \cite{chinese_phonons}, in particular for ${\rm MoS_2}$ and ${\rm MoSe_2}$ (see Table\ \ref{tab:phonon_params}), results in a large contribution to the broadening. Additional differences between the phonon-induced broadenings for the conduction and valence subbands originate from the different intersubband spacings as a function of number of layers (Figs.\ \ref{fig:spacing_vs_N_G} and \ref{fig:spacing_vs_N_Q}), and different dispersions (Figs.\ \ref{fig:Gfits} and \ref{fig:Qfits}).
As in the valence subbands case, the phonon broadening is most significant for ${\rm MoSe_2}$ due to stronger electron-phonon coupling.

\begin{figure}[t!]
	\begin{center}
		\includegraphics[width=\columnwidth]{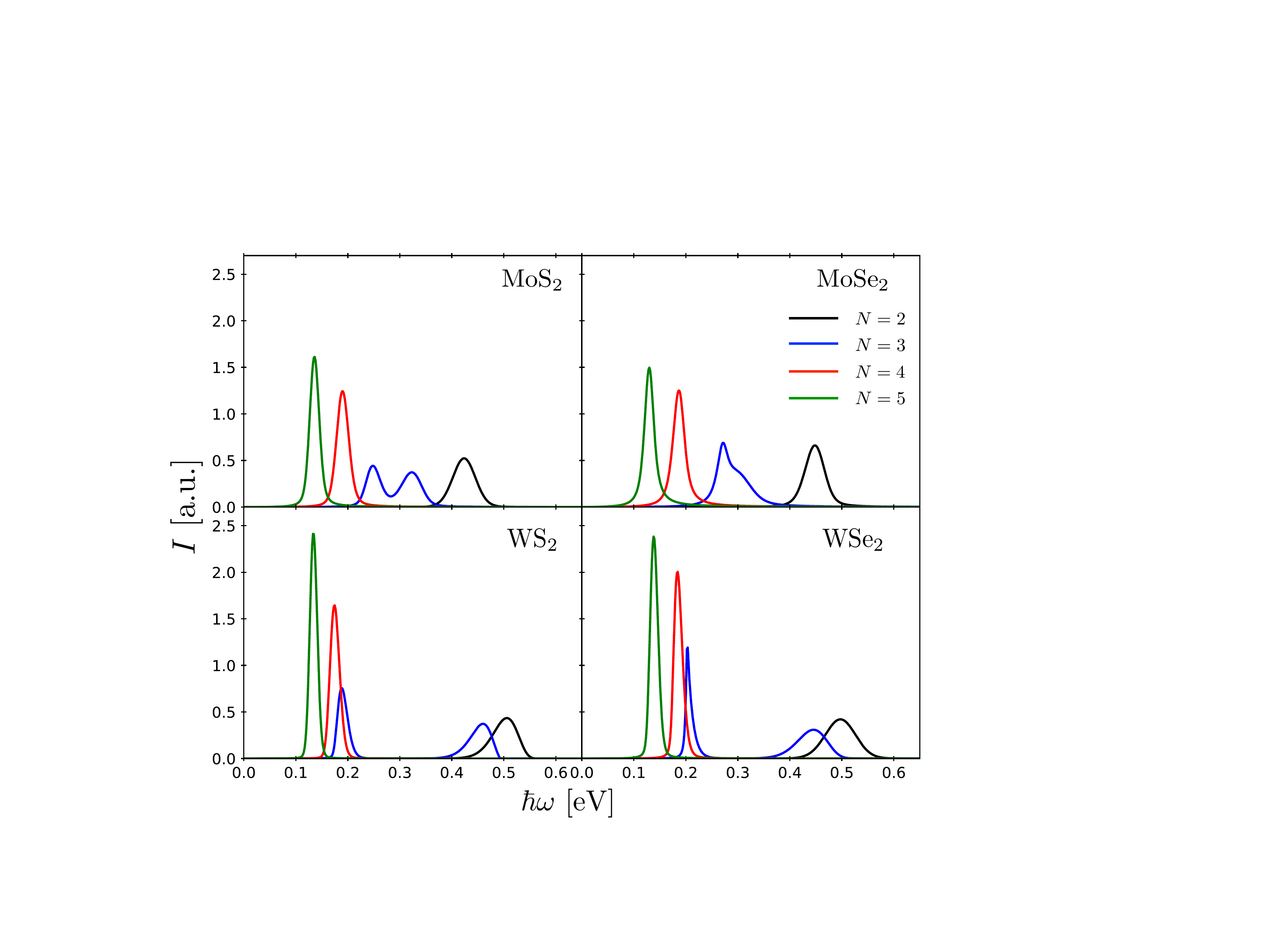}
		\caption{Optical absorption lines for $N=2$-$5$ layers of lightly $n$-doped ${\rm MoS_2, MoSe_2, WS_2}$, and ${\rm WSe_2}$, taking into account intrinsic broadening at room temperature ($T=300\ {\rm K}$). In the case of odd $N$, lines corresponding to different subband spin projections are summed. 
		}
		\label{fig:figure2Q}
	\end{center}
\end{figure}

The absorption spectra of $n$-doped TMD films calculated using Fermi's golden rule, as in Eq.\ (\ref{eq:lineshapeG}), and taking the discussions of Secs.\ \ref{sec:QmodelA},  \ref{sec:Qdispbroad}, and \ref{sec:Qrelax} into account, are shown in Fig.\ \ref{fig:figure2Q} for $N = 2$ to $5$ layers. The predicted absorption spectra for the four TMDs are seen to be more symmetric than those for the holes, primarily due to the effect of different dispersions in consecutive subbands, here aggravated by the shifts $\kappa_{n|N}^{\tau,s}$ and the BZ position of the subbands minima, in addition to the difference between the in-plane subband effective masses. The large SO splitting between the middle two subbands for $N=3$ results in two distinct lines, whereas for $N=5$ the spin-polarized subbands are nearly degenerate, resulting in the overlap of the two lines and giving a combined line with twice the amplitude.

\section{Conclusions}\label{sec:conclusions}
We have presented hybrid k$\cdot$p tight-binding models for the conduction and valence band edges of multilayer TMDs, capable of reproducing the rich low-energy subband dispersions, and allowing us to describe the intersubband optical transitions when coupled to out-of-plane polarized light.  In particular, we find the following:
\begin{itemize}
\item The subbands at the CB edge are found near the $Q$ valleys of the Brillouin zone, whereas the valence band edge is found at the $\Gamma$ point. The main differences between the two sets of subbands are due to the significant spin-orbit splitting, multi-valley structure, and anisotropic dispersions of the conduction subbands, by contrast to the valence subbands. These differences manifest themselves in the  absorption line shapes and additional selection rules, particularly for odd number of layers, where spin-orbit splitting is present. 

\item The four studied TMDs were found to have main intersubband transition energies for the conduction and valence subbands, which densely cover the spectrum range of wavelengths from $ \lambda = 2\ {\rm \mu m}$ to $ 30\ {\rm \mu m}$ ($\hbar\omega=40$ to $700\, {\rm meV}$), for $N=2$ to $7$ layers. This allows tailoring structures of a specific material, appropriate type of doping, and number of layers for a particular device application, from IR to the THz range.

\item Two contributions to the absorption line-shape broadening are identified. The first, broadening due to intersubband phonon relaxation, is found to produce a meV limit to the intersubband linewidth. This is in contrast to III-V quantum wells, where phonon broadening is found to be more damaging to the intersubband transition line quality factor \cite{qw1}. A second, elastic contribution to the line broadening caused by the different 2D masses of carriers in consecutive subbands yields a thermal broadening of the order of $k_{\rm B}T$. Similarly to inhomogeneous broadening, this effect can be reduced by coupling the transition in the film to a standing wave of light in a high-$Q$ resonator.
\end{itemize}

\begin{figure}[t!]
	\begin{center}
		\includegraphics[width=0.95\columnwidth]{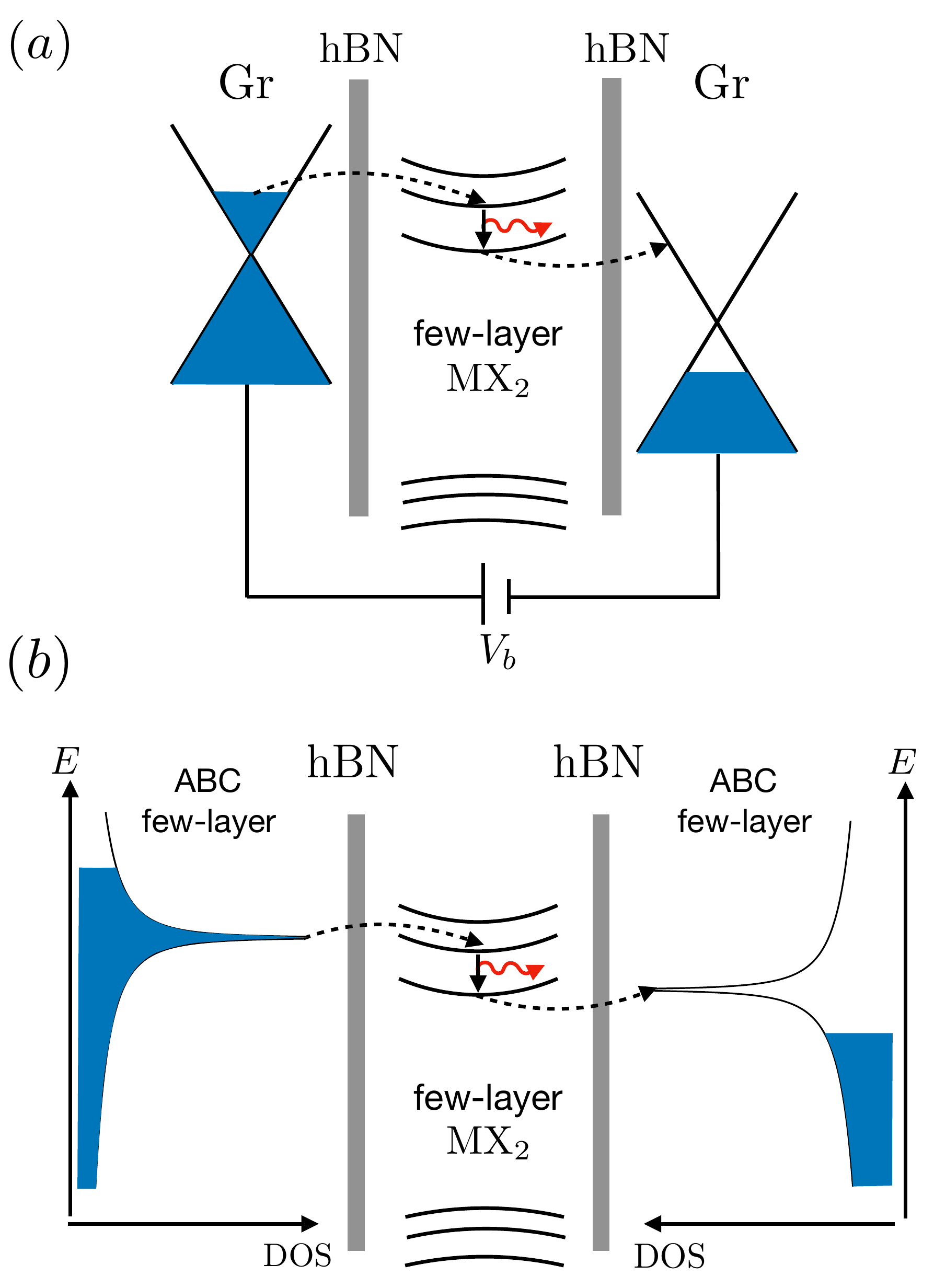}
		\caption{Proposed device application for intersubband transitions in few-layer TMDs. (a) Few-layer TMDs encapsulated between two hexagonal boron nitride ($h$BN) crystals and two graphene (G) electrodes with an applied bias voltage between them. The applied bias voltage allows to realize light emission through intersubband transitions in the few-layer TMD system, by carriers tunneling between the two graphene electrodes.
		(b) An alternative realization using few-layer ABC-stacked graphene instead of monolayer, utilizing the Van Hove singularity in the density of states. The bias voltage aligns the Van Hove singularities near the second and first subbands, making the desired emission process more favorable.
		}
		\label{fig:diag_app}
	\end{center}
\end{figure}

Finally, we propose a specific design of van der Waals multilayer structure utilizing the intersubband transitions in atomically-thin films of TMDs. The sketch in Fig.\ \ref{fig:diag_app} depicts the band configuration of a few-layer transition-metal dichalcogenide film, encapsulated by hexagonal boron nitride ($h$BN) and placed between two graphene electrodes. Applying a bias (and possibly also gate) voltage between the two electrodes results in a shift of the Dirac points relative to each other, and allows for the alignment of the Dirac point of the ``top'' graphene electrode with the lower-energy subband in the TMD, while keeping the Fermi level in graphene above the higher-energy subband. The carriers can then tunnel from the graphene electrode into the higher-energy subband. Once in the excited subband state, the carrier can undergo an intersubband transition, emitting light polarized in the out-of-plane direction, followed by tunneling to the second graphene electrode from the bottom subband state.

A potentially more favorable realization of the above process which avoids carrier loss directly from the second subband, or carrier tunneling into the bottom subband, involves using ABC-stacked few-layer graphene. The band structure of ABC few-layer graphene has a Van Hove singularity in its density of states at the edge between conduction and valence bands. Aligning the Van Hove singularities of two such electrodes with the second and first subbands, respectively, would enable one to achieve preferential injection and extraction of carriers into/from the TMD film, thus offering a new way to produce functional optical fiber cables.

The proposed ``LEGO''-type design of IR/THz emitting materials has potential for implementation as part of a composite optical fiber, where the coupling to the out-of-plane polarized photon would be supported by the wave-guide mode.

\acknowledgements{The authors acknowledge funding by the European Union's Graphene Flagship Project, ERC Synergy Grant Hetero 2D, EPSRC Doctoral Training Centre, Graphene NOWNANO, and the Lloyd Register Foundation Nanotechnology grant, as well as support from the N8 Polaris service and the use of the ARCHER supercomputer (RAP Project e547) and the Tianhe-2 supercomputer (Guangzhou, China). The authors would like to thank F.\ Koppens, K.\ Novoselov, R.\ Gorbachev, M.\ Potemski, R.\ Curry, C.\ Kocabas, S.\ Magorrian and A.\ Ceferino for fruitful discussions.}

\appendix
\section{Symmetry constraints for the bilayer Hamiltonians}\label{app:TR}
For $N=2$ (bilayer), the Hamiltonian must be invariant under spatial inversion $\mathcal{P}$, $x \rightarrow -x$ mirror symmetry $\mathcal{D}(\sigma_v)$, and time reversal $\mathcal{T}$. For the conduction-band model about the $Q$ point, this gives the conditions \cite{bernevig_hughes_book}
\begin{subequations}
\begin{equation}\label{eq:PHP}
	\mathcal{P}H_{2}^\tau(\kk)\mathcal{P}^{-1} = H_{2}^{-\tau}(-\kk)
\end{equation}
\begin{equation}\label{eq:DHD}
	\mathcal{D}(\sigma_v)H_{2}^\tau(k_x,k_y)\mathcal{D}^{-1}(\sigma_v) = H_{2}^{-\tau}(-k_x,k_y)
\end{equation}
\begin{equation}\label{eq:THT}
	\mathcal{T}H_{2}^\tau(\kk)\mathcal{T}^{-1} = H_{2}^{-\tau}(-\kk).
\end{equation}
\end{subequations}
We have $\mathcal{P}=\pi_1s_0$, $\mathcal{D}(\sigma_v)=\pi_0s_1$ and  $\mathcal{T}_2=-i\pi_0s_2 \mathcal{C}$, where $\pi_j$ ($s_j$) are Pauli matrices acting on the layer (spin) subspace, and $\mathcal{C}$ represents complex conjugation. As a result, the most general valley-spin structure for the bilayer Hamiltonian at the $\tau\QQ$ valley is
\begin{equation}
	H_2^\tau(\kk)=\sum_{i,j=0}^3A_{ij}^\tau(\kk)\pi_is_j,
\end{equation}
where the symmetry constraints (\ref{eq:PHP})-(\ref{eq:THT}) require that
\begin{subequations}
\begin{equation}
	A_{i0}^\tau(k_x,k_y)=A_{i0}^{-\tau}(-k_x,k_y)=A_{i0}^\tau(k_x,-k_y),\,i=0,1,
\end{equation}
\begin{equation}
	A_{20}^\tau(k_x,k_y)=A_{20}^{-\tau}(-k_x,k_y)=-A_{20}^\tau(k_x,-k_y),
\end{equation}
\begin{equation}
	A_{31}^\tau(k_x,k_y)=A_{31}^{-\tau}(-k_x,k_y)=-A_{31}^\tau(k_x,-k_y),
\end{equation}
\begin{equation}
	A_{32}^\tau(k_x,k_y)=-A_{32}^{-\tau}(-k_x,k_y)=A_{32}^\tau(k_x,-k_y),
\end{equation}
\begin{equation}
	A_{33}^\tau(k_x,k_y)=-A_{33}^{-\tau}(-k_x,k_y)=A_{33}^\tau(k_x,-k_y).
\end{equation}
\end{subequations}
One can check that, for $N=2$, Eq.\ (\ref{eq:Qhamil}) corresponds to $A_{00}^\tau(\kk)=\tfrac{E_{\uparrow}^+(\kk)+E_{\downarrow}^+(\kk)}{2}$, $A_{33}^\tau(\kk)=\tau\tfrac{E_{\uparrow}^+(\kk)-E_{\downarrow}^+(\kk)}{2}$, $A_{10}^\tau(\kk)=t_0+\tau t_1k_x+t_2k_x^2+u_2k_y^2$ and $A_{20}^\tau(\kk)=-u_1k_y$, and that these terms meet the symmetry requirements. Furthermore, we carried out fittings to DFT data using the additional spin-orbit terms $A_{32}^\tau(\kk) = \alpha k_x$ and $A_{31}^\tau(\kk)=\beta k_y$. The fittings give $|\alpha|,\,|\beta|, |u_1| \ll |t_1|$; hence, we conclude that these terms can be neglected.

For the interlayer hopping used in the HkpTB model for the valence band near $\Gamma$, setting $N=2$ in Eq.\ (\ref{eq:VB_bilayer}) and using a basis ordering similar to that of Eq.\ (\ref{eq:VB_bulk}), we have $\mathcal{T}=\pi_0\sigma_0\mathcal{C}$, $\mathcal{P}=\pi_1\sigma_3$ and $\mathcal{D}(\sigma_v)=\pi_0\sigma_0$, where $\sigma_i$ act on the band ($v$ and $w$) subspace. The symmetry conditions require
\begin{subequations}
\begin{equation}
	\real{t_\sigma(\kk)} = \real{t_\sigma(-\kk)} = \real{t_\sigma(-k_x,k_y)},
\end{equation}
\begin{equation}
	\imag{t_\sigma(\kk)} = -\imag{t_\sigma(-\kk)} = \imag{t_\sigma(-k_x,k_y)},
\end{equation}
\begin{equation}
	t_{vw}(\kk)=t_{vw}(-\kk)=t_{vw}(-k_x,k_y),\quad t_{vw}\in\mathbb{R}.
\end{equation}
\end{subequations}

\section{Spin-orbit-coupling induced interband coupling at the $\Gamma$ point valence bands}
\label{app:socG}

Here, we analyze the role of spin-orbit coupling and coupling to distant bands in determining parameters for the valence-band HkpTB model.

The spin-orbit coupling is given by $\hat{H}_{\rm SO}=\lambda \hat{{\bf L}}\cdot \hat{{\bf S}}$, where $\hat{\mathbf{L}}$ and $\hat{\mathbf{S}}$ are the orbital and spin angular momentum operators. This can also be written in terms of the ladder operators $L_{\pm}=L_x\pm iL_y$ and $S_{\pm}=(S_x\pm i S_y)/2$ as
\begin{equation}
\hat{H}_{SO}=\lambda(L_zS_z+L_+S_-+L_-S_+),
\end{equation}
where$L_{\pm}S_{\mp}$ describe a spin flip with corresponding change in orbital angular momentum projection. These terms couple the $v$ and $w$ bands with the bands $v_1$ and $v_3$ [Fig.\ \ref{fig:figure1}(d)], which in the absence of SO coupling are doubly degenerate. Band $v_1$ ($E''$ Irrep of $C_{3d}$) has basis functions which are odd under $z\rightarrow -z$ (metal $d_{\pm 1}$ orbitals being the dominant component \cite{kdotp}, as well as chalcogen $p_{\pm 1}$). Band $v_3$ ($E'$ Irrep of $C_{3d}$) has basis functions even under $z\rightarrow -z$ (metal $d_{\pm 2}$ orbitals and chalcogen $p_{\pm 1}$). Including SO coupling results in the splitting of these bands into new bands denoted by the orbital and spin angular momentum projections along $\hat{z}$, $v_1(\pm3/2), v_3(\pm3/2)$, and $v_1(\pm1/2), v_3(\pm1/2)$, corresponding to total angular momentum projections of $J_z=\pm 3/2$ and $J_z=\pm 1/2$, respectively.

The $v$-band belongs to the one-dimensional $A_1'$ 
Irrep (even under $z\rightarrow -z$, with metal $d_{0}$ dominant and chalcogen $p_0$), and has $L_z=0$ and $s_z=\pm 1/2$.
Similarly, the $w$-band belongs to the one-dimensional $A_2''$ Irrep, and is dominated by the odd (under $z\rightarrow-z$) chalcogen $p_0$ orbitals, giving two states with $L_z=0$ and $s_z=\pm 1/2$.

Therefore, in the bilayer, where $z\rightarrow -z$ symmetry is broken, the $v$ and $w$-bands can couple to $v_1$ and $v_3$ bands, with the appropriate spin-flip terms. In the second-order perturbation theory, this coupling produces corrections to the on-site energy 
\begin{equation}
\delta_{\sigma} = \sum_{\substack{L_z, s_z\\i=1,3\\\sigma=v,w }}\frac{|\langle v_i(L_z,s_z)|\lambda L_{\pm}S_{\mp}|  \sigma(L_z=0,s_z=\pm 1/2)\rangle|^2}{E_{\sigma}-E_{v_i(L_z,s_z)}}.
\end{equation}
Note that these corrections are the same for both spin components of the $v$ or $w$ bands, with only one of the terms $L_{\pm}S_{\mp}$ contributing for a given spin state.

An additional SO induced interband coupling with a spin-flip may be present in the multilayer case, affecting the interlayer coupling
\begin{equation}
\hat{H}'_{\rm SO} = \mu \hat{z}\cdot({\bf k} \times {\bf S})=i\mu (S_-k_+-S_+k_-),
\label{eq:htso}
\end{equation}
where the pre-factor $\mu$ is related to the gradient of the interlayer pseudo-potential $\mu\propto \partial_z V$, and we defined $k_{\pm}=k_x\pm ik_y$. 
In contrast to the previous coupling, this coupling has a $k$-dependence, which affects the dispersions. 
The coupling in Eq.\ (\ref{eq:htso}) is odd under spatial inversion. Due to the 2H-stacked bilayer having spatial inversion symmetry,  the coupling is non-zero only between different bands in the two layers. In second-order perturbation theory, we get a nominal redefinition of the 2D mass used in the HkpTB model, by adding the term 
\begin{equation}
\begin{split}
\mu_{\sigma}(\kk) = \sum_{\substack{v_i\\ \sigma=v,w }}\frac{|\langle v_i| \mu S_{\mp}k_{\pm}|  \sigma\rangle|^2}{E_{\sigma}-E_{v_i}}
=\mu_{\sigma} k^2,
\end{split}
\label{eq:del2}
\end{equation}
with $\mu_{\sigma}$ a fitting parameter.

\section{Spin-split bands at the Brillouin zone edge for odd number of layers}\label{app:Nodd}
The effective $Q$-point Hamiltonians $H_{NQ}^\tau(\kk)$ for $N$ odd can be split into two decoupled blocks of different spin projection as $H_{NQ}^\tau(\kk) = \diag{h_{N}^{\tau,\uparrow}(\kk), h_{N}^{\tau,\downarrow}(\kk)}$, where the blocks have the alternating $N\times N$ matrix form
\begin{widetext}
\begin{equation}\label{eq:hodd}
	h_{N}^{\tau,s}(\kk)=\begin{pmatrix}
	\varepsilon_0(\kk) + s\tau \Delta(\kk) & t_\tau(\kk) & 0 & 0 & \cdots & 0\\
	t_\tau^*(\kk) & \varepsilon_0(\kk) - s\tau \Delta(\kk) & t_\tau^*(\kk) & 0 &\cdots & 0\\
	0 & t_\tau(\kk) & \varepsilon_0(\kk) + s\tau \Delta(\kk) & t_\tau(\kk) & \cdots & 0\\
	0 & 0 & t_\tau^*(\kk) & \varepsilon_0(\kk) - s\tau \Delta(\kk) & \cdots & 0\\
	\vdots & \vdots & \vdots & \vdots &  \ddots & t_\tau(\kk)\\
	0 & 0 & 0 & \cdots &  t_\tau^*(\kk) & \varepsilon_0(\kk) + s\tau \Delta(\kk) 
	\end{pmatrix},
\end{equation}
\end{widetext}
and we have defined
\begin{subequations}
\begin{equation}
	\varepsilon_0(\kk) = \tfrac{E_\uparrow^+(\kk)+E_\downarrow^+(\kk)}{2},
\end{equation}
\begin{equation}
	\Delta(\kk) = \tfrac{E_\uparrow^+(\kk)-E_\downarrow^+(\kk)}{2}.
\end{equation}
\end{subequations}
Defining the even-dimensional $(N-1)\times(N-1)$ matrix
\begin{widetext}
\begin{equation}\label{eq:heven}
	\tilde{h}_{N-1}^{\tau,s}(\kk)=\begin{pmatrix}
	\varepsilon_0(\kk) - s\tau \Delta(\kk) & t_\tau^*(\kk) & 0 &\cdots & 0\\
	t_\tau(\kk) & \varepsilon_0(\kk) + s\tau \Delta(\kk) & t_\tau(\kk) & \cdots & 0\\
	 0 & t_\tau^*(\kk) & \varepsilon_0(\kk) - s\tau \Delta(\kk) & \cdots & 0\\
	 \vdots & \vdots & \vdots &  \ddots & t_\tau(\kk)\\
	 0 & 0 & \cdots &  t_\tau^*(\kk) & \varepsilon_0(\kk) + s\tau \Delta(\kk) 
	\end{pmatrix},
\end{equation}
\end{widetext}
the eigenvalues $\varepsilon$ of (\ref{eq:hodd}) are given by a secular equation
\begin{equation}\label{eq:expand}
\begin{split}
	&\det{\{\varepsilon - h_{N}^{\tau,s}\}} = [\varepsilon - \varepsilon_0(\kk) - s\tau\Delta(\kk)]\det{\{\varepsilon - \tilde{h}_{N-1}^{\tau,s}\}}\\
	&\quad-\abs{t_\tau(\kk)}^2\det{\{\varepsilon  - h_{N-2}^{\tau,s}\}}\\
	&=[\varepsilon - \varepsilon_0(\kk) - s\tau\Delta(\kk)]\det{\{\varepsilon - \tilde{h}_{N-1}^{\tau,s}\}}\\
	&\quad-\abs{t_\tau(\kk)}^2\left([\varepsilon - \varepsilon_0(\kk) - s\tau\Delta(\kk)]\det{\{\varepsilon-\tilde{h}_{N-3}^{\tau,s} \}}\right.\\
	&\qquad -\left. \abs{t_\tau(\kk)}^2\det{\{\varepsilon - h_{N-4}^{\tau,s} \}} \right)=\cdots .
\end{split}
\end{equation}
Using the fact that $\det{\{\varepsilon - h_{1}^{\tau,s}\}}=\varepsilon - \varepsilon_0(\kk) - s\tau\Delta(\kk)$, we can continue expanding Eq.\ (\ref{eq:expand}) to obtain
\begin{equation}
\begin{split}
	&\det{\{\varepsilon - h_{N}^{\tau,s}\}}=[\varepsilon - \varepsilon_0(\kk) - s\tau\Delta(\kk)]\\
	&\times\Bigg(\sum_{m=0}^{\tfrac{N-3}{2}}(-1)^m\abs{t_\tau(\kk)}^{2m}\det{\{\varepsilon-\tilde{h}_{N-(2m+1)}\}}\\
	&\qquad\quad+ (-1)^{\tfrac{N-1}{2}}\abs{t_\tau(\kk)}^{N-1} \Bigg),
\end{split}
\end{equation}
which explicitly shows that $[\varepsilon - \varepsilon_0(\kk) - s\tau\Delta(\kk)]$ is an overall factor, and thus $\varepsilon=\varepsilon_0(\kk) + s\tau\Delta(\kk)\equiv \varepsilon_s^\tau(\kk)$ is always an eigenvalue, regardless of the (odd) value of $N$. For a given $\tau$, the different $s$ quantum numbers give two spin-split monolayer dispersions $\varepsilon_s^\tau(\kk)$ about the $\tau C_3^{m}\QQ$ ($m=0,1,2$) points, corresponding to the features observed in Fig.\ \ref{fig:Qfits}. The fact that this prediction is verified in the DFT band structures clearly confirms the validity of our hybrid model.

\begin{figure}[h!]
\begin{center}
\includegraphics[width=0.9\columnwidth]{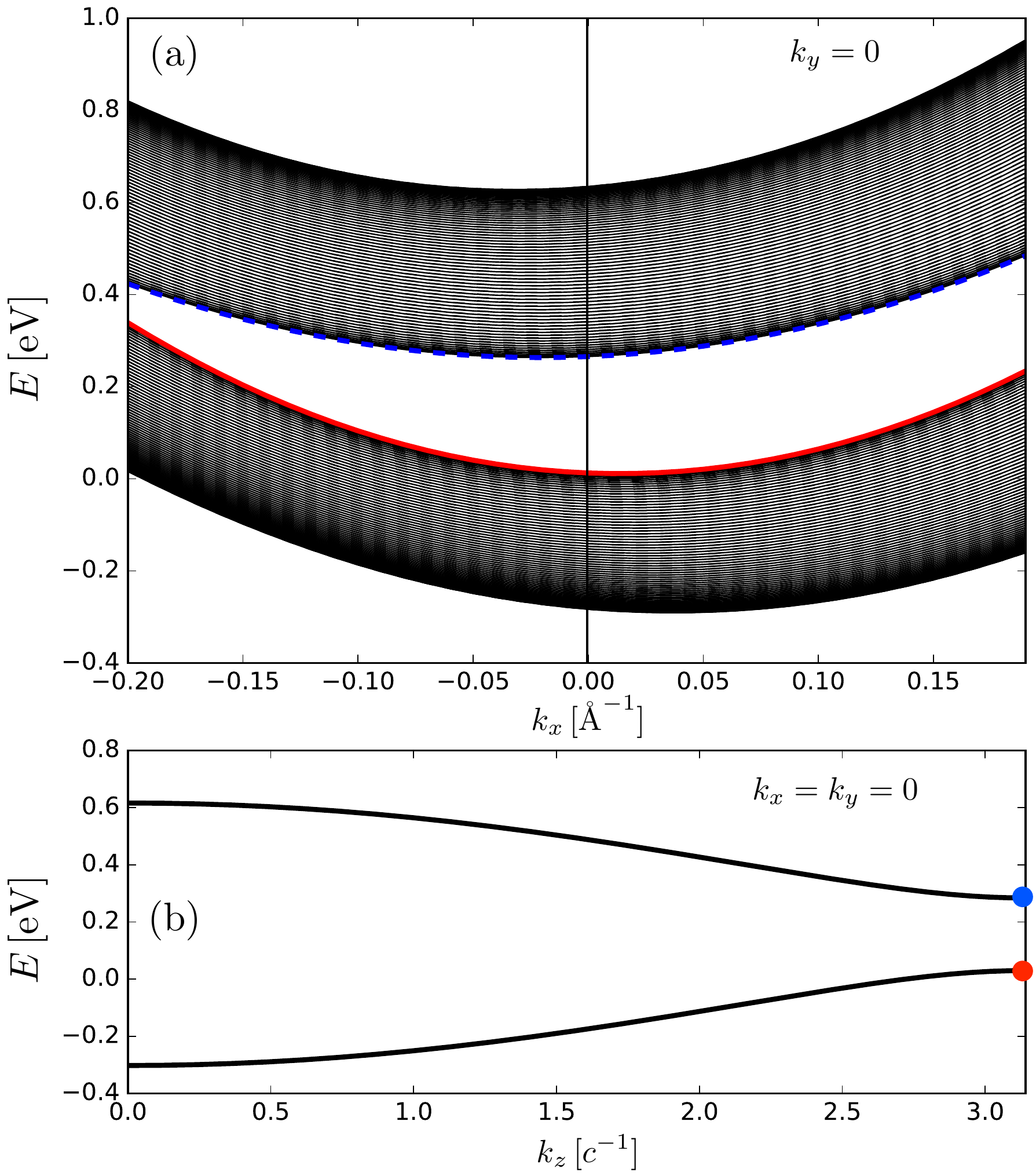}
\caption{(a) Subband structure of 101-layer WS${}_2$ near $\tau\QQ$ ($k_x=0$), along the $\overline{\Gamma K}$ line. Spin-up (-down) bands are shown with solid (dashed) curves. The spin-split bands $\varepsilon_s^\tau(\kk)$, pinned in the middle of the odd $N$ subband structure, are shown in blue and red. (b) Bulk band structure for WS${}_2$ along the 3D BZ line indicated in the inset of Fig.\ \ref{fig:CBbulk_main}. Blue and red dots mark the position of the spin-split bands in the Brillouin zone.}
\label{fig:CBbulk}
\end{center}
\end{figure}

For large odd $N$, nearly spin-degenerate bands grow denser on either side of the spin-split bands $\varepsilon_s^\tau(\kk)$ without crossing them, as shown in Fig.\ \ref{fig:CBbulk}(a). The reason for this becomes clear when we take the bulk limit, and find that the spin-split states form the band edges around a central gap in the subband structure. This is shown in Fig.\ \ref{fig:CBbulk}(b). Indeed, in the limit of large $N$ the Hamiltonian (\ref{eq:hodd}) corresponds to the bulk Hamiltonian at $k_z = \pi/c$, since $\varepsilon_0(\kk) = \varepsilon_0(\kk,k_z=\tfrac{\pi}{c})$ [see Eq.\ (\ref{eq:QH_bulk})].

\section{Electron-phonon coupling for LO phonon in multilayer system}
\label{sec:lo_coup}
In this appendix, we derive the expression used for the electron-phonon coupling with LO phonon in a multilayer system. As described in the text, we treat the LO phonon in each layer as independent and degenerate. However, in the LO phonon case, the generated electrostatic potential due LO phonon in one layer interacts with the electrons in all the other layers in the system, following similar steps as in Ref.\ [\onlinecite{danish_LO_phonon}].

Within a monolayer, the LO phonon-induced in-plane polarization is given by its in-plane Fourier component,
\begin{equation}
{\bf P}_{\bf q}(z) = \frac{eZ}{\epsilon(q) A}{\bf u}_{\bf q}\delta(z),
\end{equation}
where $Z$ is the Born effective charge on the metal and chalcogens, $A$ is the unit-cell area, ${\bf u}_{\bf q} = \sqrt{\frac{\hbar}{2M_rN_{\text{cell}} \omega_{\text{LO}}}}\hat{\bf e}$ is the phonon-induced atomic displacement in the direction connecting the metal and chalcogens in the unit cell, with $M_r$ the reduced mass of the metal and chalcogens, $N$ the number of unit cells in the sample, and $\omega_{\text{LO}}$ the LO phonon frequency. $\epsilon(q)$ is the dielectric function characterizing the response of the material to the phonon induced electric field.

The induced charge density in the layer is given by $\rho = -\nabla\cdot {\bf P}$, with the Fourier component
\begin{equation}
\rho_{\bf q} = -i{\bf q}\cdot P_{\bf q}.
\end{equation}

The potential resulting from the charge distribution is given by Poisson's equation $\nabla^2\phi = -4\pi \rho$. Fourier-transforming in three dimensions gives,
\begin{equation}
\phi_{\bf q}(k)=\frac{-4\pi i e Z}{\epsilon(q)A}\frac{{\bf q}\cdot {\bf u}_{\bf q}}{q^2+k^2},
\end{equation} 
where $k$ is the Fourier parameter in the $z$ direction.
Inverse Fourier transforming in $k$ gives the $z$ dependence of the potential with in-plane Fourier component ${\bf q}$
\begin{equation}
\phi_{\bf q}(z) = -i \frac{2\pi eZ u_{\bf q}}{\epsilon(q)A }e^{-q|z|}.
\label{eq:lo_pot}
\end{equation}
The electron-phonon coupling for an electron localized in an isolated monolayer is given by $g(q) = e\phi_q(0) = -i\frac{2\pi e^2 Z u_q}{\epsilon A}$. This form of the coupling is similar to the form derived in Refs.\ [\onlinecite{calandra_lo,danovich_phonons}], where the polarizability of a two-dimensional dielectric was taken into account by the replacement $\epsilon(q)\rightarrow 1+r_*q$, with $r_*$ the screening length in the material.
For multilayer 2H-stacked TMDs, as the polarization in subsequent layers alternates in its sign, the resulting electrostatic potential also alternates in its sign.

\section{Electron-phonon coupling for ZO phonon in multilayer system}
\label{sec:zo_coup}

The atomic vibrations for the ZO optical phonon mode result in a polarization in the out-of-plane direction due to the opposite motions of the metal and two chalcogens, and the finite Born effective charges in the $z$-direction.
The interaction energy in the multilayer system between charges and phonon-induced out-of-plane polarizations in all layers is given by \cite{ganchev}
\begin{equation}
\begin{split}
&E_{\text{int}}= \sum_{n,m}\int d^2 r d^2r' \frac{\rho_n(\rr)P_{z,m}(\rr')d(n-m)}{\Delta r^3},
\\
&\Delta r = [(\rr-\rr')^2+d^2(n-m)^2]^{1/2},
\end{split}
\end{equation}
where $\rho_n(\rr)$ is the charge density on layer $n$, $P_{z,m}(\rr')$ is the out of plane polarization in layer $m$ caused by the ZO optical phonon, $d$ the interlayer separation, and we sum over all layer pairs.
Fourier transforming the charge density and polarization in the in-plane momentum components gives
\begin{equation}
\rho_n(\rr)=\int \frac{d^2q}{(2\pi)^2}e^{i\qq\cdot \rr}\rho_m(\qq),
\end{equation}
and similarly for the polarization.
The interaction energy then takes the form,
\begin{equation}
\begin{split}
&E_{\text{int}}=\sum_{n,m}\int d^2rd^2r' \frac{d(n-m)}{\Delta r^3}
\\
&\times\int \frac{d^2q d^2q'}{(2\pi)^4}
e^{i\qq\cdot \rr}e^{i\qq'\cdot \rr'}\rho_n(\qq)P_{z,m}(\qq').
\end{split}
\end{equation}
Defining the new variables $\tilde{\rr} = \rr-\rr'$, $\RR=\rr+\rr'$, and integrating over $\RR$ gives $\delta_{q,-q'}$,
\begin{equation}
\begin{split}
&E_{\text{e-ph}}=\sum_{n,m}\int d^2\tilde{r}\int \frac{d^2q}{(2\pi)^2} \frac{d(n-m)}{\Delta \tilde{r}^3}
e^{i\qq\cdot \tilde{\rr}}
\\
&\times\rho_n^*(\qq)P_{z,m}(\qq),
\end{split}
\end{equation}
where in the last row we used $\rho^*_n(\qq)=\rho_n(-\qq)$, since the density is real.
Carrying out the integration over $\tilde{\rr}$ gives
\begin{equation}
E_{\text{e-ph}}=\sum'_{n,m} \frac{2\pi(n-m)}{|n-m|}\int \frac{d^2q}{(2\pi)^2}
e^{-qd|n-m|}\rho_n^*(\qq)P_{z,m}(\qq),
\end{equation}
where the prime over the sum means that the summation excludes the term with $n=m$.
Quantizing the phonon polarization and the carrier density gives
\begin{equation}
\begin{split}
&\rho_n^*(\qq)=e\sum_{\kk} c^{\dagger}_{\kk,n}c_{\kk+\qq,n},
\\
&P_{z,m}(\qq)=\frac{e Z_z}{A}\sqrt{\frac{\hbar}{2N_{\text{cell}}M_r \omega}}(a_{-\qq,m}+a^{\dagger}_{\qq,m}),
\end{split}
\end{equation}
where $c_{\kk,n}\,(c^{\dagger}_{\kk,n})$ is the annihilation (creation) operator for an electron in state $\kk$ in layer $n$, and $a_{\qq,m}\,(a_{\qq,m}^{\dagger})$ is the annihilation (creation) operator for a phonon with in-plane wave vector $\qq$.  The phonon-induced polarization is given, similarly to the LO phonon case, by the Born effective charge and the phonon displacement.

The electron-phonon interaction Hamiltonian is then given by
\begin{equation}
\begin{split}
&H_{\text{e-ph}}=\frac{2\pi e^2 Z_z}{A}\sqrt{\frac{\hbar}{2N_{\text{cell}}M_r \omega_{\rm ZO}}} \sum'_{n,m}\sum_{\kk,\qq}
\frac{n-m}{|n-m|}e^{-qd|n-m|}
\\
&\times c^{\dagger}_{\kk,n}c_{\kk+\qq,n}(a_{-\qq,m}+a^{\dagger}_{\qq,m}).
\end{split}
\end{equation}

\bibliographystyle{apsrev4-1} 
\bibliography{refs}

\begin{thebibliography}{51}%
\makeatletter
\providecommand \@ifxundefined [1]{%
 \@ifx{#1\undefined}
}%
\providecommand \@ifnum [1]{%
 \ifnum #1\expandafter \@firstoftwo
 \else \expandafter \@secondoftwo
 \fi
}%
\providecommand \@ifx [1]{%
 \ifx #1\expandafter \@firstoftwo
 \else \expandafter \@secondoftwo
 \fi
}%
\providecommand \natexlab [1]{#1}%
\providecommand \enquote  [1]{``#1''}%
\providecommand \bibnamefont  [1]{#1}%
\providecommand \bibfnamefont [1]{#1}%
\providecommand \citenamefont [1]{#1}%
\providecommand \href@noop [0]{\@secondoftwo}%
\providecommand \href [0]{\begingroup \@sanitize@url \@href}%
\providecommand \@href[1]{\@@startlink{#1}\@@href}%
\providecommand \@@href[1]{\endgroup#1\@@endlink}%
\providecommand \@sanitize@url [0]{\catcode `\\12\catcode `\$12\catcode
  `\&12\catcode `\#12\catcode `\^12\catcode `\_12\catcode `\%12\relax}%
\providecommand \@@startlink[1]{}%
\providecommand \@@endlink[0]{}%
\providecommand \url  [0]{\begingroup\@sanitize@url \@url }%
\providecommand \@url [1]{\endgroup\@href {#1}{\urlprefix }}%
\providecommand \urlprefix  [0]{URL }%
\providecommand \Eprint [0]{\href }%
\providecommand \doibase [0]{http://dx.doi.org/}%
\providecommand \selectlanguage [0]{\@gobble}%
\providecommand \bibinfo  [0]{\@secondoftwo}%
\providecommand \bibfield  [0]{\@secondoftwo}%
\providecommand \translation [1]{[#1]}%
\providecommand \BibitemOpen [0]{}%
\providecommand \bibitemStop [0]{}%
\providecommand \bibitemNoStop [0]{.\EOS\space}%
\providecommand \EOS [0]{\spacefactor3000\relax}%
\providecommand \BibitemShut  [1]{\csname bibitem#1\endcsname}%
\let\auto@bib@innerbib\@empty
\bibitem [{\citenamefont {Jariwala}\ \emph {et~al.}(2014)\citenamefont
  {Jariwala}, \citenamefont {Sangwan}, \citenamefont {Lauhon}, \citenamefont
  {Marks},\ and\ \citenamefont {Hersam}}]{jariwala_applications}%
  \BibitemOpen
  \bibfield  {author} {\bibinfo {author} {\bibfnamefont {D.}~\bibnamefont
  {Jariwala}}, \bibinfo {author} {\bibfnamefont {V.~K.}\ \bibnamefont
  {Sangwan}}, \bibinfo {author} {\bibfnamefont {L.~J.}\ \bibnamefont {Lauhon}},
  \bibinfo {author} {\bibfnamefont {T.~J.}\ \bibnamefont {Marks}}, \ and\
  \bibinfo {author} {\bibfnamefont {M.~C.}\ \bibnamefont {Hersam}},\ }\href
  {\doibase 10.1021/nn500064s} {\bibfield  {journal} {\bibinfo  {journal} {ACS
  Nano}\ }\textbf {\bibinfo {volume} {8}},\ \bibinfo {pages} {1102} (\bibinfo
  {year} {2014})}\BibitemShut {NoStop}%
\bibitem [{\citenamefont {Choi}\ \emph {et~al.}(2017)\citenamefont {Choi},
  \citenamefont {Choudhary}, \citenamefont {Han}, \citenamefont {Park},
  \citenamefont {Akinwande},\ and\ \citenamefont {Lee}}]{tmds_applications}%
  \BibitemOpen
  \bibfield  {author} {\bibinfo {author} {\bibfnamefont {W.}~\bibnamefont
  {Choi}}, \bibinfo {author} {\bibfnamefont {N.}~\bibnamefont {Choudhary}},
  \bibinfo {author} {\bibfnamefont {G.~H.}\ \bibnamefont {Han}}, \bibinfo
  {author} {\bibfnamefont {J.}~\bibnamefont {Park}}, \bibinfo {author}
  {\bibfnamefont {D.}~\bibnamefont {Akinwande}}, \ and\ \bibinfo {author}
  {\bibfnamefont {Y.~H.}\ \bibnamefont {Lee}},\ }\href {\doibase
  10.1016/j.mattod.2016.10.002} {\bibfield  {journal} {\bibinfo  {journal}
  {Materials Today}\ }\textbf {\bibinfo {volume} {20}},\ \bibinfo {pages} {116
  } (\bibinfo {year} {2017})}\BibitemShut {NoStop}%
\bibitem [{\citenamefont {Zhao}\ \emph {et~al.}(2018)\citenamefont {Zhao},
  \citenamefont {Yu},\ and\ \citenamefont {Ouyang}}]{hetro_tmds_bands}%
  \BibitemOpen
  \bibfield  {author} {\bibinfo {author} {\bibfnamefont {Y.}~\bibnamefont
  {Zhao}}, \bibinfo {author} {\bibfnamefont {W.}~\bibnamefont {Yu}}, \ and\
  \bibinfo {author} {\bibfnamefont {G.}~\bibnamefont {Ouyang}},\ }\href
  {http://stacks.iop.org/0022-3727/51/i=1/a=015111} {\bibfield  {journal}
  {\bibinfo  {journal} {Journal of Physics D: Applied Physics}\ }\textbf
  {\bibinfo {volume} {51}},\ \bibinfo {pages} {015111} (\bibinfo {year}
  {2018})}\BibitemShut {NoStop}%
\bibitem [{\citenamefont {Wang}\ \emph {et~al.}(2012)\citenamefont {Wang},
  \citenamefont {Kalantar-{Z}adeh}, \citenamefont {Kis}, \citenamefont
  {Coleman},\ and\ \citenamefont {Strano}}]{wang_tmds_props}%
  \BibitemOpen
  \bibfield  {author} {\bibinfo {author} {\bibfnamefont {Q.~H.}\ \bibnamefont
  {Wang}}, \bibinfo {author} {\bibfnamefont {K.}~\bibnamefont
  {Kalantar-{Z}adeh}}, \bibinfo {author} {\bibfnamefont {A.}~\bibnamefont
  {Kis}}, \bibinfo {author} {\bibfnamefont {J.~N.}\ \bibnamefont {Coleman}}, \
  and\ \bibinfo {author} {\bibfnamefont {M.~S.}\ \bibnamefont {Strano}},\
  }\href {\doibase 10.1038/nnano.2012.193} {\bibfield  {journal} {\bibinfo
  {journal} {Nat. Nanotechnol.}\ }\textbf {\bibinfo {volume} {7}},\ \bibinfo
  {pages} {699} (\bibinfo {year} {2012})}\BibitemShut {NoStop}%
\bibitem [{\citenamefont {Liu}\ \emph {et~al.}(2015)\citenamefont {Liu},
  \citenamefont {Xiao}, \citenamefont {Yao}, \citenamefont {Xu},\ and\
  \citenamefont {Yao}}]{wangyao_review}%
  \BibitemOpen
  \bibfield  {author} {\bibinfo {author} {\bibfnamefont {G.-B.}\ \bibnamefont
  {Liu}}, \bibinfo {author} {\bibfnamefont {D.}~\bibnamefont {Xiao}}, \bibinfo
  {author} {\bibfnamefont {Y.}~\bibnamefont {Yao}}, \bibinfo {author}
  {\bibfnamefont {X.}~\bibnamefont {Xu}}, \ and\ \bibinfo {author}
  {\bibfnamefont {W.}~\bibnamefont {Yao}},\ }\href {\doibase
  10.1039/C4CS00301B} {\bibfield  {journal} {\bibinfo  {journal} {Chem. Soc.
  Rev.}\ }\textbf {\bibinfo {volume} {44}},\ \bibinfo {pages} {2643} (\bibinfo
  {year} {2015})}\BibitemShut {NoStop}%
\bibitem [{\citenamefont {Mak}\ and\ \citenamefont
  {Shan}(2016)}]{mak_optoelectronics_tmds}%
  \BibitemOpen
  \bibfield  {author} {\bibinfo {author} {\bibfnamefont {K.~F.}\ \bibnamefont
  {Mak}}\ and\ \bibinfo {author} {\bibfnamefont {J.}~\bibnamefont {Shan}},\
  }\href {\doibase 10.1038/nphoton.2015.282} {\bibfield  {journal} {\bibinfo
  {journal} {Nat. Photonics}\ }\textbf {\bibinfo {volume} {10}},\ \bibinfo
  {pages} {216} (\bibinfo {year} {2016})}\BibitemShut {NoStop}%
\bibitem [{\citenamefont {Wurstbauer}\ \emph {et~al.}(2017)\citenamefont
  {Wurstbauer}, \citenamefont {Miller}, \citenamefont {Parzinger},\ and\
  \citenamefont {Holleitner}}]{tmds_light_matter}%
  \BibitemOpen
  \bibfield  {author} {\bibinfo {author} {\bibfnamefont {U.}~\bibnamefont
  {Wurstbauer}}, \bibinfo {author} {\bibfnamefont {B.}~\bibnamefont {Miller}},
  \bibinfo {author} {\bibfnamefont {E.}~\bibnamefont {Parzinger}}, \ and\
  \bibinfo {author} {\bibfnamefont {A.~W.}\ \bibnamefont {Holleitner}},\ }\href
  {http://stacks.iop.org/0022-3727/50/i=17/a=173001} {\bibfield  {journal}
  {\bibinfo  {journal} {Journal of Physics D: Applied Physics}\ }\textbf
  {\bibinfo {volume} {50}},\ \bibinfo {pages} {173001} (\bibinfo {year}
  {2017})}\BibitemShut {NoStop}%
\bibitem [{\citenamefont {Mak}\ \emph {et~al.}(2010)\citenamefont {Mak},
  \citenamefont {Lee}, \citenamefont {Hone}, \citenamefont {Shan},\ and\
  \citenamefont {Heinz}}]{mak_prl_2010}%
  \BibitemOpen
  \bibfield  {author} {\bibinfo {author} {\bibfnamefont {K.~F.}\ \bibnamefont
  {Mak}}, \bibinfo {author} {\bibfnamefont {C.}~\bibnamefont {Lee}}, \bibinfo
  {author} {\bibfnamefont {J.}~\bibnamefont {Hone}}, \bibinfo {author}
  {\bibfnamefont {J.}~\bibnamefont {Shan}}, \ and\ \bibinfo {author}
  {\bibfnamefont {T.~F.}\ \bibnamefont {Heinz}},\ }\href {\doibase
  10.1103/PhysRevLett.105.136805} {\bibfield  {journal} {\bibinfo  {journal}
  {Phys. Rev. Lett.}\ }\textbf {\bibinfo {volume} {105}},\ \bibinfo {pages}
  {136805} (\bibinfo {year} {2010})}\BibitemShut {NoStop}%
\bibitem [{\citenamefont {Aivazian}\ \emph {et~al.}(2015)\citenamefont
  {Aivazian}, \citenamefont {Gong}, \citenamefont {Jones}, \citenamefont {Chu},
  \citenamefont {Yan}, \citenamefont {Mandrus}, \citenamefont {Zhang},
  \citenamefont {Cobden}, \citenamefont {Yao},\ and\ \citenamefont
  {Xu}}]{pseudospin_control}%
  \BibitemOpen
  \bibfield  {author} {\bibinfo {author} {\bibfnamefont {G.}~\bibnamefont
  {Aivazian}}, \bibinfo {author} {\bibfnamefont {Z.}~\bibnamefont {Gong}},
  \bibinfo {author} {\bibfnamefont {A.~M.}\ \bibnamefont {Jones}}, \bibinfo
  {author} {\bibfnamefont {R.-L.}\ \bibnamefont {Chu}}, \bibinfo {author}
  {\bibfnamefont {J.}~\bibnamefont {Yan}}, \bibinfo {author} {\bibfnamefont
  {D.~G.}\ \bibnamefont {Mandrus}}, \bibinfo {author} {\bibfnamefont
  {C.}~\bibnamefont {Zhang}}, \bibinfo {author} {\bibfnamefont
  {D.}~\bibnamefont {Cobden}}, \bibinfo {author} {\bibfnamefont
  {W.}~\bibnamefont {Yao}}, \ and\ \bibinfo {author} {\bibfnamefont
  {X.}~\bibnamefont {Xu}},\ }\href {\doibase 10.1038/nphys3201} {\bibfield
  {journal} {\bibinfo  {journal} {Nat. Phys.}\ }\textbf {\bibinfo {volume}
  {11}},\ \bibinfo {pages} {148} (\bibinfo {year} {2015})}\BibitemShut
  {NoStop}%
\bibitem [{\citenamefont {MacNeill}\ \emph {et~al.}(2015)\citenamefont
  {MacNeill}, \citenamefont {Heikes}, \citenamefont {Mak}, \citenamefont
  {Anderson}, \citenamefont {Korm\'anyos}, \citenamefont {Z\'olyomi},
  \citenamefont {Park},\ and\ \citenamefont {Ralph}}]{breaking_valley_mose2}%
  \BibitemOpen
  \bibfield  {author} {\bibinfo {author} {\bibfnamefont {D.}~\bibnamefont
  {MacNeill}}, \bibinfo {author} {\bibfnamefont {C.}~\bibnamefont {Heikes}},
  \bibinfo {author} {\bibfnamefont {K.~F.}\ \bibnamefont {Mak}}, \bibinfo
  {author} {\bibfnamefont {Z.}~\bibnamefont {Anderson}}, \bibinfo {author}
  {\bibfnamefont {A.}~\bibnamefont {Korm\'anyos}}, \bibinfo {author}
  {\bibfnamefont {V.}~\bibnamefont {Z\'olyomi}}, \bibinfo {author}
  {\bibfnamefont {J.}~\bibnamefont {Park}}, \ and\ \bibinfo {author}
  {\bibfnamefont {D.~C.}\ \bibnamefont {Ralph}},\ }\href {\doibase
  10.1103/PhysRevLett.114.037401} {\bibfield  {journal} {\bibinfo  {journal}
  {Phys. Rev. Lett.}\ }\textbf {\bibinfo {volume} {114}},\ \bibinfo {pages}
  {037401} (\bibinfo {year} {2015})}\BibitemShut {NoStop}%
\bibitem [{\citenamefont {Zhu}\ \emph {et~al.}(2014)\citenamefont {Zhu},
  \citenamefont {Zeng}, \citenamefont {Dai},\ and\ \citenamefont
  {Cui}}]{spinvalleystudy}%
  \BibitemOpen
  \bibfield  {author} {\bibinfo {author} {\bibfnamefont {B.}~\bibnamefont
  {Zhu}}, \bibinfo {author} {\bibfnamefont {H.}~\bibnamefont {Zeng}}, \bibinfo
  {author} {\bibfnamefont {J.}~\bibnamefont {Dai}}, \ and\ \bibinfo {author}
  {\bibfnamefont {X.}~\bibnamefont {Cui}},\ }\href {\doibase
  10.1002/adma.201305367} {\bibfield  {journal} {\bibinfo  {journal} {Advanced
  Materials}\ }\textbf {\bibinfo {volume} {26}},\ \bibinfo {pages} {5504}
  (\bibinfo {year} {2014})}\BibitemShut {NoStop}%
\bibitem [{\citenamefont {Yang}\ \emph {et~al.}(2015)\citenamefont {Yang},
  \citenamefont {Sinitsyn}, \citenamefont {Chen}, \citenamefont {Yuan},
  \citenamefont {Zhang}, \citenamefont {Lou},\ and\ \citenamefont
  {Crooker}}]{longlived}%
  \BibitemOpen
  \bibfield  {author} {\bibinfo {author} {\bibfnamefont {L.}~\bibnamefont
  {Yang}}, \bibinfo {author} {\bibfnamefont {N.~A.}\ \bibnamefont {Sinitsyn}},
  \bibinfo {author} {\bibfnamefont {W.}~\bibnamefont {Chen}}, \bibinfo {author}
  {\bibfnamefont {J.}~\bibnamefont {Yuan}}, \bibinfo {author} {\bibfnamefont
  {J.}~\bibnamefont {Zhang}}, \bibinfo {author} {\bibfnamefont
  {J.}~\bibnamefont {Lou}}, \ and\ \bibinfo {author} {\bibfnamefont {S.~A.}\
  \bibnamefont {Crooker}},\ }\href {\doibase 10.1038/nphys3419} {\bibfield
  {journal} {\bibinfo  {journal} {Nat. Phys.}\ }\textbf {\bibinfo {volume}
  {11}},\ \bibinfo {pages} {830} (\bibinfo {year} {2015})}\BibitemShut
  {NoStop}%
\bibitem [{\citenamefont {Yu}\ \emph {et~al.}(2014)\citenamefont {Yu},
  \citenamefont {Liu}, \citenamefont {Gong}, \citenamefont {Xu},\ and\
  \citenamefont {Yao}}]{berryexcitons}%
  \BibitemOpen
  \bibfield  {author} {\bibinfo {author} {\bibfnamefont {H.}~\bibnamefont
  {Yu}}, \bibinfo {author} {\bibfnamefont {G.-B.}\ \bibnamefont {Liu}},
  \bibinfo {author} {\bibfnamefont {P.}~\bibnamefont {Gong}}, \bibinfo {author}
  {\bibfnamefont {X.}~\bibnamefont {Xu}}, \ and\ \bibinfo {author}
  {\bibfnamefont {W.}~\bibnamefont {Yao}},\ }\href {\doibase
  10.1038/ncomms4876} {\bibfield  {journal} {\bibinfo  {journal} {Nat.
  Commun.}\ }\textbf {\bibinfo {volume} {5}},\ \bibinfo {pages} {3876}
  (\bibinfo {year} {2014})}\BibitemShut {NoStop}%
\bibitem [{\citenamefont {Gunlycke}\ and\ \citenamefont
  {Tseng}(2016)}]{pccp2016}%
  \BibitemOpen
  \bibfield  {author} {\bibinfo {author} {\bibfnamefont {D.}~\bibnamefont
  {Gunlycke}}\ and\ \bibinfo {author} {\bibfnamefont {F.}~\bibnamefont
  {Tseng}},\ }\href {\doibase 10.1039/C6CP00205F} {\bibfield  {journal}
  {\bibinfo  {journal} {Phys. Chem. Chem. Phys.}\ }\textbf {\bibinfo {volume}
  {18}},\ \bibinfo {pages} {8579} (\bibinfo {year} {2016})}\BibitemShut
  {NoStop}%
\bibitem [{\citenamefont {Cheiwchanchamnangij}\ and\ \citenamefont
  {Lambrecht}(2012)}]{lambrecht_prb_2012}%
  \BibitemOpen
  \bibfield  {author} {\bibinfo {author} {\bibfnamefont {T.}~\bibnamefont
  {Cheiwchanchamnangij}}\ and\ \bibinfo {author} {\bibfnamefont {W.~R.~L.}\
  \bibnamefont {Lambrecht}},\ }\href {\doibase 10.1103/PhysRevB.85.205302}
  {\bibfield  {journal} {\bibinfo  {journal} {Phys. Rev. B}\ }\textbf {\bibinfo
  {volume} {85}},\ \bibinfo {pages} {205302} (\bibinfo {year}
  {2012})}\BibitemShut {NoStop}%
\bibitem [{\citenamefont {Cappelluti}\ \emph {et~al.}(2013)\citenamefont
  {Cappelluti}, \citenamefont {Rold\'an}, \citenamefont {Silva-Guill\'en},
  \citenamefont {Ordej\'on},\ and\ \citenamefont
  {Guinea}}]{cappelluti_prb_2013}%
  \BibitemOpen
  \bibfield  {author} {\bibinfo {author} {\bibfnamefont {E.}~\bibnamefont
  {Cappelluti}}, \bibinfo {author} {\bibfnamefont {R.}~\bibnamefont
  {Rold\'an}}, \bibinfo {author} {\bibfnamefont {J.~A.}\ \bibnamefont
  {Silva-Guill\'en}}, \bibinfo {author} {\bibfnamefont {P.}~\bibnamefont
  {Ordej\'on}}, \ and\ \bibinfo {author} {\bibfnamefont {F.}~\bibnamefont
  {Guinea}},\ }\href {\doibase 10.1103/PhysRevB.88.075409} {\bibfield
  {journal} {\bibinfo  {journal} {Phys. Rev. B}\ }\textbf {\bibinfo {volume}
  {88}},\ \bibinfo {pages} {075409} (\bibinfo {year} {2013})}\BibitemShut
  {NoStop}%
\bibitem [{\citenamefont {Debbichi}\ \emph {et~al.}(2014)\citenamefont
  {Debbichi}, \citenamefont {Eriksson},\ and\ \citenamefont
  {Leb\`egue}}]{debbichi_prb_2014}%
  \BibitemOpen
  \bibfield  {author} {\bibinfo {author} {\bibfnamefont {L.}~\bibnamefont
  {Debbichi}}, \bibinfo {author} {\bibfnamefont {O.}~\bibnamefont {Eriksson}},
  \ and\ \bibinfo {author} {\bibfnamefont {S.}~\bibnamefont {Leb\`egue}},\
  }\href {\doibase 10.1103/PhysRevB.89.205311} {\bibfield  {journal} {\bibinfo
  {journal} {Phys. Rev. B}\ }\textbf {\bibinfo {volume} {89}},\ \bibinfo
  {pages} {205311} (\bibinfo {year} {2014})}\BibitemShut {NoStop}%
\bibitem [{\citenamefont {Padilha}\ \emph {et~al.}(2014)\citenamefont
  {Padilha}, \citenamefont {Peelaers}, \citenamefont {Janotti},\ and\
  \citenamefont {Van~de Walle}}]{padilha_prb_2014}%
  \BibitemOpen
  \bibfield  {author} {\bibinfo {author} {\bibfnamefont {J.~E.}\ \bibnamefont
  {Padilha}}, \bibinfo {author} {\bibfnamefont {H.}~\bibnamefont {Peelaers}},
  \bibinfo {author} {\bibfnamefont {A.}~\bibnamefont {Janotti}}, \ and\
  \bibinfo {author} {\bibfnamefont {C.~G.}\ \bibnamefont {Van~de Walle}},\
  }\href {\doibase 10.1103/PhysRevB.90.205420} {\bibfield  {journal} {\bibinfo
  {journal} {Phys. Rev. B}\ }\textbf {\bibinfo {volume} {90}},\ \bibinfo
  {pages} {205420} (\bibinfo {year} {2014})}\BibitemShut {NoStop}%
\bibitem [{\citenamefont {Chang}\ \emph {et~al.}(2014)\citenamefont {Chang},
  \citenamefont {Lin}, \citenamefont {Jeng},\ and\ \citenamefont
  {Bansil}}]{chang_scirep_2014}%
  \BibitemOpen
  \bibfield  {author} {\bibinfo {author} {\bibfnamefont {T.-R.}\ \bibnamefont
  {Chang}}, \bibinfo {author} {\bibfnamefont {H.}~\bibnamefont {Lin}}, \bibinfo
  {author} {\bibfnamefont {H.-T.}\ \bibnamefont {Jeng}}, \ and\ \bibinfo
  {author} {\bibfnamefont {A.}~\bibnamefont {Bansil}},\ }\href
  {http://dx.doi.org/10.1038/srep06270} {\bibfield  {journal} {\bibinfo
  {journal} {Scientific Reports}\ }\textbf {\bibinfo {volume} {4}},\ \bibinfo
  {pages} {6270 EP } (\bibinfo {year} {2014})}\BibitemShut {NoStop}%
\bibitem [{\citenamefont {Fang}\ \emph {et~al.}(2015)\citenamefont {Fang},
  \citenamefont {Kuate~Defo}, \citenamefont {Shirodkar}, \citenamefont {Lieu},
  \citenamefont {Tritsaris},\ and\ \citenamefont
  {Kaxiras}}]{abinitioTB_prb_2015}%
  \BibitemOpen
  \bibfield  {author} {\bibinfo {author} {\bibfnamefont {S.}~\bibnamefont
  {Fang}}, \bibinfo {author} {\bibfnamefont {R.}~\bibnamefont {Kuate~Defo}},
  \bibinfo {author} {\bibfnamefont {S.~N.}\ \bibnamefont {Shirodkar}}, \bibinfo
  {author} {\bibfnamefont {S.}~\bibnamefont {Lieu}}, \bibinfo {author}
  {\bibfnamefont {G.~A.}\ \bibnamefont {Tritsaris}}, \ and\ \bibinfo {author}
  {\bibfnamefont {E.}~\bibnamefont {Kaxiras}},\ }\href {\doibase
  10.1103/PhysRevB.92.205108} {\bibfield  {journal} {\bibinfo  {journal} {Phys.
  Rev. B}\ }\textbf {\bibinfo {volume} {92}},\ \bibinfo {pages} {205108}
  (\bibinfo {year} {2015})}\BibitemShut {NoStop}%
\bibitem [{\citenamefont {Bradley}\ \emph {et~al.}(2015)\citenamefont
  {Bradley}, \citenamefont {M.~Ugeda}, \citenamefont {da~Jornada},
  \citenamefont {Qiu}, \citenamefont {Ruan}, \citenamefont {Zhang},
  \citenamefont {Wickenburg}, \citenamefont {Riss}, \citenamefont {Lu},
  \citenamefont {Mo}, \citenamefont {Hussain}, \citenamefont {Shen},
  \citenamefont {Louie},\ and\ \citenamefont
  {Crommie}}]{bradley_nanolett_2015}%
  \BibitemOpen
  \bibfield  {author} {\bibinfo {author} {\bibfnamefont {A.~J.}\ \bibnamefont
  {Bradley}}, \bibinfo {author} {\bibfnamefont {M.}~\bibnamefont {M.~Ugeda}},
  \bibinfo {author} {\bibfnamefont {F.~H.}\ \bibnamefont {da~Jornada}},
  \bibinfo {author} {\bibfnamefont {D.~Y.}\ \bibnamefont {Qiu}}, \bibinfo
  {author} {\bibfnamefont {W.}~\bibnamefont {Ruan}}, \bibinfo {author}
  {\bibfnamefont {Y.}~\bibnamefont {Zhang}}, \bibinfo {author} {\bibfnamefont
  {S.}~\bibnamefont {Wickenburg}}, \bibinfo {author} {\bibfnamefont
  {A.}~\bibnamefont {Riss}}, \bibinfo {author} {\bibfnamefont {J.}~\bibnamefont
  {Lu}}, \bibinfo {author} {\bibfnamefont {S.-K.}\ \bibnamefont {Mo}}, \bibinfo
  {author} {\bibfnamefont {Z.}~\bibnamefont {Hussain}}, \bibinfo {author}
  {\bibfnamefont {Z.-X.}\ \bibnamefont {Shen}}, \bibinfo {author}
  {\bibfnamefont {S.~G.}\ \bibnamefont {Louie}}, \ and\ \bibinfo {author}
  {\bibfnamefont {M.~F.}\ \bibnamefont {Crommie}},\ }\href {\doibase
  10.1021/acs.nanolett.5b00160} {\bibfield  {journal} {\bibinfo  {journal}
  {Nano Letters}\ }\textbf {\bibinfo {volume} {15}},\ \bibinfo {pages} {2594}
  (\bibinfo {year} {2015})},\ \bibinfo {note} {pMID: 25775022},\ \Eprint
  {http://arxiv.org/abs/http://dx.doi.org/10.1021/acs.nanolett.5b00160}
  {http://dx.doi.org/10.1021/acs.nanolett.5b00160} \BibitemShut {NoStop}%
\bibitem [{\citenamefont {Sun}\ \emph {et~al.}(2016)\citenamefont {Sun},
  \citenamefont {Wang},\ and\ \citenamefont {Shuai}}]{sun_jchemphys_2016}%
  \BibitemOpen
  \bibfield  {author} {\bibinfo {author} {\bibfnamefont {Y.}~\bibnamefont
  {Sun}}, \bibinfo {author} {\bibfnamefont {D.}~\bibnamefont {Wang}}, \ and\
  \bibinfo {author} {\bibfnamefont {Z.}~\bibnamefont {Shuai}},\ }\href
  {\doibase 10.1021/acs.jpcc.6b08748} {\bibfield  {journal} {\bibinfo
  {journal} {The Journal of Physical Chemistry C}\ }\textbf {\bibinfo {volume}
  {120}},\ \bibinfo {pages} {21866} (\bibinfo {year} {2016})},\ \Eprint
  {http://arxiv.org/abs/http://dx.doi.org/10.1021/acs.jpcc.6b08748}
  {http://dx.doi.org/10.1021/acs.jpcc.6b08748} \BibitemShut {NoStop}%
\bibitem [{\citenamefont {Pisoni}\ \emph {et~al.}(2017)\citenamefont {Pisoni},
  \citenamefont {Lee}, \citenamefont {Overweg}, \citenamefont {Eich},
  \citenamefont {Simonet}, \citenamefont {Watanabe}, \citenamefont {Taniguchi},
  \citenamefont {Gorbachev}, \citenamefont {Ihn},\ and\ \citenamefont
  {Ensslin}}]{shubnikov}%
  \BibitemOpen
  \bibfield  {author} {\bibinfo {author} {\bibfnamefont {R.}~\bibnamefont
  {Pisoni}}, \bibinfo {author} {\bibfnamefont {Y.}~\bibnamefont {Lee}},
  \bibinfo {author} {\bibfnamefont {H.}~\bibnamefont {Overweg}}, \bibinfo
  {author} {\bibfnamefont {M.}~\bibnamefont {Eich}}, \bibinfo {author}
  {\bibfnamefont {P.}~\bibnamefont {Simonet}}, \bibinfo {author} {\bibfnamefont
  {K.}~\bibnamefont {Watanabe}}, \bibinfo {author} {\bibfnamefont
  {T.}~\bibnamefont {Taniguchi}}, \bibinfo {author} {\bibfnamefont
  {R.}~\bibnamefont {Gorbachev}}, \bibinfo {author} {\bibfnamefont
  {T.}~\bibnamefont {Ihn}}, \ and\ \bibinfo {author} {\bibfnamefont
  {K.}~\bibnamefont {Ensslin}},\ }\bibfield  {booktitle} {\emph {\bibinfo
  {booktitle} {Nano Letters}},\ }\href {\doibase 10.1021/acs.nanolett.7b02186}
  {\bibfield  {journal} {\bibinfo  {journal} {Nano Letters}\ }\textbf {\bibinfo
  {volume} {17}},\ \bibinfo {pages} {5008} (\bibinfo {year}
  {2017})}\BibitemShut {NoStop}%
\bibitem [{\citenamefont {Magorrian}\ \emph {et~al.}(2016)\citenamefont
  {Magorrian}, \citenamefont {Z\'olyomi},\ and\ \citenamefont
  {Fal'ko}}]{sam_inse}%
  \BibitemOpen
  \bibfield  {author} {\bibinfo {author} {\bibfnamefont {S.~J.}\ \bibnamefont
  {Magorrian}}, \bibinfo {author} {\bibfnamefont {V.}~\bibnamefont
  {Z\'olyomi}}, \ and\ \bibinfo {author} {\bibfnamefont {V.~I.}\ \bibnamefont
  {Fal'ko}},\ }\href {\doibase 10.1103/PhysRevB.94.245431} {\bibfield
  {journal} {\bibinfo  {journal} {Phys. Rev. B}\ }\textbf {\bibinfo {volume}
  {94}},\ \bibinfo {pages} {245431} (\bibinfo {year} {2016})}\BibitemShut
  {NoStop}%
\bibitem [{\citenamefont {Bandurin}\ \emph {et~al.}(2016)\citenamefont
  {Bandurin}, \citenamefont {Tyurnina}, \citenamefont {Yu}, \citenamefont
  {Mishchenko}, \citenamefont {Z{\'o}lyomi}, \citenamefont {Morozov},
  \citenamefont {Kumar}, \citenamefont {Gorbachev}, \citenamefont {Kudrynskyi},
  \citenamefont {Pezzini}, \citenamefont {Kovalyuk}, \citenamefont {Zeitler},
  \citenamefont {Novoselov}, \citenamefont {Patan{\`e}}, \citenamefont {Eaves},
  \citenamefont {Grigorieva}, \citenamefont {Fal'ko}, \citenamefont {Geim},\
  and\ \citenamefont {Cao}}]{bandurin}%
  \BibitemOpen
  \bibfield  {author} {\bibinfo {author} {\bibfnamefont {D.~A.}\ \bibnamefont
  {Bandurin}}, \bibinfo {author} {\bibfnamefont {A.~V.}\ \bibnamefont
  {Tyurnina}}, \bibinfo {author} {\bibfnamefont {G.~L.}\ \bibnamefont {Yu}},
  \bibinfo {author} {\bibfnamefont {A.}~\bibnamefont {Mishchenko}}, \bibinfo
  {author} {\bibfnamefont {V.}~\bibnamefont {Z{\'o}lyomi}}, \bibinfo {author}
  {\bibfnamefont {S.~V.}\ \bibnamefont {Morozov}}, \bibinfo {author}
  {\bibfnamefont {R.~K.}\ \bibnamefont {Kumar}}, \bibinfo {author}
  {\bibfnamefont {R.~V.}\ \bibnamefont {Gorbachev}}, \bibinfo {author}
  {\bibfnamefont {Z.~R.}\ \bibnamefont {Kudrynskyi}}, \bibinfo {author}
  {\bibfnamefont {S.}~\bibnamefont {Pezzini}}, \bibinfo {author} {\bibfnamefont
  {Z.~D.}\ \bibnamefont {Kovalyuk}}, \bibinfo {author} {\bibfnamefont
  {U.}~\bibnamefont {Zeitler}}, \bibinfo {author} {\bibfnamefont {K.~S.}\
  \bibnamefont {Novoselov}}, \bibinfo {author} {\bibfnamefont {A.}~\bibnamefont
  {Patan{\`e}}}, \bibinfo {author} {\bibfnamefont {L.}~\bibnamefont {Eaves}},
  \bibinfo {author} {\bibfnamefont {I.~V.}\ \bibnamefont {Grigorieva}},
  \bibinfo {author} {\bibfnamefont {V.~I.}\ \bibnamefont {Fal'ko}}, \bibinfo
  {author} {\bibfnamefont {A.~K.}\ \bibnamefont {Geim}}, \ and\ \bibinfo
  {author} {\bibfnamefont {Y.}~\bibnamefont {Cao}},\ }\href {\doibase
  10.1038/nnano.2016.242} {\bibfield  {journal} {\bibinfo  {journal} {Nat.
  Nanotechnol.}\ }\textbf {\bibinfo {volume} {12}},\ \bibinfo {pages} {223}
  (\bibinfo {year} {2016})}\BibitemShut {NoStop}%
\bibitem [{\citenamefont {Korm\'anyos}\ \emph {et~al.}(2013)\citenamefont
  {Korm\'anyos}, \citenamefont {Z\'olyomi}, \citenamefont {Drummond},
  \citenamefont {Rakyta}, \citenamefont {Burkard},\ and\ \citenamefont
  {Fal'ko}}]{kormanyos_prb_2013}%
  \BibitemOpen
  \bibfield  {author} {\bibinfo {author} {\bibfnamefont {A.}~\bibnamefont
  {Korm\'anyos}}, \bibinfo {author} {\bibfnamefont {V.}~\bibnamefont
  {Z\'olyomi}}, \bibinfo {author} {\bibfnamefont {N.~D.}\ \bibnamefont
  {Drummond}}, \bibinfo {author} {\bibfnamefont {P.}~\bibnamefont {Rakyta}},
  \bibinfo {author} {\bibfnamefont {G.}~\bibnamefont {Burkard}}, \ and\
  \bibinfo {author} {\bibfnamefont {V.~I.}\ \bibnamefont {Fal'ko}},\ }\href
  {\doibase 10.1103/PhysRevB.88.045416} {\bibfield  {journal} {\bibinfo
  {journal} {Phys. Rev. B}\ }\textbf {\bibinfo {volume} {88}},\ \bibinfo
  {pages} {045416} (\bibinfo {year} {2013})}\BibitemShut {NoStop}%
\bibitem [{\citenamefont {Korm{\'a}nyos}\ \emph {et~al.}(2015)\citenamefont
  {Korm{\'a}nyos}, \citenamefont {Burkard}, \citenamefont {Gmitra},
  \citenamefont {Fabian}, \citenamefont {Z{\'o}lyomi}, \citenamefont
  {Drummond},\ and\ \citenamefont {Fal'ko}}]{kdotp}%
  \BibitemOpen
  \bibfield  {author} {\bibinfo {author} {\bibfnamefont {A.}~\bibnamefont
  {Korm{\'a}nyos}}, \bibinfo {author} {\bibfnamefont {G.}~\bibnamefont
  {Burkard}}, \bibinfo {author} {\bibfnamefont {M.}~\bibnamefont {Gmitra}},
  \bibinfo {author} {\bibfnamefont {J.}~\bibnamefont {Fabian}}, \bibinfo
  {author} {\bibfnamefont {V.}~\bibnamefont {Z{\'o}lyomi}}, \bibinfo {author}
  {\bibfnamefont {N.~D.}\ \bibnamefont {Drummond}}, \ and\ \bibinfo {author}
  {\bibfnamefont {V.}~\bibnamefont {Fal'ko}},\ }\href
  {http://stacks.iop.org/2053-1583/2/i=2/a=022001} {\bibfield  {journal}
  {\bibinfo  {journal} {2D Materials}\ }\textbf {\bibinfo {volume} {2}},\
  \bibinfo {pages} {022001} (\bibinfo {year} {2015})}\BibitemShut {NoStop}%
\bibitem [{\citenamefont {Danovich}\ \emph {et~al.}(2017)\citenamefont
  {Danovich}, \citenamefont {Aleiner}, \citenamefont {Drummond},\ and\
  \citenamefont {Fal'ko}}]{danovich_phonons}%
  \BibitemOpen
  \bibfield  {author} {\bibinfo {author} {\bibfnamefont {M.}~\bibnamefont
  {Danovich}}, \bibinfo {author} {\bibfnamefont {I.~L.}\ \bibnamefont
  {Aleiner}}, \bibinfo {author} {\bibfnamefont {N.~D.}\ \bibnamefont
  {Drummond}}, \ and\ \bibinfo {author} {\bibfnamefont {V.~I.}\ \bibnamefont
  {Fal'ko}},\ }\href {\doibase 10.1109/JSTQE.2016.2583059} {\bibfield
  {journal} {\bibinfo  {journal} {IEEE Journal of Selected Topics in Quantum
  Electronics}\ }\textbf {\bibinfo {volume} {23}},\ \bibinfo {pages} {168}
  (\bibinfo {year} {2017})}\BibitemShut {NoStop}%
\bibitem [{\citenamefont {Kozawa}\ \emph {et~al.}(2014)\citenamefont {Kozawa},
  \citenamefont {Kumar}, \citenamefont {Carvalho}, \citenamefont {Kumar~Amara},
  \citenamefont {Zhao}, \citenamefont {Wang}, \citenamefont {Toh},
  \citenamefont {Ribeiro}, \citenamefont {Castro~Neto}, \citenamefont
  {Matsuda},\ and\ \citenamefont {Eda}}]{relax_exp}%
  \BibitemOpen
  \bibfield  {author} {\bibinfo {author} {\bibfnamefont {D.}~\bibnamefont
  {Kozawa}}, \bibinfo {author} {\bibfnamefont {R.}~\bibnamefont {Kumar}},
  \bibinfo {author} {\bibfnamefont {A.}~\bibnamefont {Carvalho}}, \bibinfo
  {author} {\bibfnamefont {K.}~\bibnamefont {Kumar~Amara}}, \bibinfo {author}
  {\bibfnamefont {W.}~\bibnamefont {Zhao}}, \bibinfo {author} {\bibfnamefont
  {S.}~\bibnamefont {Wang}}, \bibinfo {author} {\bibfnamefont {M.}~\bibnamefont
  {Toh}}, \bibinfo {author} {\bibfnamefont {R.~M.}\ \bibnamefont {Ribeiro}},
  \bibinfo {author} {\bibfnamefont {A.~H.}\ \bibnamefont {Castro~Neto}},
  \bibinfo {author} {\bibfnamefont {K.}~\bibnamefont {Matsuda}}, \ and\
  \bibinfo {author} {\bibfnamefont {G.}~\bibnamefont {Eda}},\ }\href {\doibase
  10.1038/ncomms5543} {\bibfield  {journal} {\bibinfo  {journal} {Nat.
  Commun.}\ }\textbf {\bibinfo {volume} {5}},\ \bibinfo {pages} {4543}
  (\bibinfo {year} {2014})}\BibitemShut {NoStop}%
\bibitem [{\citenamefont {Unuma}\ \emph {et~al.}(2003)\citenamefont {Unuma},
  \citenamefont {Yoshita}, \citenamefont {Noda}, \citenamefont {Sakaki},\ and\
  \citenamefont {Akiyama}}]{qw1}%
  \BibitemOpen
  \bibfield  {author} {\bibinfo {author} {\bibfnamefont {T.}~\bibnamefont
  {Unuma}}, \bibinfo {author} {\bibfnamefont {M.}~\bibnamefont {Yoshita}},
  \bibinfo {author} {\bibfnamefont {T.}~\bibnamefont {Noda}}, \bibinfo {author}
  {\bibfnamefont {H.}~\bibnamefont {Sakaki}}, \ and\ \bibinfo {author}
  {\bibfnamefont {H.}~\bibnamefont {Akiyama}},\ }\bibfield  {booktitle} {\emph
  {\bibinfo {booktitle} {Journal of Applied Physics}},\ }\href {\doibase
  10.1063/1.1535733} {\bibfield  {journal} {\bibinfo  {journal} {Journal of
  Applied Physics}\ }\textbf {\bibinfo {volume} {93}},\ \bibinfo {pages} {1586}
  (\bibinfo {year} {2003})}\BibitemShut {NoStop}%
\bibitem [{\citenamefont {Mattheiss}(1973)}]{mattheiss_tdms_1973}%
  \BibitemOpen
  \bibfield  {author} {\bibinfo {author} {\bibfnamefont {L.~F.}\ \bibnamefont
  {Mattheiss}},\ }\href {\doibase 10.1103/PhysRevB.8.3719} {\bibfield
  {journal} {\bibinfo  {journal} {Phys. Rev. B}\ }\textbf {\bibinfo {volume}
  {8}},\ \bibinfo {pages} {3719} (\bibinfo {year} {1973})}\BibitemShut
  {NoStop}%
\bibitem [{sup()}]{supplement}%
  \BibitemOpen
  \href@noop {} {}\bibinfo {note} {See
  \href{http://link.aps.org/supplemental/10.1103/PhysRevB.98.035411}{Supplemental
  Material} for detailed DFT band structure results of 2H-stacked $N$-layer
  films ($N=3$ to $6$) of all four transition-metal dichalcogenides discussed
  in this paper, as well as fittings of our hybrid
  $\mathbf{k}\cdot\mathbf{p}$-tight-binding models to DFT data.}\BibitemShut
  {Stop}%
\bibitem [{\citenamefont {Giannozzi}\ \emph {et~al.}(2009)\citenamefont
  {Giannozzi}, \citenamefont {Baroni}, \citenamefont {Bonini}, \citenamefont
  {Calandra}, \citenamefont {Car}, \citenamefont {Cavazzoni}, \citenamefont
  {Ceresoli}, \citenamefont {Chiarotti}, \citenamefont {Cococcioni},
  \citenamefont {Dabo}, \citenamefont {Corso}, \citenamefont {de~Gironcoli},
  \citenamefont {Fabris}, \citenamefont {Fratesi}, \citenamefont {Gebauer},
  \citenamefont {Gerstmann}, \citenamefont {Gougoussis}, \citenamefont
  {Kokalj}, \citenamefont {Lazzeri}, \citenamefont {Martin-Samos},
  \citenamefont {Marzari}, \citenamefont {Mauri}, \citenamefont {Mazzarello},
  \citenamefont {Paolini}, \citenamefont {Pasquarello}, \citenamefont
  {Paulatto}, \citenamefont {Sbraccia}, \citenamefont {Scandolo}, \citenamefont
  {Sclauzero}, \citenamefont {Seitsonen}, \citenamefont {Smogunov},
  \citenamefont {Umari},\ and\ \citenamefont
  {Wentzcovitch}}]{quantum_espresso}%
  \BibitemOpen
  \bibfield  {author} {\bibinfo {author} {\bibfnamefont {P.}~\bibnamefont
  {Giannozzi}}, \bibinfo {author} {\bibfnamefont {S.}~\bibnamefont {Baroni}},
  \bibinfo {author} {\bibfnamefont {N.}~\bibnamefont {Bonini}}, \bibinfo
  {author} {\bibfnamefont {M.}~\bibnamefont {Calandra}}, \bibinfo {author}
  {\bibfnamefont {R.}~\bibnamefont {Car}}, \bibinfo {author} {\bibfnamefont
  {C.}~\bibnamefont {Cavazzoni}}, \bibinfo {author} {\bibfnamefont
  {D.}~\bibnamefont {Ceresoli}}, \bibinfo {author} {\bibfnamefont {G.~L.}\
  \bibnamefont {Chiarotti}}, \bibinfo {author} {\bibfnamefont {M.}~\bibnamefont
  {Cococcioni}}, \bibinfo {author} {\bibfnamefont {I.}~\bibnamefont {Dabo}},
  \bibinfo {author} {\bibfnamefont {A.~D.}\ \bibnamefont {Corso}}, \bibinfo
  {author} {\bibfnamefont {S.}~\bibnamefont {de~Gironcoli}}, \bibinfo {author}
  {\bibfnamefont {S.}~\bibnamefont {Fabris}}, \bibinfo {author} {\bibfnamefont
  {G.}~\bibnamefont {Fratesi}}, \bibinfo {author} {\bibfnamefont
  {R.}~\bibnamefont {Gebauer}}, \bibinfo {author} {\bibfnamefont
  {U.}~\bibnamefont {Gerstmann}}, \bibinfo {author} {\bibfnamefont
  {C.}~\bibnamefont {Gougoussis}}, \bibinfo {author} {\bibfnamefont
  {A.}~\bibnamefont {Kokalj}}, \bibinfo {author} {\bibfnamefont
  {M.}~\bibnamefont {Lazzeri}}, \bibinfo {author} {\bibfnamefont
  {L.}~\bibnamefont {Martin-Samos}}, \bibinfo {author} {\bibfnamefont
  {N.}~\bibnamefont {Marzari}}, \bibinfo {author} {\bibfnamefont
  {F.}~\bibnamefont {Mauri}}, \bibinfo {author} {\bibfnamefont
  {R.}~\bibnamefont {Mazzarello}}, \bibinfo {author} {\bibfnamefont
  {S.}~\bibnamefont {Paolini}}, \bibinfo {author} {\bibfnamefont
  {A.}~\bibnamefont {Pasquarello}}, \bibinfo {author} {\bibfnamefont
  {L.}~\bibnamefont {Paulatto}}, \bibinfo {author} {\bibfnamefont
  {C.}~\bibnamefont {Sbraccia}}, \bibinfo {author} {\bibfnamefont
  {S.}~\bibnamefont {Scandolo}}, \bibinfo {author} {\bibfnamefont
  {G.}~\bibnamefont {Sclauzero}}, \bibinfo {author} {\bibfnamefont {A.~P.}\
  \bibnamefont {Seitsonen}}, \bibinfo {author} {\bibfnamefont {A.}~\bibnamefont
  {Smogunov}}, \bibinfo {author} {\bibfnamefont {P.}~\bibnamefont {Umari}}, \
  and\ \bibinfo {author} {\bibfnamefont {R.~M.}\ \bibnamefont {Wentzcovitch}},\
  }\href {http://stacks.iop.org/0953-8984/21/i=39/a=395502} {\bibfield
  {journal} {\bibinfo  {journal} {Journal of Physics: Condensed Matter}\
  }\textbf {\bibinfo {volume} {21}},\ \bibinfo {pages} {395502} (\bibinfo
  {year} {2009})}\BibitemShut {NoStop}%
\bibitem [{\citenamefont {Perdew}\ and\ \citenamefont {Zunger}(1981)}]{pbe}%
  \BibitemOpen
  \bibfield  {author} {\bibinfo {author} {\bibfnamefont {J.~P.}\ \bibnamefont
  {Perdew}}\ and\ \bibinfo {author} {\bibfnamefont {A.}~\bibnamefont
  {Zunger}},\ }\href {\doibase 10.1103/PhysRevB.23.5048} {\bibfield  {journal}
  {\bibinfo  {journal} {Phys. Rev. B}\ }\textbf {\bibinfo {volume} {23}},\
  \bibinfo {pages} {5048} (\bibinfo {year} {1981})}\BibitemShut {NoStop}%
\bibitem [{\citenamefont {Dal~Corso}(2014)}]{pseudo}%
  \BibitemOpen
  \bibfield  {author} {\bibinfo {author} {\bibfnamefont {A.}~\bibnamefont
  {Dal~Corso}},\ }\href {\doibase
  https://doi.org/10.1016/j.commatsci.2014.07.043} {\bibfield  {journal}
  {\bibinfo  {journal} {Computational Materials Science}\ }\textbf {\bibinfo
  {volume} {95}},\ \bibinfo {pages} {337} (\bibinfo {year} {2014})}\BibitemShut
  {NoStop}%
\bibitem [{\citenamefont {Monkhorst}\ and\ \citenamefont
  {Pack}(1976)}]{kspace}%
  \BibitemOpen
  \bibfield  {author} {\bibinfo {author} {\bibfnamefont {H.~J.}\ \bibnamefont
  {Monkhorst}}\ and\ \bibinfo {author} {\bibfnamefont {J.~D.}\ \bibnamefont
  {Pack}},\ }\href {\doibase 10.1103/PhysRevB.13.5188} {\bibfield  {journal}
  {\bibinfo  {journal} {Phys. Rev. B}\ }\textbf {\bibinfo {volume} {13}},\
  \bibinfo {pages} {5188} (\bibinfo {year} {1976})}\BibitemShut {NoStop}%
\bibitem [{\citenamefont {Methfessel}\ and\ \citenamefont
  {Paxton}(1989)}]{smearing}%
  \BibitemOpen
  \bibfield  {author} {\bibinfo {author} {\bibfnamefont {M.}~\bibnamefont
  {Methfessel}}\ and\ \bibinfo {author} {\bibfnamefont {A.~T.}\ \bibnamefont
  {Paxton}},\ }\href {\doibase 10.1103/PhysRevB.40.3616} {\bibfield  {journal}
  {\bibinfo  {journal} {Phys. Rev. B}\ }\textbf {\bibinfo {volume} {40}},\
  \bibinfo {pages} {3616} (\bibinfo {year} {1989})}\BibitemShut {NoStop}%
\bibitem [{\citenamefont {B\"oker}\ \emph {et~al.}(2001)\citenamefont
  {B\"oker}, \citenamefont {Severin}, \citenamefont {M\"uller}, \citenamefont
  {Janowitz}, \citenamefont {Manzke}, \citenamefont {Vo\ss{}}, \citenamefont
  {Kr\"uger}, \citenamefont {Mazur},\ and\ \citenamefont {Pollmann}}]{mos2_d}%
  \BibitemOpen
  \bibfield  {author} {\bibinfo {author} {\bibfnamefont {T.}~\bibnamefont
  {B\"oker}}, \bibinfo {author} {\bibfnamefont {R.}~\bibnamefont {Severin}},
  \bibinfo {author} {\bibfnamefont {A.}~\bibnamefont {M\"uller}}, \bibinfo
  {author} {\bibfnamefont {C.}~\bibnamefont {Janowitz}}, \bibinfo {author}
  {\bibfnamefont {R.}~\bibnamefont {Manzke}}, \bibinfo {author} {\bibfnamefont
  {D.}~\bibnamefont {Vo\ss{}}}, \bibinfo {author} {\bibfnamefont
  {P.}~\bibnamefont {Kr\"uger}}, \bibinfo {author} {\bibfnamefont
  {A.}~\bibnamefont {Mazur}}, \ and\ \bibinfo {author} {\bibfnamefont
  {J.}~\bibnamefont {Pollmann}},\ }\href {\doibase 10.1103/PhysRevB.64.235305}
  {\bibfield  {journal} {\bibinfo  {journal} {Phys. Rev. B}\ }\textbf {\bibinfo
  {volume} {64}},\ \bibinfo {pages} {235305} (\bibinfo {year}
  {2001})}\BibitemShut {NoStop}%
\bibitem [{\citenamefont {Al-Hilli}\ and\ \citenamefont
  {Evans}(1972)}]{mose2_d}%
  \BibitemOpen
  \bibfield  {author} {\bibinfo {author} {\bibfnamefont {A.~A.}\ \bibnamefont
  {Al-Hilli}}\ and\ \bibinfo {author} {\bibfnamefont {B.~L.}\ \bibnamefont
  {Evans}},\ }\href {\doibase https://doi.org/10.1016/0022-0248(72)90129-7}
  {\bibfield  {journal} {\bibinfo  {journal} {Journal of Crystal Growth}\
  }\textbf {\bibinfo {volume} {15}},\ \bibinfo {pages} {93} (\bibinfo {year}
  {1972})}\BibitemShut {NoStop}%
\bibitem [{\citenamefont {Wieting}(1970)}]{ws2_d}%
  \BibitemOpen
  \bibfield  {author} {\bibinfo {author} {\bibfnamefont {T.~J.}\ \bibnamefont
  {Wieting}},\ }\href {\doibase https://doi.org/10.1016/0022-3697(70)90017-X}
  {\bibfield  {journal} {\bibinfo  {journal} {Journal of Physics and Chemistry
  of Solids}\ }\textbf {\bibinfo {volume} {31}},\ \bibinfo {pages} {2148}
  (\bibinfo {year} {1970})}\BibitemShut {NoStop}%
\bibitem [{\citenamefont {Hicks}(1964)}]{wse2_d}%
  \BibitemOpen
  \bibfield  {author} {\bibinfo {author} {\bibfnamefont {W.~T.}\ \bibnamefont
  {Hicks}},\ }\href {\doibase 10.1149/1.2426317} {\bibfield  {journal}
  {\bibinfo  {journal} {Journal of The Electrochemical Society}\ }\textbf
  {\bibinfo {volume} {111}},\ \bibinfo {pages} {1058} (\bibinfo {year}
  {1964})},\ \Eprint
  {http://arxiv.org/abs/http://jes.ecsdl.org/content/111/9/1058.full.pdf+html}
  {http://jes.ecsdl.org/content/111/9/1058.full.pdf+html} \BibitemShut
  {NoStop}%
\bibitem [{\citenamefont {Magorrian}\ \emph {et~al.}(2018)\citenamefont
  {Magorrian}, \citenamefont {Ceferino}, \citenamefont {Z\'olyomi},\ and\
  \citenamefont {Fal'ko}}]{sam_subbands_2018}%
  \BibitemOpen
  \bibfield  {author} {\bibinfo {author} {\bibfnamefont {S.~J.}\ \bibnamefont
  {Magorrian}}, \bibinfo {author} {\bibfnamefont {A.}~\bibnamefont {Ceferino}},
  \bibinfo {author} {\bibfnamefont {V.}~\bibnamefont {Z\'olyomi}}, \ and\
  \bibinfo {author} {\bibfnamefont {V.~I.}\ \bibnamefont {Fal'ko}},\ }\href
  {\doibase 10.1103/PhysRevB.97.165304} {\bibfield  {journal} {\bibinfo
  {journal} {Phys. Rev. B}\ }\textbf {\bibinfo {volume} {97}},\ \bibinfo
  {pages} {165304} (\bibinfo {year} {2018})}\BibitemShut {NoStop}%
\bibitem [{\citenamefont {Sohier}\ \emph {et~al.}(2016)\citenamefont {Sohier},
  \citenamefont {Calandra},\ and\ \citenamefont {Mauri}}]{calandra_lo}%
  \BibitemOpen
  \bibfield  {author} {\bibinfo {author} {\bibfnamefont {T.}~\bibnamefont
  {Sohier}}, \bibinfo {author} {\bibfnamefont {M.}~\bibnamefont {Calandra}}, \
  and\ \bibinfo {author} {\bibfnamefont {F.}~\bibnamefont {Mauri}},\ }\href
  {\doibase 10.1103/PhysRevB.94.085415} {\bibfield  {journal} {\bibinfo
  {journal} {Phys. Rev. B}\ }\textbf {\bibinfo {volume} {94}},\ \bibinfo
  {pages} {085415} (\bibinfo {year} {2016})}\BibitemShut {NoStop}%
\bibitem [{\citenamefont {Froehlicher}\ \emph {et~al.}(2015)\citenamefont
  {Froehlicher}, \citenamefont {Lorchat}, \citenamefont {Fernique},
  \citenamefont {Joshi}, \citenamefont {Molina-S\'anchez}, \citenamefont
  {Wirtz},\ and\ \citenamefont {Berciaud}}]{multilayer_phonons}%
  \BibitemOpen
  \bibfield  {author} {\bibinfo {author} {\bibfnamefont {G.}~\bibnamefont
  {Froehlicher}}, \bibinfo {author} {\bibfnamefont {E.}~\bibnamefont
  {Lorchat}}, \bibinfo {author} {\bibfnamefont {F.}~\bibnamefont {Fernique}},
  \bibinfo {author} {\bibfnamefont {C.}~\bibnamefont {Joshi}}, \bibinfo
  {author} {\bibfnamefont {A.}~\bibnamefont {Molina-S\'anchez}}, \bibinfo
  {author} {\bibfnamefont {L.}~\bibnamefont {Wirtz}}, \ and\ \bibinfo {author}
  {\bibfnamefont {S.}~\bibnamefont {Berciaud}},\ }\href {\doibase
  10.1021/acs.nanolett.5b02683} {\bibfield  {journal} {\bibinfo  {journal}
  {Nano Letters}\ }\textbf {\bibinfo {volume} {15}},\ \bibinfo {pages} {6481}
  (\bibinfo {year} {2015})},\ \bibinfo {note} {pMID: 26371970},\ \Eprint
  {http://arxiv.org/abs/http://dx.doi.org/10.1021/acs.nanolett.5b02683}
  {http://dx.doi.org/10.1021/acs.nanolett.5b02683} \BibitemShut {NoStop}%
\bibitem [{\citenamefont {Jin}\ \emph {et~al.}(2014)\citenamefont {Jin},
  \citenamefont {Li}, \citenamefont {Mullen},\ and\ \citenamefont
  {Kim}}]{chinese_phonons}%
  \BibitemOpen
  \bibfield  {author} {\bibinfo {author} {\bibfnamefont {Z.}~\bibnamefont
  {Jin}}, \bibinfo {author} {\bibfnamefont {X.}~\bibnamefont {Li}}, \bibinfo
  {author} {\bibfnamefont {J.~T.}\ \bibnamefont {Mullen}}, \ and\ \bibinfo
  {author} {\bibfnamefont {K.~W.}\ \bibnamefont {Kim}},\ }\href {\doibase
  10.1103/PhysRevB.90.045422} {\bibfield  {journal} {\bibinfo  {journal} {Phys.
  Rev. B}\ }\textbf {\bibinfo {volume} {90}},\ \bibinfo {pages} {045422}
  (\bibinfo {year} {2014})}\BibitemShut {NoStop}%
\bibitem [{\citenamefont {Gu}\ and\ \citenamefont
  {Yang}(2014)}]{chinese_phonons2}%
  \BibitemOpen
  \bibfield  {author} {\bibinfo {author} {\bibfnamefont {X.}~\bibnamefont
  {Gu}}\ and\ \bibinfo {author} {\bibfnamefont {R.}~\bibnamefont {Yang}},\
  }\href {\doibase 10.1063/1.4896685} {\bibfield  {journal} {\bibinfo
  {journal} {Applied Physics Letters}\ }\textbf {\bibinfo {volume} {105}},\
  \bibinfo {pages} {131903} (\bibinfo {year} {2014})}\BibitemShut {NoStop}%
\bibitem [{\citenamefont {Gibertini}\ \emph {et~al.}(2014)\citenamefont
  {Gibertini}, \citenamefont {Pellegrino}, \citenamefont {Marzari},\ and\
  \citenamefont {Polini}}]{polini_prb_2014}%
  \BibitemOpen
  \bibfield  {author} {\bibinfo {author} {\bibfnamefont {M.}~\bibnamefont
  {Gibertini}}, \bibinfo {author} {\bibfnamefont {F.~M.~D.}\ \bibnamefont
  {Pellegrino}}, \bibinfo {author} {\bibfnamefont {N.}~\bibnamefont {Marzari}},
  \ and\ \bibinfo {author} {\bibfnamefont {M.}~\bibnamefont {Polini}},\ }\href
  {\doibase 10.1103/PhysRevB.90.245411} {\bibfield  {journal} {\bibinfo
  {journal} {Phys. Rev. B}\ }\textbf {\bibinfo {volume} {90}},\ \bibinfo
  {pages} {245411} (\bibinfo {year} {2014})}\BibitemShut {NoStop}%
\bibitem [{\citenamefont {Wu}\ \emph {et~al.}(2016)\citenamefont {Wu},
  \citenamefont {Xu}, \citenamefont {Lu}, \citenamefont {Khamoshi},
  \citenamefont {Liu}, \citenamefont {Han}, \citenamefont {Wu}, \citenamefont
  {Lin}, \citenamefont {Long}, \citenamefont {He}, \citenamefont {Cai},
  \citenamefont {Yao}, \citenamefont {Zhang},\ and\ \citenamefont
  {Wang}}]{evenodd}%
  \BibitemOpen
  \bibfield  {author} {\bibinfo {author} {\bibfnamefont {Z.}~\bibnamefont
  {Wu}}, \bibinfo {author} {\bibfnamefont {S.}~\bibnamefont {Xu}}, \bibinfo
  {author} {\bibfnamefont {H.}~\bibnamefont {Lu}}, \bibinfo {author}
  {\bibfnamefont {A.}~\bibnamefont {Khamoshi}}, \bibinfo {author}
  {\bibfnamefont {G.-B.}\ \bibnamefont {Liu}}, \bibinfo {author} {\bibfnamefont
  {T.}~\bibnamefont {Han}}, \bibinfo {author} {\bibfnamefont {Y.}~\bibnamefont
  {Wu}}, \bibinfo {author} {\bibfnamefont {J.}~\bibnamefont {Lin}}, \bibinfo
  {author} {\bibfnamefont {G.}~\bibnamefont {Long}}, \bibinfo {author}
  {\bibfnamefont {Y.}~\bibnamefont {He}}, \bibinfo {author} {\bibfnamefont
  {Y.}~\bibnamefont {Cai}}, \bibinfo {author} {\bibfnamefont {Y.}~\bibnamefont
  {Yao}}, \bibinfo {author} {\bibfnamefont {F.}~\bibnamefont {Zhang}}, \ and\
  \bibinfo {author} {\bibfnamefont {N.}~\bibnamefont {Wang}},\ }\href {\doibase
  10.1038/ncomms12955} {\bibfield  {journal} {\bibinfo  {journal} {Nat.
  Commun.}\ }\textbf {\bibinfo {volume} {7}},\ \bibinfo {pages} {12955}
  (\bibinfo {year} {2016})}\BibitemShut {NoStop}%
\bibitem [{\citenamefont {Bernevig}\ and\ \citenamefont
  {Hugues}(2013)}]{bernevig_hughes_book}%
  \BibitemOpen
  \bibfield  {author} {\bibinfo {author} {\bibfnamefont {A.~B.}\ \bibnamefont
  {Bernevig}}\ and\ \bibinfo {author} {\bibfnamefont {T.~L.}\ \bibnamefont
  {Hugues}},\ }\href@noop {} {\emph {\bibinfo {title} {Topological insulators
  and topological superconductors}}}\ (\bibinfo  {publisher} {Princeton
  University Press, Princeton, NJ},\ \bibinfo {year} {2013})\BibitemShut
  {NoStop}%
\bibitem [{\citenamefont {Kaasbjerg}\ \emph {et~al.}(2012)\citenamefont
  {Kaasbjerg}, \citenamefont {Thygesen},\ and\ \citenamefont
  {Jacobsen}}]{danish_LO_phonon}%
  \BibitemOpen
  \bibfield  {author} {\bibinfo {author} {\bibfnamefont {K.}~\bibnamefont
  {Kaasbjerg}}, \bibinfo {author} {\bibfnamefont {K.~S.}\ \bibnamefont
  {Thygesen}}, \ and\ \bibinfo {author} {\bibfnamefont {K.~W.}\ \bibnamefont
  {Jacobsen}},\ }\href {\doibase 10.1103/PhysRevB.85.115317} {\bibfield
  {journal} {\bibinfo  {journal} {Phys. Rev. B}\ }\textbf {\bibinfo {volume}
  {85}},\ \bibinfo {pages} {115317} (\bibinfo {year} {2012})}\BibitemShut
  {NoStop}%
\bibitem [{\citenamefont {Ganchev}\ \emph {et~al.}(2015)\citenamefont
  {Ganchev}, \citenamefont {Drummond}, \citenamefont {Aleiner},\ and\
  \citenamefont {Fal'ko}}]{ganchev}%
  \BibitemOpen
  \bibfield  {author} {\bibinfo {author} {\bibfnamefont {B.}~\bibnamefont
  {Ganchev}}, \bibinfo {author} {\bibfnamefont {N.}~\bibnamefont {Drummond}},
  \bibinfo {author} {\bibfnamefont {I.}~\bibnamefont {Aleiner}}, \ and\
  \bibinfo {author} {\bibfnamefont {V.}~\bibnamefont {Fal'ko}},\ }\href
  {\doibase 10.1103/PhysRevLett.114.107401} {\bibfield  {journal} {\bibinfo
  {journal} {Phys. Rev. Lett.}\ }\textbf {\bibinfo {volume} {114}},\ \bibinfo
  {pages} {107401} (\bibinfo {year} {2015})}\BibitemShut {NoStop}%
\end{thebibliography}%

\end{document}